\documentclass[twocolumn,apj]{openjournal}

\usepackage{xcolor}
\usepackage{color,colortbl}
\usepackage[utf8]{inputenc}
\usepackage{hyperref}
\definecolor{linkcolor}{rgb}{0.0,0.3,0.5}
\hypersetup{
    colorlinks=true,
    linkcolor=linkcolor,
    citecolor=linkcolor,
    filecolor=linkcolor,
    urlcolor=linkcolor,
}
\usepackage{verbatim}
\usepackage{orcidlink}
\graphicspath{{./figures/}}

\newcommand{\sklearn}{\texttt{scikit-learn}}
\newcommand{\ugrizy}{$ugrizy$}
\newcommand{\newtext}[1]{#1}
\providecommand{\software}[1]{\vskip1ex\noindent\textbf{Software:} #1}  

\shorttitle{Multi-Method ML Classification with Rubin DP1 \& DP2}
\shortauthors{Plazas Malag\'on \& Ordenes-Briceño et al.}

\begin{document}

\title{Multi-Method Machine Learning Classification of Astronomical Sources with Rubin Data Previews: a Spectroscopy-Free Approach with DP1 and DP2}

\author{Andr\'es Alejandro Plazas Malag\'on\orcidlink{0000-0002-2598-0514}}
\email{plazas@stanford.edu}
\affiliation{SLAC National Accelerator Laboratory, 2575 Sand Hill Rd., Menlo Park, CA 94025, USA}
\affiliation{Kavli Institute for Particle Astrophysics and Cosmology, Stanford University, Stanford, CA, USA}

\author{Yasna Ordenes-Brice\~no\orcidlink{0000-0001-7966-7606}}
\email{yasna.ordenes@mail.udp.cl}
\affiliation{Instituto de Estudios Astrof\'isicos, Facultad de Ingenier\'ia y Ciencias, Universidad Diego Portales, Av.\ Ej\'ercito Libertador 441, Santiago, Chile}

\author{Dan S. Taranu\orcidlink{0000-0001-6268-1882}}
\email{dtaranu@astro.princeton.edu}
\affiliation{Department of Astrophysical Sciences, Princeton University, Princeton, NJ 08544, USA}

\author{Alex Broughton\orcidlink{0000-0001-6966-5316}}
\email{abrought@stanford.edu}
\affiliation{SLAC National Accelerator Laboratory, 2575 Sand Hill Rd., Menlo Park, CA 94025, USA}
\affiliation{Kavli Institute for Particle Astrophysics and Cosmology, Stanford University, Stanford, CA, USA}

\author{Yumi Choi\orcidlink{0000-0003-1680-1884}}
\email{yumi.choi@noirlab.edu}
\affiliation{NSF National Optical-Infrared Astronomy Research Laboratory, 950 North Cherry Avenue, Tucson, AZ 85719, USA}

\author{Jeffrey L. Carlin\orcidlink{0000-0002-3936-9628}}
\email{jeff.carlin@noirlab.edu}
\affiliation{NSF--DOE Vera C. Rubin Observatory/NSF NOIRLab, 950 N. Cherry Ave., Tucson, AZ 85719, USA}

\author{Sophie Reed\orcidlink{0000-0002-4422-0553}}
\email{sophiereed@princeton.edu} 
\affiliation{Department of Astrophysical Sciences, Princeton University, Princeton, NJ 08544, USA}

\author{Karla Pe\~na Ram\'irez\orcidlink{0000-0002-5855-401X}}
\email{karla.pena@noirlab.edu}
\affiliation{NSF--DOE Vera C. Rubin Observatory/NSF NOIRLab, La Serena, Chile}

\author{Sean Patrick MacBride\orcidlink{0000-0002-9514-7245}}
\email{sean.macbride@physik.uzh.ch}
\affiliation{Physik-Institut, University of Zurich, Winterthurerstrasse 190, 8057 Zurich, Switzerland}

\author{Hurum Maksora Tohfa\orcidlink{0000-0002-8766-6627}}
\email{htohfa@uw.edu}
\affiliation{DIRAC Institute, University of Washington, Seattle, WA 98195, USA}

\author{Arianna Ranabhat\orcidlink{0000-0002-6112-9778}}
\email{ariranabhat@gmail.com}
\affiliation{Australian Astronomical Optics--Macquarie, Macquarie University} 

\begin{abstract}

We present a multi-method machine-learning study for the binary photometric classification of point-like and extended sources in the Rubin Observatory Data Preview~1 (DP1) and Data Preview~2 (DP2). The workflow runs entirely on the Rubin Science Platform using DP1 and DP2 \ugrizy{} photometry, the DP1 ancillary photometric-redshift catalog, and the Monster astrometric and photometric reference catalog, without relying on external spectroscopy. Six methods (SVM, Random Forest, XGBoost, $k$-Nearest Neighbors, an Artificial Neural Network, and a soft Voting ensemble) are benchmarked in the deepest DP1 field, the Extended Chandra Deep Field South (ECDFS), using 97{,}383 objects. These methods reach 10-fold cross-validation macro-$F_{1}$ scores of 0.95--0.97 on the balanced training subset and test macro-$F_{1}$ scores of 0.87--0.92 on the imbalanced held-out set, with the two tree ensembles, Random Forest and the soft Voting ensemble, tied at the top (0.918); the differences among the leading methods are within the statistical uncertainty. Permutation importance identifies full-width at half-maximum as the dominant feature, and the six methods agree on 79.5\% of the full catalog. The same pipeline is then applied without retuning the settings to the two other DP1 fields with full \ugrizy{} coverage and photo-$z$ support, Euclid Deep Field South (EDFS) and the Low Galactic Latitude Field. Together, the three fields span a factor of $\sim 7$ in $E(B-V)$ and $\sim 10$ in stellar density, yet headline macro-$F_{1}$ remains above 0.87 in every field. We then apply the same pipeline to four DP2 fields (ECDFS, COSMOS, EDFS, and ELAIS-S1). Using colors and morphology, the best method reaches a test macro-$F_{1}$ of 0.83 to 0.88 across these fields (highest in EDFS, lowest in the deepest field, COSMOS), and the six methods agree on 77 to 86 per cent of each catalog. Adding the per-band photometric uncertainties as features raises the best macro-$F_{1}$ to about 0.95 in ECDFS DP1 and 0.97 in ECDFS DP2 (and to $\sim0.97$--$0.98$ in the other fields).
The workflow establishes a reproducible, spectroscopy-free baseline for source classification in subsequent Rubin and LSST data releases.

\end{abstract}


\maketitle

\section{Introduction} \label{sec:intro}
The classification of astronomical sources into point-like and extended populations is a basic step in the construction and scientific use of large imaging-survey catalogs \citep{Vasconcellos2011,Fadely2012,Krakowski2016,Nakoneczny2019}. This task is especially important for the Vera C. Rubin Observatory's Legacy Survey of Space and Time (LSST), which will repeatedly image the southern sky in six broad optical bands (\ugrizy{}) over a 10-year survey and produce catalogs of billions of sources \citep{Ivezic2019}. Before the first annual LSST data release, Rubin's Early Science Program will release commissioning data products through a series of Data Previews \citep{RTN011}. Data Preview~1 (DP1), based on observations with the 9-CCD LSST Commissioning Camera (LSSTComCam, $\sim0.4$~deg$^2$) and processed with the LSST Science Pipelines, provides the first opportunity to benchmark source-classification workflows using on-sky Rubin imaging data \citep{RubinDP1,LSSTComCam,LSSTPipelines}. Data Preview~2 (DP2) data products are based on images taken primarily during ``on-sky commissioning", supplemented by overlapping observations through early Jan 2026 with the full 189-CCD LSST Science Camera (LSSTCam, $\sim9.6$~deg$^2$) \cite{RTN115,2025rubn.rept...77P}.

Accurate star--galaxy separation is a cross-cutting requirement for Rubin/LSST science. Many analyses require clean stellar samples, clean galaxy samples, or well-characterized contamination between the two. For example, searches for ultra-faint dwarf galaxies (UFDs) rely on identifying sparse resolved stellar populations against a background of compact and unresolved galaxies \citep{Simon2019,MutluPakdil2021}, and reliable star--galaxy separation at faint magnitudes is essential to the UFD detection efficiency, as shown for the Virgo~III satellite in DP2 by \citet{2026arXiv260725044P}. Conversely, weak-lensing and cosmic-shear analyses require source-galaxy samples with controlled stellar contamination and selection effects, while studies of galactic structure, stellar streams, variable sources, transients, and photometric redshifts can all be affected by source-type misclassification \citep{Soumagnac2015,Mandelbaum2018}.

A common starting point is morphology, because point sources and resolved galaxies differ in size and profile shape. However, purely morphological cuts become increasingly incomplete or contaminated near the survey limit, where compact galaxies, atmospheric seeing, blending, and lower signal-to-noise make the distinction ambiguous \citep{Fadely2012,Soumagnac2015}. Machine-learning (ML) methods offer a complementary approach by combining colors, fluxes, uncertainties, and morphological measurements in a single multi-dimensional feature space. Yet supervised ML also faces a practical limitation: spectroscopic labels are highly non-uniform across the LSST footprint, so classifiers trained on externally assembled spectroscopic samples may inherit field-dependent selection biases.

In this work, we present a spectroscopy-free, cloud-native ML pipeline for binary classification of point-like and extended \newtext{DP1 \& DP2} sources. The workflow runs entirely on the Rubin Science Platform (RSP) at \url{data.lsst.cloud}, the web-based integrated platform where Rubin Data Rights Holders can access Rubin and LSST data. It uses RSP-resident or RSP-accessible data products (and third-party software that can be installed at the RSP): DP1 \newtext{\& DP2} \ugrizy{} photometry from the \texttt{dp1.Object} \newtext{and \texttt{dp2.Object}} table, the DP1 ancillary photometric-redshift (photo-$z$) catalog of \citet{Charles2025} and, for DP2, the RAIL photo-$z$ catalog of \citet{RTN124}, accessed as a Hierarchical Adaptive Tiling Scheme (HATS) catalog with the Large Scale Database (LSDB) package, and the Gaia-anchored Monster astrometric and photometric reference catalog. This design is intended to be reproducible by Rubin users and transferable to LSST fields with adequate multi-band photometry and photo-$z$ support, without requiring local spectroscopic training sets.

We benchmark six widely used classifiers: a Support Vector Machine (SVM), Random Forest (RF), XGBoost, $k$-Nearest Neighbors (KNN), an Artificial Neural Network (ANN), and a Soft Voting ensemble. The primary analysis uses \newtext{the Deep Drilling Fields (DDF)} Extended Chandra Deep Field South (ECDFS), \newtext{available in DP1 \& DP2}, and we then apply the same workflow without retuning the settings for DDF Euclid Deep Field South (EDFS) \newtext{with DP1 \& DP2}, \newtext{DDF COSMOS in DP2}, the Rubin Science Validation (SV) Low Galactic Latitude field \newtext{with DP1, and the ELAIS-S1 Field with DP2. We add the 47~Tucanae crowded field DP1 to test the performance in the star classification}. This cross-field test probes how well the pipeline transfers across different extinction, depth, and stellar-density regimes, \newtext{and the inherent differences between Rubin DP1 and DP2 datasets}. Our study complements recent Rubin analyses, including the spectroscopically trained RF study of \citet{Gatto2026} in ECDFS, the DP1 47~Tucanae analysis of \citet{Choi2025}, which also highlights the limitations of simple morphology based classification in crowded or faint regimes, and the DP2 ultra-faint dwarf study of \citet{2026arXiv260725044P}, whose science depends on faint-end star--galaxy separation.

The goals of this paper are to establish a reproducible DP1 \newtext{\& DP2} baseline for spectroscopy-free source classification, compare the six ML methods on a common Rubin feature set, identify the most informative photometric and morphological features, and quantify cross-field transfer performance. Section~\ref{sec:data} describes the DP1 \newtext{\& DP2} data, quality cuts, extinction corrections, morphology measurements, and feature engineering. Section~\ref{sec:training} presents the training-label construction, and Section~\ref{sec:method} describes the ML methods and validation strategy. Sections~\ref{sec:results}--\ref{sec:xfield} present the ECDFS results, full-catalog application, and cross-field tests. Section~\ref{sec:discussion} discusses limitations, comparison with related work, and implications for future Rubin/LSST releases.

\section{Data and Feature Engineering} \label{sec:data}

\subsection{Field, Observations \& Data Products}
\label{sec:fielddata}

The Vera C.\ Rubin Observatory Data Preview~1 \citep[DP1;][]{RubinDP1} is the first on-sky data release from the NSF--DOE Vera C.\ Rubin Observatory, comprising 1{,}792 science-grade 1$\times$30~s LSSTComCam \citep{10.71929/rubin/2561361} exposures acquired between October 24, 2024 and December 11, 2024 over seven non-contiguous on-sky fields totaling roughly 15~deg$^2$ in all six LSST bands \ugrizy{}. The release contains raw and instrumentally-calibrated single-epoch images, 2{,}644 deep coadds, template coadds, difference images, and source, object, forced-source, difference-image, and Solar System object catalogs, totaling $\sim$3.5~TB and roughly 2.3~million distinct astrophysical objects of which 1.6~million are extended in at least one band.

Of the seven DP1 fields, the Extended Chandra Deep Field South (ECDFS), centered at (RA, Dec) = ($53.13^\circ$, $-28.10^\circ$) \citep{SITCOMTN149}, is the deepest field with full \ugrizy{} coverage; the deep-coadd 5$\sigma$ point-source depths in ECDFS reach $u/g/r/i/z/y \approx 24.5/26.2/26.0/25.7/25.1/23.1$~mag.  All DP1 data were processed by the LSST Science Pipelines \citep{LSSTPipelines, Bosch2018}, with instrument-signature removal \citep{plazas2025}, PSF modeling, astrometric solution against Gaia DR3 \citep{GaiaCollab2021} via The Monster reference catalog \citep{DMTN277}, and the Forward Global Calibration Method \citep[FGCM,][]{burke2018}; the resulting \texttt{deep\_coadd} object catalogs are what we query in Section~\ref{sec:query} \citep[see][for the full pipeline details]{RubinDP1}. DP1 is a commissioning-era preview \citep{RubinDP1}, part of the Rubin Early Science Program \citep{RTN011}, and is made available to the community on a shared-risk basis, with caveats. For example,  \citet{Charles2025} note that the $u$- and $y$-band photometric errors are underestimated and their depths overestimated in this release, and other studies using this data, such as \citet{Crenshaw2026},  drop $u$ and $y$ from their colors in the same data.  In this work we keep $u$ and $y$ in the base feature set because they carry the strongest discriminating power for point/extended classification (Section~\ref{sec:importance}); the caveat is most relevant for the per-band \texttt{psfMagErr} features in the +Errors ablation of Section~\ref{sec:featconfig_results}.

\newtext{Data Preview~2 \citep[DP2;][]{RTN115} is the first Rubin data release based on the full LSSTCam. It reprocesses the science-grade exposures from the LSSTCam on-sky commissioning campaign and the start of early operations, and provides deep coadds and object catalogs in all six \ugrizy{} bands over a substantially larger area than DP1. Because DP2 is a commissioning-era release, the delivered image quality is broader and the depth is less uniform across each field than the design LSST survey \citep{RTN115}. We use three DP2 Deep Drilling Fields that have full \ugrizy{} coverage and a photometric-redshift catalog: ECDFS, which is also a DP1 field and so supports a same-field comparison between the two releases and is the basis of our main analysis; COSMOS, the deepest DP2 field, our primary DP2 benchmark; and ELAIS-S1, an additional transfer field. The DP2 ECDFS depths reach $24.4/25.8/25.5/25.9/24.8/22.2$~mag, so DP2 ECDFS is shallower than DP1 ECDFS in $g$, $r$, $z$, and $y$ and marginally deeper in $i$. The COSMOS deep-coadd has 5$\sigma$ point-source depths reach $u/g/r/i/z/y \approx 25.8/26.7/26.4/25.9/25.0/23.7$~mag. DP2 also adds a continuous model-based classifier, \texttt{model\_extendedness}, provided per band and for the combined $griz$ flux (\texttt{griz\_model\_extendedness}), which we use as an additional morphology baseline in the DP2 analysis (Section~\ref{sec:dp2}).}

\begin{figure*}[t]
\centering
\includegraphics[trim=0cm -0.3cm 0cm 0cm,clip,width=0.95\textwidth]{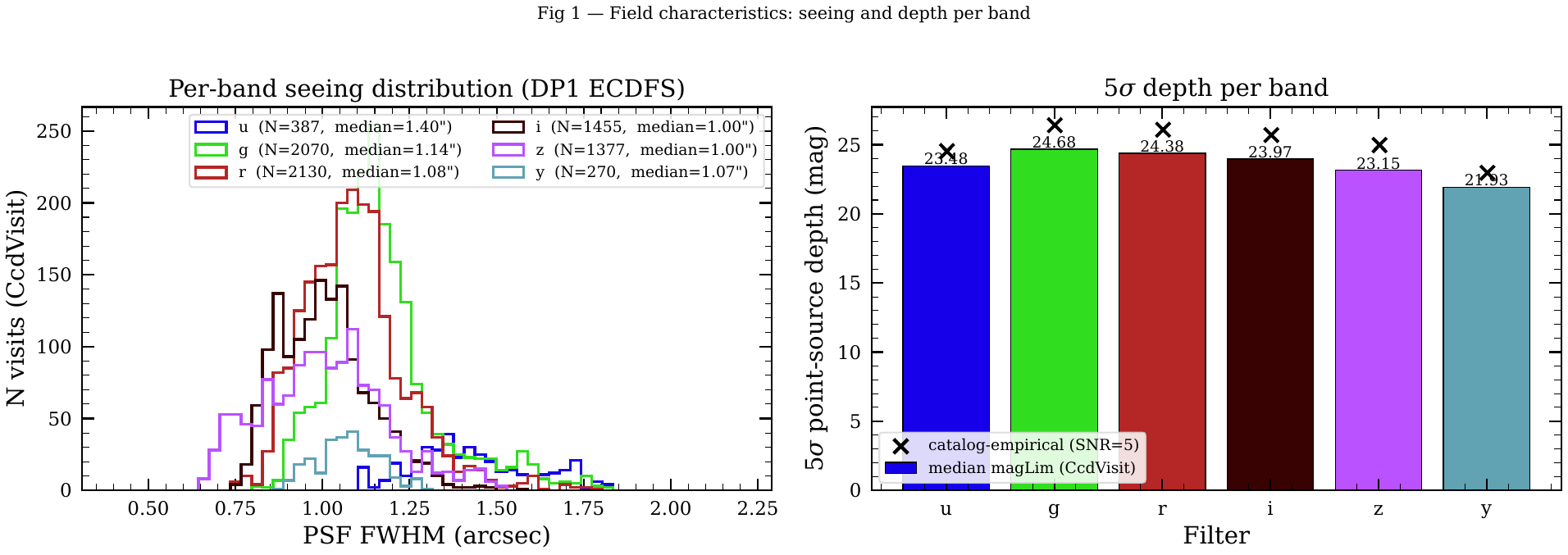}
\includegraphics[width=0.95\textwidth]{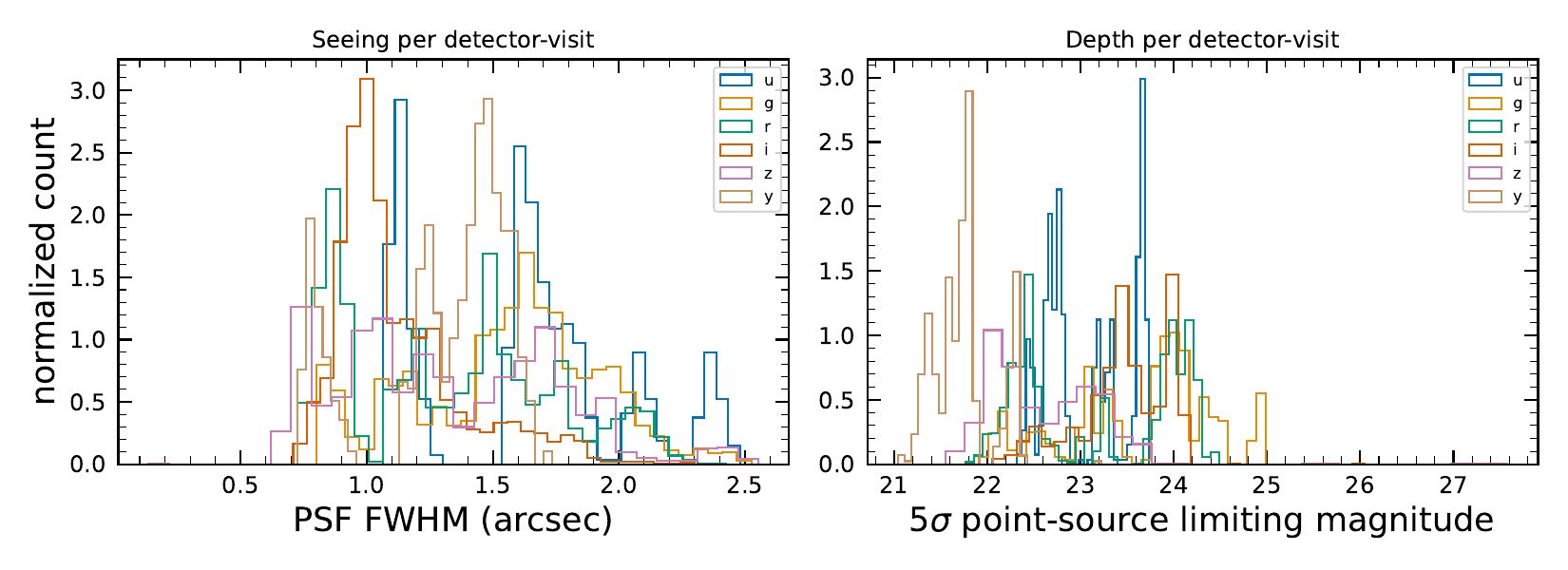}
\caption{ECDFS field characteristics from DP1 \texttt{dp1.CcdVisit} \citep{RubinDP1} top panels and from DP2 \texttt{dp2.CcdVisitDetector} \citep{RTN115}.  Left panels show per-band distribution of PSF FWHM in arcsec, with per-band median annotated in the legend.  Right panels show per-band 5$\sigma$ point-source depth, with bars showing the median \texttt{magLim} column from the visit table and black crosses showing the catalog-empirical estimate (median PSF magnitude where S/N=5).  The figure follows the layout of Figure~1 of \citet{Choi2025} and the ECDFS exploration recipe in DP1 tutorial 301.4 \citep{LSSTTutorials}. The FWHM panels show the \emph{per-visit} delivered image quality; many of the poorest-seeing visits are down-weighted or excluded when the deep coadds are built, so the coadd image quality is better than the tail of the per-visit distribution suggests.}
\label{fig:field_characteristics}
\end{figure*}

Figure~\ref{fig:field_characteristics} characterizes the ECDFS visit history retrieved from the \newtext{DP1 \texttt{dp1.CcdVisit} (top row) and DP2 \texttt{dp2.CcdVisitDetector} (bottom row) tables}. The left panels show the per-band distribution of delivered image quality as the Point Spread Function full-width at half maximum (PSF FWHM, arcsec): per-visit FWHM clusters between $0.8''$ and $2.0''$ with a peak around $1.1''$--$1.3''$ in all bands, consistent with the LSSTComCam on-sky campaign performance reported by \citet{Choi2025} for 47~Tuc (DP1).  In our ECDFS data the $i$-band has the best median seeing of any of the six filters \newtext{for the DP1 and DP2 datasets}; we adopt $i$ as the reference filter for morphological measurements derived from adaptive moments throughout this work, consistent with the \texttt{refBand} assignment for the great majority of sources in the cleaned catalog (see also Section~\ref{sec:morphology_params}).  The right panels of Figure~\ref{fig:field_characteristics} show the per-band 5$\sigma$ point-source depth, computed two ways: median \texttt{magLim} from \texttt{dp1.CcdVisit} or \texttt{dp2.CcdVisitDetector}, and an empirical estimate from the cleaned object catalog (median PSF magnitude of sources with $4.5 <$~S/N~$< 5.5$).  The two estimates agree to better than 0.1~mag in all bands, validating the published per-visit values for our cleaned sample.

\subsection{Data Query and Quality Cuts} \label{sec:query}

We query the \texttt{dp1.Object} table \citep{ObjectTableDOI} via the Table Access Protocol (TAP) service on the RSP for sources within a $1.0^\circ$ radius of the ECDFS field center at (RA, Dec) = ($53.13^\circ$, $-28.10^\circ$).
We retrieve PSF and cModel magnitudes in all six \ugrizy{} bands, along with the two DP1 morphology classifiers \texttt{refExtendedness} (binary, derived from PSF/cModel flux ratios) and \texttt{refSizeExtendedness} (continuous, moments-based; \citealt{LSSTTutorials}) (see Section~\ref{sec:morphology_params}), $i$-band adaptive moments ($I_{xx}$, $I_{yy}$, $I_{xy}$), Kron radius (\texttt{i\_kronRad}), and blendedness (\texttt{i\_blendedness}).
We also retrieve PSF fluxes and flux errors in all bands for signal-to-noise ratio (S/N) computation, as well as quality flags for photometric and pixel-level data quality.
The full Astronomical Data Query Language (ADQL) query run in the Notebook aspect of the RSP is provided in Appendix~\ref{sec:adql}.
For DP2 we run the same query on the \texttt{dp2.Object} table with the same quality cuts, and additionally retrieve the per-band input-visit count (\texttt{\{band\}\_inputCount}) used for the coverage cut of Section~\ref{sec:dp2} and the \texttt{griz\_model\_extendedness} classifier.

We apply quality cuts at the TAP query stage, based on the DP1 baseline flag recommendations in \citet{RubinDP1}; see also the DP1 documentation flag reference\footnote{\url{https://dp1.lsst.io/products/catalogs/flags.html}} (similar recommendations apply for DP2). The cuts are the following:
\begin{itemize}
    \item Non-null PSF flux detections in all six bands;
    \item Clean photometric flags: \texttt{\{band\}\_psfFlux\_flag = 0} for each of the six bands and \texttt{i\_cModel\_flag = 0};
    \item Pixel quality flags: no saturated, cosmic-ray, interpolated, sensor-edge, or coadd-clipped center pixels in the $i$ band (\texttt{i\_pixelFlags\_saturatedCenter}, \texttt{\_crCenter}, \texttt{\_interpolatedCenter}, \texttt{\_sensor\_edgeCenter}, \texttt{\_clippedCenter} all equal to FALSE);
    \item Valid moments-based extendedness flag in the reference band: \texttt{\{refBand\}\_sizeExtendedness\_flag = 0};
    \item Bright-end saturation cuts following \citet{Choi2025}: $g \geq 15.70$, $r \geq 15.08$, $i \geq 15.19$ mag;
\end{itemize}

The initial query in ECDFS retrieves 177{,}097 objects for DP1 and 922{,}265 for DP2. Although DP2 is shallower in ECDFS (Section~\ref{sec:data}), it returns more objects because the wide-field LSSTCam fills the $1.0^\circ$ cone, whereas the small LSSTComCam footprint covered only part of it in DP1.
Following the bright-end saturation cuts (\citet{Choi2025} -style: $g\geq 15.70$, $r\geq 15.08$, $i\geq 15.19$ mag), 164{,}654 (DP1) and 846{,}784 (DP2) objects remain.
Requiring finite PSF magnitudes in all six \ugrizy{} bands yields the clean 18-feature catalog used for all headline results: 97{,}383 objects for DP1 and 489{,}503 for DP2. Relaxing the requirement to $grizy$ (no $u$) or $griz$ (no $u$, $y$), for the band ablation of Section~\ref{sec:ablation}, retains progressively more sources. The analysis thus spans the magnitude range set by these cuts: brighter than the per-band saturation limits ($g\geq 15.70$, $r\geq 15.08$, $i\geq 15.19$) and, for the labeled training sample, to $i_\mathrm{cModel}<23$ for galaxies and the Monster bright limit ($r\lesssim 19$--$21$) for stars. The same cuts are applied to DP1 and DP2, and the per-band $5\sigma$ coadd depths for every field are listed in Table~\ref{tab:crossfield_t1}--\ref{tab:crossfield_t2}.

\subsection{Evaluation Metric} \label{sec:macro-f1}

Throughout this paper, we evaluate performance with the macro-averaged $F_1$: for each class considered, $F_1$ is the harmonic mean of precision (fraction of predicted members that are correct) and recall (fraction of true members recovered), Eq.~\ref{eq:f1}:
\begin{equation}
\mathrm{F1} = \frac{2 \times \mathrm{Precision} \times \mathrm{Recall}}{\mathrm{Precision} + \mathrm{Recall}}. \label{eq:f1}
\end{equation}
The macro average is the unweighted mean of the per-class $F_1$, weighting both classes equally regardless of prevalence.

\subsection{Photometric Error Features} \label{sec:errors}

In addition to the base feature set, we include photometric errors as extended features \citep{Gatto2026}.
We take the per-band PSF magnitude errors \texttt{\{band\}\_psfMagErr} directly from the Object catalog in all six \ugrizy{} bands. For point sources these are equivalent to propagating the PSF flux and its uncertainty via
\begin{equation}
\sigma_m = \frac{2.5}{\ln 10} \frac{\sigma_f}{f},
\end{equation}
where $f$ is the PSF flux and $\sigma_f$ is its uncertainty.
These 6 photometric error features encode the signal-to-noise information of each source, providing the classifiers with an additional handle on source brightness and photometric quality that can improve classification, particularly at faint magnitudes where errors are larger.

\subsection{Galactic Extinction Correction} \label{sec:extinction}

We correct all PSF magnitudes for Galactic extinction using the $E(B-V)$ values from the \citet{SchlegelFinkbeinerDavis1998} dust maps as recalibrated by \citet{SchlaflyFinkbeiner2011}, with the underlying extinction-curve normalization following \citet{Cardelli1989}.
The $E(B-V)$ values are retrieved directly from the \texttt{dp1.Object} and the \texttt{dp2.Object} table via the \texttt{ebv\_sfd98} column.
For the ECDFS field, the extinction is modest, with a mean $E(B-V) \simeq 0.008$~mag and a range of $[0.0061, 0.0145]$~mag.
We compute the extinction in each band as $ A_\lambda = R_\lambda \times E(B-V)$, using the standard Milky Way extinction coefficients $R_\lambda$ from \citet{SchlaflyFinkbeiner2011}.
Extinction-corrected magnitudes are used for all color index computations.
We verified that this correction has a negligible effect on the classification: repeating the base Random Forest with raw, uncorrected colors changes the test macro-$F_{1}$ by only $-0.001$ in ECDFS ($E(B-V) \approx 0.008$) and $+0.002$ in the Low Galactic Latitude field ($E(B-V) \approx 0.044$). Within a $1^\circ$ field the $E(B-V)$ spread is small, so de-reddening applies a nearly uniform color shift to which the tree-based classifier is insensitive, and the morphological features are extinction-independent by construction. We retain the correction for physical accuracy and for cross-field color comparability.

\subsection{Morphology Parameters}
\label{sec:morphology_params}

DP1 and DP2 have morphology classifiers per object \citep{LSSTTutorials}, called \texttt{refExtendedness} and \texttt{refSizeExtendedness} for the different bands available.  The recommended usage is to take the parameters as measured by the reference band. \texttt{refExtendedness} is a binary flag (0 = point-like, 1 = extended) derived from the ratio of PSF to cModel flux in the reference band; it is the traditional flux-ratio star-galaxy separator.  \texttt{refSizeExtendedness} is a continuous score in $[0,1]$ defined as $1 - \exp(-\chi^2/2)$, where $\chi^2$ compares the adaptive-moments trace of the source to the local PSF trace\footnote{The adaptive-moments \emph{trace} is $I_{xx}+I_{yy}$, the sum of the object's second-order spatial light-distribution moments; it measures the object's size, so comparing it to the local PSF trace tests whether the source is more spatially extended than a point source.}; DP1 Notebook tutorial~309.3 \footnote{\url{https://dp1.lsst.io/tutorials}} reports it to be more robust at faint magnitudes than the flux-ratio classifier, because it operates on the source moments rather than on integrated flux differences.  We retain both classifiers as quantities on the object table but do not use either to define galaxy training labels in this work (Section~\ref{sec:training}). DP2 additionally provides \texttt{model\_extendedness}, a continuous classifier that combines the ratio of the S\'ersic-model to PSF-model flux with the exponential-model effective radius, both per band and for the summed $griz$ flux; we use the $griz$ version, \texttt{$griz$\_model\_extendedness}, as a morphology baseline in the DP2 analysis (Section~\ref{sec:dp2}).

In addition we compute three light-profile morphological descriptors from $i$-band measurements, chosen as a reference band with good seeing (see Figure~\ref{fig:field_characteristics}).  The FWHM is derived from the adaptive moments as $\mathrm{FWHM} = 2\sqrt{2\ln 2} \times \sqrt{0.5(I_{xx} + I_{yy})} \times p$, where $p = 0.2''$/pixel is the LSSTComCam and LSSTCam pixel scale in \texttt{deep\_coadd} images.  The simplified ellipticity is $e = 1 - \sqrt{I_{yy}/I_{xx}}$. The Kron radius is a proxy for object size, using an aperture that adapts to the size and light distribution of the source \citep{Kron1980}; in \texttt{SExtractor}-like implementations \citep{BertinArnouts1996} the Kron aperture is defined in units of the semi-major axis, computed from the first moment of the light distribution.
These three descriptors enter to the base 18-feature set (Section~\ref{sec:featconfig}); the two DP1 extendedness scores are used solely for ablation tests\footnote{An \emph{ablation} test removes or adds a single ingredient (one feature, or a group of features) while holding everything else fixed, and measures the resulting change in performance; it isolates how much that ingredient contributes.} (i.e., training with and without the features) to assess their impact in model performance, and are never used as label sources.

\subsection{Color Indices and Color-Color Diagrams} \label{sec:colors}

Following \citet{OrdenesBriceno2026} and \citet{Zeraatgari2024}, we construct all 15 unique color indices from the 6 extinction-corrected \ugrizy{} PSF magnitudes:
$(u\!-\!g)$, $(u\!-\!r)$, $(u\!-\!i)$, $(u\!-\!z)$, $(u\!-\!y)$, $(g\!-\!r)$, $(g\!-\!i)$, $(g\!-\!z)$, $(g\!-\!y)$, $(r\!-\!i)$, $(r\!-\!z)$, $(r\!-\!y)$, $(i\!-\!z)$, $(i\!-\!y)$, and $(z\!-\!y)$.
If $u$- or $y$-band photometry is unavailable for a subset of sources, the corresponding 5-band filter set yields \newtext{10} unique colors, and the $griz$ subset yields \newtext{6}.  Color counts decrease as $n(n-1)/2$ with the number of usable filters; we explore the impact of dropping the $u$- and $y$-band colors in the feature-ablation study of Section~\ref{sec:ablation}.
We use PSF magnitudes for colors because point sources (stars and quasars, QSOs) are unresolved at LSST resolution, and PSF photometry provides the most robust flux estimates for compact objects.

\label{sec:morpho}

\begin{figure*}[t]
\includegraphics[width=0.65\textwidth]{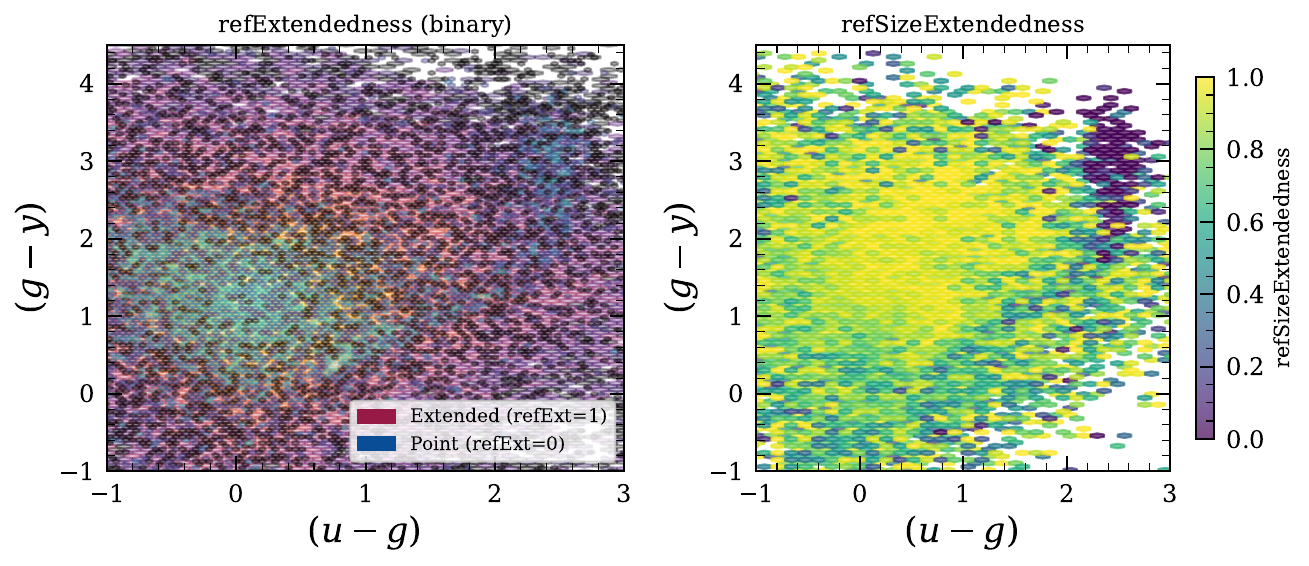}\\
\includegraphics[width=\textwidth]{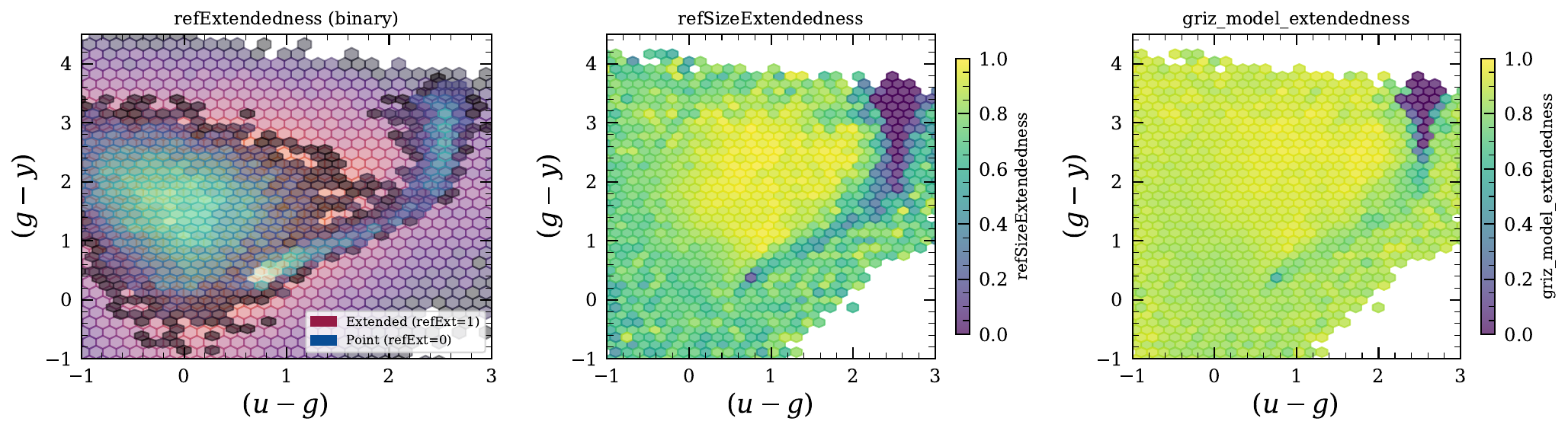}
\caption{$(g-y)$ vs $(u-g)$ color-color diagrams of the cleaned ECDFS catalog for DP1 (top row) and DP2 (bottom row). \textcolor{black}{The panels are color coded by the different parameters available in Rubin DP1 and DP2 to distinguish between point and extended sources}. Left panels shows sources colored by the binary \texttt{refExtendedness} (orange = point, blue = extended).  Middle panels show the same plane colored by the continuous \texttt{refSizeExtendedness} $\in [0,1]$. The bottom right panel is color coded by the continuous \texttt{$griz$\_model\_extendedness} $\in [0,1]$ only available in DP2. In the left column point and extended sources are shown as two overlaid hexbin \emph{density} layers (not a single majority-class color); in the remaining columns each hex cell is colored by the \emph{median} extendedness of the objects it contains. The hex grid size is matched across the DP1 and DP2 panels.}
\label{fig:cc_extcoloring}
\end{figure*}

\begin{figure*}[t]
\centering
\includegraphics[width=0.95\textwidth]{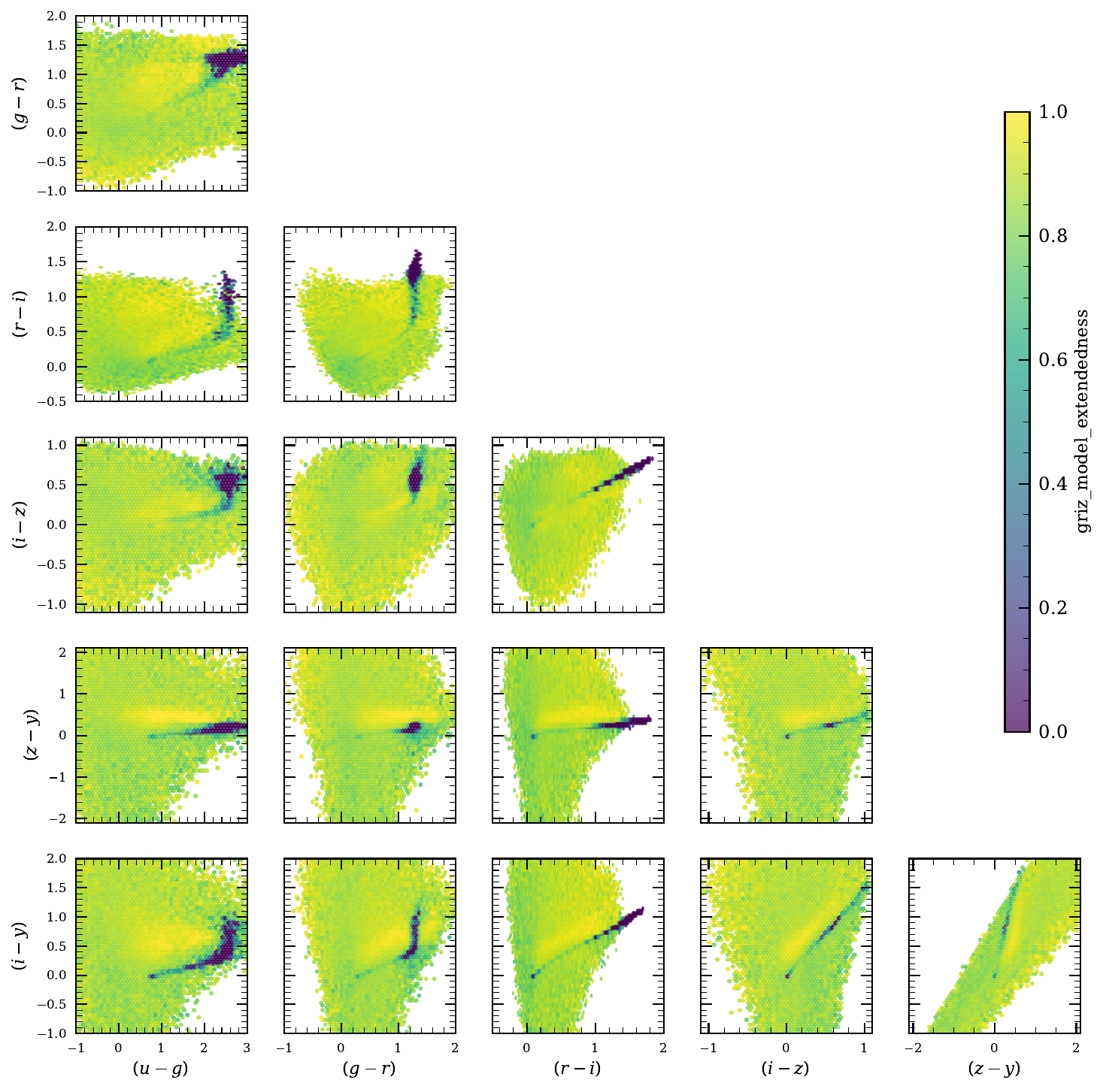}
\caption{Lower-triangular grid of adjacent-band color-color planes for the ECDFS DP2 catalog, colored by \texttt{griz\_model\_extendedness}.  Adjacent-band pairs are $(u-g, g-r, r-i, i-z, z-y, i-y)$; each panel shows the (column, row) plane. }
\label{fig:cc_grid_refsize}
\end{figure*}

Figure~\ref{fig:cc_extcoloring} shows the $(u-g)$ vs $(g-y)$ plane of the cleaned ECDFS catalog (DP1 top, DP2 bottom), with point area encoding the inverse-photometric-error tercile. Each row is colored three ways by the point/extended indicators available in that release: the binary \texttt{refExtendedness} (left; orange point, blue extended) and the continuous \texttt{refSizeExtendedness} and, for DP2, \texttt{$griz$\_model\_extendedness} (each hex colored by the median indicator). In the LSST analog of the $(u'-g')$ vs $(g'-K_s)$ pair of \citet{OrdenesBriceno2026}, stars occupy a narrow curved sequence from $(u-g, g-y) \approx (0.7, 0.5)$ to $(2.7, 3.7)$ while extended sources fill the bluer, more diffuse region and overlap the sequence near its blue end. The sequence is sharpest in the continuous DP2 indicators, confirming the $u$-band as a strong discriminator.

Figure~\ref{fig:cc_grid_refsize} provides a complementary view: a lower-triangular grid of the adjacent-band colors $(u-g, g-r, r-i, i-z, z-y)$ plus $(i-y)$, colored by \texttt{refSizeExtendedness}. The stellar sequence is again the most prominent feature in every panel, tightest in the redder pairs and collapsing to a near-straight line in $(z-y, i-y)$. The $u$-anchored panels show the broader optical-UV locus and a horizontal pile-up at $(u-g) \approx 2$--$3$ from the thick/thin-disk clump---the structure that makes $u$-band colors dominate the permutation importance of Section~\ref{sec:importance}. The $u$-free panels show that some locus/cloud separation survives without the $u$-band, though the sequence is geometrically simpler there and more easily contaminated by faint galaxies lying close to it.

\section{Features and Training Sample Construction} \label{sec:training}

\subsection{Features} \label{sec:featconfig}

The feature engineering for this paper uses the six-filter \ugrizy{} catalog (Section~\ref{sec:query}); ablations (without the presence of some filters) with the $grizy$ and $griz$ subsets are reported in Section~\ref{sec:ablation}. We evaluate four feature configurations that share a common 18-feature \emph{base} set (the 15 unique pairwise \ugrizy{} color indices plus three light-profile morphological descriptors (FWHM, ellipticity, Kron radius) and differ only in which optional features are added on top:

\begin{itemize}
    \item \emph{Base} (18 features): the 15 colors + 3 morphology descriptors. This is the primary configuration used for all headline results unless otherwise stated.
    \item \emph{+errors} (24 features): base + the six per-band PSF magnitude errors, following \citet{Gatto2026}.
    \item \emph{+size\_extendedness} (19 features): base + \texttt{refSizeExtendedness}, treated as an ablation feature (galaxy labels come from the photo-$z$ catalog, not from \texttt{refSizeExtendedness}); this holds for both DP1 and DP2.
    \item \emph{full} (25 features): base + the six magnitude errors + \texttt{refSizeExtendedness}.
\end{itemize}

All four configurations use color indices rather than apparent magnitudes as the photometric features: an object's apparent magnitude depends on its distance and intrinsic luminosity and so does not by itself separate point-like from extended sources, whereas colors are distance-independent spectral energy distribution (SED) proxies. The per-band photometric errors in the \emph{+errors} configuration enter as a signal-to-noise / depth handle, not as brightness features.

\subsection{Training Sample} \label{sec:training_sample}

The training sample combines two RSP-available data products: The Monster reference catalog \citep{DMTN277} for the point-source class, and the DP1 ancillary photometric-redshift catalog \citep{Charles2025} for the extended-source class.
No external spectroscopic catalogs are used. Anchoring galaxy labels on the SED-based photo-z PDF rather than on the moments-based morphology classifier removes the direct label-leakage path through the morphology-correlated features (FWHM, Kron radius) that dominate the base feature set. The morphology classifiers---\texttt{refExtendedness}, \texttt{refSizeExtendedness} (both releases), and \texttt{$griz$\_model\_extendedness} (DP2 only)---are never used as a label source; we instead test \texttt{refSizeExtendedness} as an ablation feature (Section~\ref{sec:featconfig_results}) and use \texttt{$griz$\_model\_extendedness} as a DP2 baseline (Section~\ref{sec:dp2}).

\subsubsection{Class 0: Point Sources (Stars)} \label{sec:stars}

The ECDFS catalog itself is retrieved from \texttt{dp1.Object} via a $1.0^\circ$ cone search around the field center
(Section~\ref{sec:query}); the point-source label is then assigned by cross-matching each retrieved object against The Monster reference catalog within a $1.0''$ match radius.

The Monster reference catalog \citep{DMTN277} contains only stellar sources, combining data from multiple previous reference catalogs. For $grizy$ bands, it uses data from the Dark Energy Survey \citep[DES Y6 - calibration stars;][]{Rykoff2023}, Gaia-XP Synthetic Magnitudes \citep{GaiaCollab2023}, Pan-STARRS1 3pi survey \citep[PS1;][]{Chambers2016}, SkyMapper DR2 \citep{Onken2019} and VST ATLAS survey \citep[VST DR4;][]{Shanks2015}. For $u$-band, it uses Sloan Digital Sky Survey \citep[SDSS DR16 - standard stars;][]{Ahumada2020} with S/N$>5$, Gaia-XP Synthetic Magnitudes \citep{GaiaCollab2023} and Stellar Locus Regression-based magnitudes (SLR) (using SDSS~$u$, DES~$g$ and DES~$g-r$). They are all transformed to a common system. The Monster catalog is used as the all-sky photometric and astrometric reference used to bootstrap the calibration of Rubin Observatory data. Each output band carries a flux, flux-error, and source-flag triplet, so the originating reference catalog for every measurement is traceable. The stellar references are bright (per-band ranges spanning roughly the saturation limit to $r \lesssim 19.7$, $u \lesssim 21.3$); see \citet{DMTN277} for the full per-band ranges and construction details.

The cross-match between the Monster catalog and ECDFS DP1 yields 1{,}284 sources and for DP2 yields 6{,}525 sources.

\subsubsection{Class 1: Extended Sources (Galaxies)} \label{sec:galaxies}

Galaxy labels come from the DP1 photometric-redshift catalog
\citep{Charles2025}, which is hosted in HATS format on the Rubin Science Platform cloud file system and accessed via the LSDB software package \citep{LSDB}, following the example demonstrated in DP1 tutorial notebook 310.1 \footnote{\url{dp1.lsst.io/tutorials}} \citep{LSSTTutorials}.
The catalog publishes point estimates and PDF percentiles for eight
photo-z estimators on a per-source basis; \citet{Charles2025} report FlexZBoost as the best-performing estimator on the ECDFS validation set (scatter $\sigma_{\Delta z/(1+z)} = 0.026$, zero bias on 2{,}905 spec-z galaxies), and recommend a per-estimator quality cut $\sigma_{p(z)} < 0.15$.
We compute $\sigma_{p(z)}$ as half the central 68\% PDF interval,
$\sigma_{p(z)} = (z_{\rm err68\,high} - z_{\rm err68\,low})/2$,
using the FlexZBoost columns of the photo-z catalog. We adopt the \citet{Charles2025} recommended value $\sigma_{p(z)} < 0.15$; stacking-based DP1 analyses such as \citet{Crenshaw2026} use the stricter $\sigma_{p(z)} < 0.10$ on each of two photo-$z$ methods, appropriate for high-purity foreground/background separation in cosmology contexts, while the ML-training use case here benefits from retaining more labeled galaxies.

A source is labeled as an extended training object if it appears in the DP1 photo-z catalog with a reliable estimate
($\sigma_{p(z)}^{\rm FlexZBoost} < 0.15$), lies within the training-set reliability range
$0.02 < z_{\rm mode}^{\rm FlexZBoost} < 1.5$ (the $z > 0.02$ floor excludes the $z \approx 0$ photo-z failure-mode spike, and the $z < 1.5$ ceiling matches the spec-z training depth quoted by \citealt{Charles2025}), and has $i_{\rm cModel} < 23$.

FlexZBoost and $k$NN photo-z estimators agree on the redshift mode within $\Delta z < 0.1$ for 77.3\% of the reliable subset, consistent with the single-estimator scatter and supporting the chosen quality cut. This is an internal consistency check in the spirit of the two-method ($\texttt{FlexZBoost}$\,+\,$\texttt{LePhare}$) agreement cut used by \citet{Crenshaw2026}, although we do not adopt a hard inter-method veto because \citet{Charles2025} report FlexZBoost as the single best-performing DP1 estimator.

For DP2 we use the analogous photometric-redshift catalog of \citet{RTN124}, produced with the RAIL pipeline. It provides six estimators (BPZ, DNF, FlexZBoost, GPz, kNN, TPZ) computed from $1.0''$ Gaussian-aperture (GAaP) fluxes, with separate six-band and four-band models to accommodate the inhomogeneous DP2 coverage; \citet{RTN124} report FlexZBoost as the best-scatter estimator ($\sigma_\mathrm{NMAD} = 0.033$ in six bands), so we again adopt FlexZBoost and the same reliability cuts used for DP1 ($\sigma_{p(z)} < 0.15$, $0.02 < z_\mathrm{mode} < 1.5$, $i_\mathrm{cModel} < 23$) for a like-for-like comparison. The $z < 1.5$ ceiling matches the redshift beyond which \citet{RTN124} caution that the training sample becomes sparse. The catalog is served on the RSP\footnote{Through LSDB at \texttt{/rubin/lsdb\_data/dp2/object\_photoz}, at the time of writing.}, using the same FlexZBoost mode and 68\% interval columns as DP1.

After cross-matching the photo-z catalog to the cleaned \ugrizy{}
object table, applying the FlexZBoost quality cut ($\sigma_{p(z)}<0.15$, $0.02<z_{\rm mode}<1.5$) and the $i_{\rm cModel}<23$ magnitude floor, 12{,}867 sources in DP1 and 83{,}827 in DP2 are labeled as extended.

\subsubsection{Final Training sample}

The resulting training sample for DP1 comprises 14{,}151 labeled objects (1{,}284 stars + 12{,}867 galaxies) and 83{,}232 unclassified objects, for a $\sim 10{:}1$ class imbalance. For DP2 the sample comprises 90{,}352 labeled objects (6{,}525 stars + 83{,}827 galaxies) and 399{,}151 unclassified objects, for a $\sim 13{:}1$ class imbalance. We retain random undersampling of the majority class as the default balancing strategy for consistency with \citet{Zeraatgari2024}; we have verified on the full labeled sample that \texttt{class\_weight = `balanced'} and SMOTE (see Section~\ref{sec:split} for its definition) yield RF macro-$F_{1}$ of $0.941$ and $0.937$ respectively, both slightly above the $0.918$ obtained under the headline undersampling baseline. This $\sim +0.02$ shift is within the classifier-to-classifier spread and changes neither the method ordering nor any conclusion. When an object matches both label sources (a Monster star that also has a reliable photo-$z$), it is assigned the star label; the star and galaxy training sets are disjoint. Because the photo-$z$ catalog is computed for all objects and the only point-source veto is the bright, astrometric Monster match, a faint star or QSO below the Monster limit with a confident but incorrect low-$z$ photo-$z$ can still receive a galaxy label; we bound this residual contamination with an independent Euclid cross-match in Section~\ref{sec:limitations}. In ECDFS DP2, $\sim 41\%$ of the Monster star labels also satisfy the photo-$z$ galaxy criterion and are resolved to star by this precedence rule (equivalently $\sim 3\%$ of the galaxy labels), consistent with the independent Euclid stellar-contamination estimate of Section~\ref{sec:limitations}.

\begin{figure}[t]
\centering
\includegraphics[width=\columnwidth]{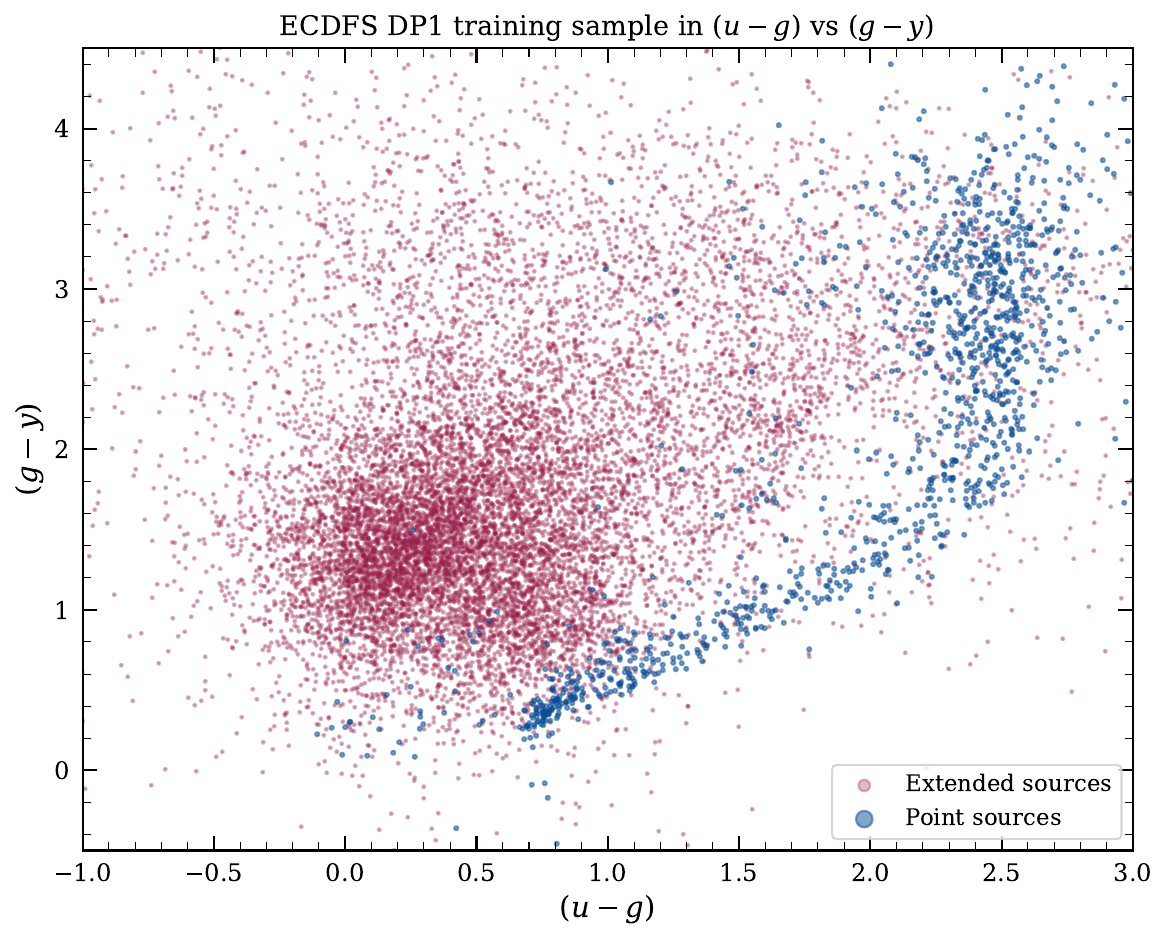}\\
\includegraphics[width=\columnwidth]{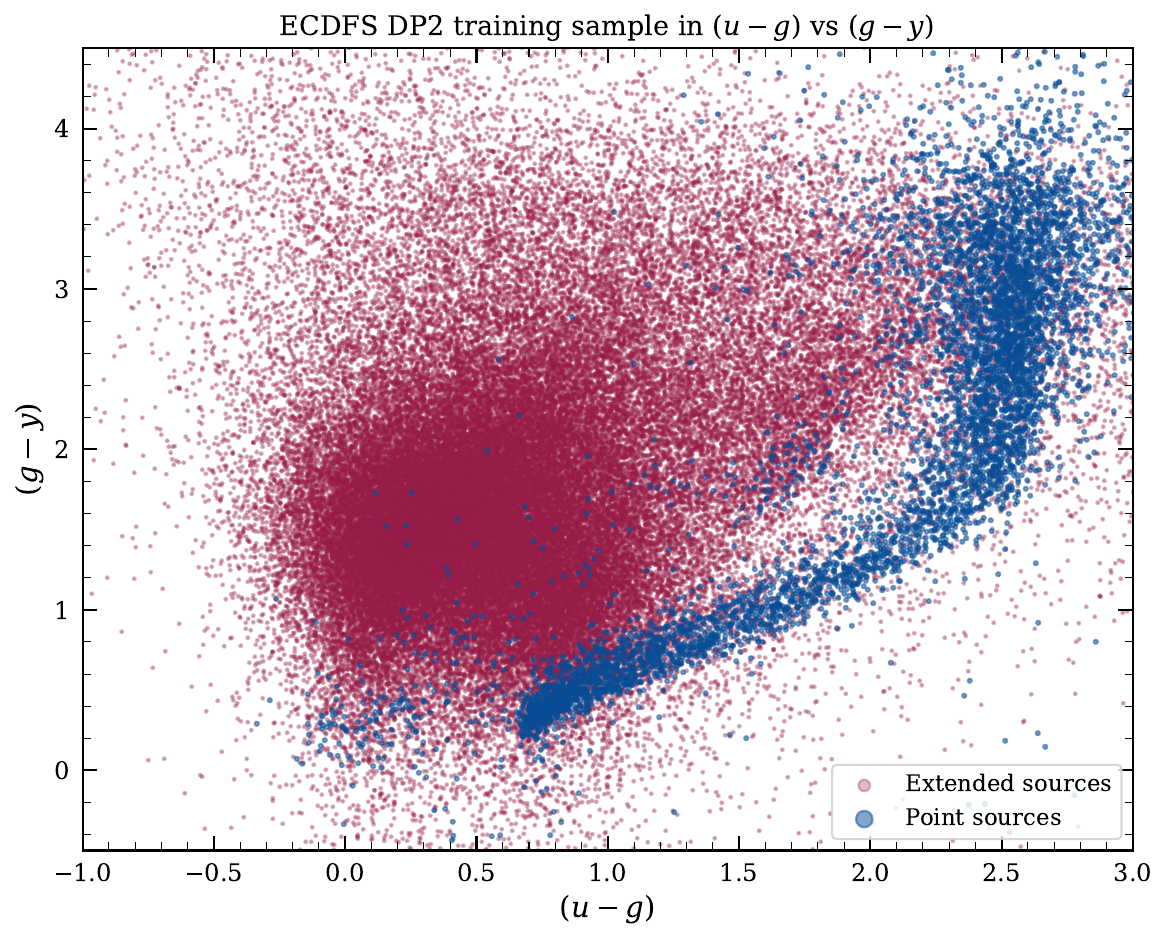}
\caption{Training sample for ECDFS for DP1 (top panel) and DP2 (bottom panel), displayed in the $(u-g)$ vs.\ $(g-y)$ color-color space. The Point sources (orange, cross-match with The Monster catalog) and extended sources (blue, photo-z reliable with $\sigma_{p(z)}<0.15$, $0.02<z<1.5$, $i_{\rm cModel}<23$) are shown, see details in Sect.\ref{sec:training_sample}.
The two label sets are disjoint by construction: any object that satisfies both the Monster (star) and photo-$z$ (galaxy) criteria is assigned the star label (stars take precedence in the labeling code), so no object receives both labels.}
\label{fig:training}
\end{figure}

The training sample is visualized in Figure~\ref{fig:training} in the $(u-g)$ vs.\ $(g-y)$ color-color space, \newtext{in the top panel DP1 dataset, and bottom panel DP2 dataset}.

\section{Multi-Method ML Classification Pipeline} \label{sec:method}

We implement a systematic comparison of six ML classifiers following the methodology of \citet{Zeraatgari2024}, using the Python \sklearn{} library \citep{Pedregosa2011} and XGBoost \citep{ChenGuestrin2016}.

\subsection{Classifiers} \label{sec:classifiers}

The six methods are:
\begin{enumerate}
    \item \textbf{SVM}: Support Vector Machine with RBF kernel \citep{CortesVapnik1995}, with probability estimates via Platt scaling \citep{Platt1999}.
    \item \textbf{RF}: Random Forest, an ensemble of decision trees \citep{Breiman2001}.
    \item \textbf{XGBoost}: Extreme Gradient Boosting \citep{ChenGuestrin2016}, an optimized gradient-boosted tree algorithm that builds trees sequentially, each correcting errors of the previous ensemble.
    \item \textbf{KNN}: $K$-Nearest Neighbors \citep{Cover1967}, a non-parametric instance-based classifier that assigns the majority class among the $K$ nearest training examples in feature space.
    \item \textbf{ANN}: Artificial Neural Network (Multi-Layer Perceptron) with early stopping.
    \item \textbf{Voting}: A soft-voting ensemble combining RF, XGBoost, and SVM with weights [2, 2, 1] (the weighted average of their predicted class probabilities), following \citet{Zeraatgari2024}.
\end{enumerate}
The hyperparameters tuned for each method, and their selected values, are given in Section~\ref{sec:gridsearch} and Table~\ref{tab:bestparams}.

\subsection{Train/Test Split and Class Balancing} \label{sec:split}

The labeled sample are 14{,}151 objects for DP1 and 90{,}352 objects for DP2. They are split 70/30 into training set (DP1: 9{,}905, DP2: 63{,}246) and test set (DP1: 4{,}246, DP2: 27{,}106), using stratified random sampling.
To address the strong class imbalance (Section~\ref{sec:training_sample}; $\sim 10{:}1$ in DP1, $\sim 13{:}1$ in DP2), we explore three balancing strategies:
\begin{itemize}
    \item \textbf{Random undersampling}: Reducing the majority class to match the minority class count (DP1: 899 per class, 1{,}798 total. DP2: 4{,}567 per class, 9{,}134 total), using \texttt{RandomUnderSampler} \citep{Lemaitre2017}.
    This is the default strategy used throughout unless otherwise noted.
    \item \textbf{Class weighting}: Using the parameter \texttt{class\_weight='balanced'} available in SVM, RF, and other classifiers to assign higher weight to the minority class during training.
    \item \textbf{SMOTE}: Synthetic Minority Over-sampling Technique, which generates synthetic minority-class examples by interpolating between existing minority-class instances \citep{Lemaitre2017}.
\end{itemize}
Features are standardized using \texttt{StandardScaler}, fitted on the balanced training set and applied to the test set.

\subsection{Dimensionality Reduction Visualization} \label{sec:dimred}

Following \citet{Zeraatgari2024}, we visualize the training data (from ECDFS DP1 and DP2) in two dimensions using Principal Component Analysis, PCA, \citep{Jolliffe2002} and Uniform Manifold Approximation and Projection, UMAP, \citep{McInnes2018} to assess class separability in the 18-dimensional feature space.

\begin{figure*}[t]
\centering
\includegraphics[trim=0cm 0cm -1cm 0cm,clip,width=0.92\columnwidth]{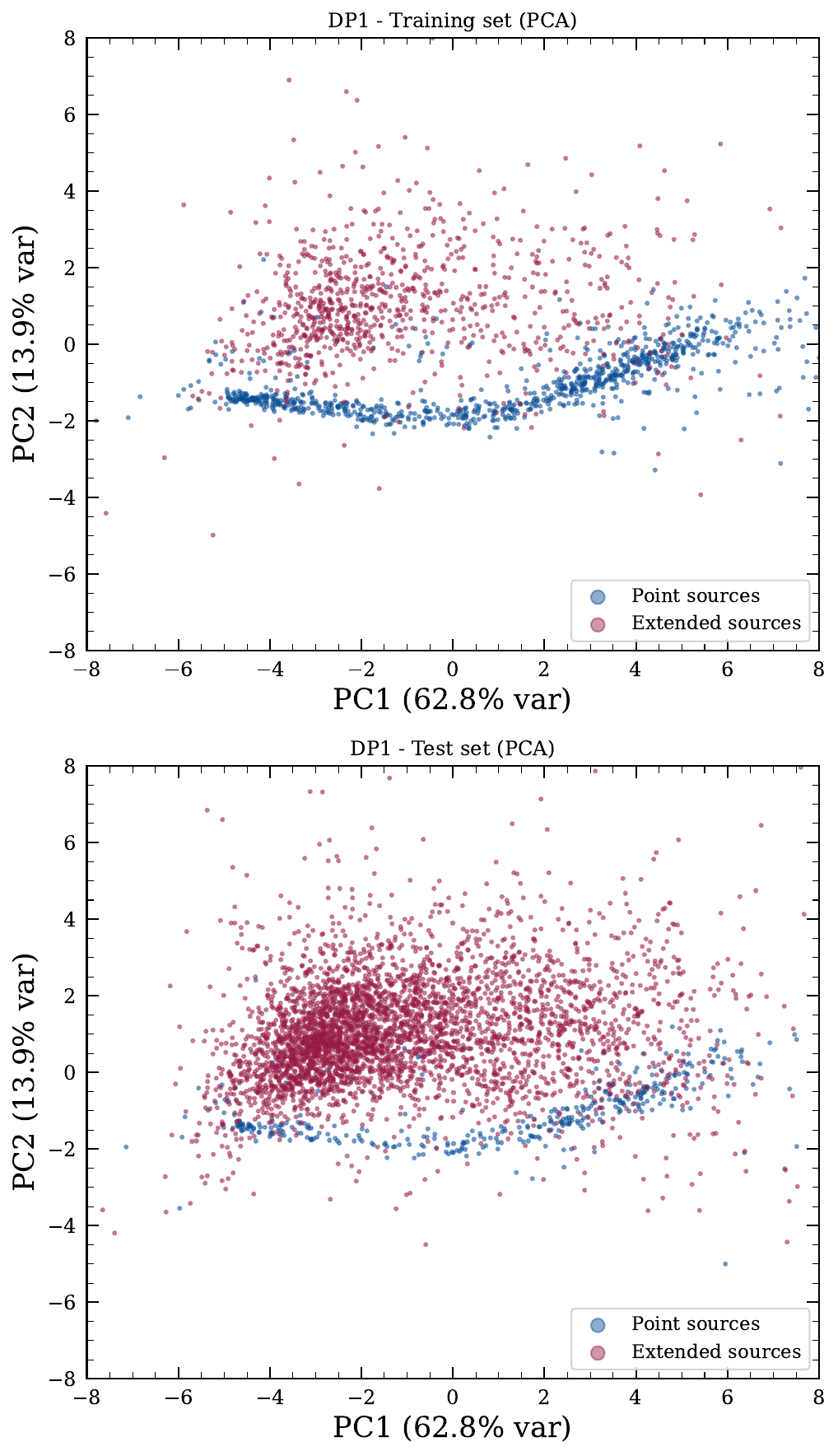}
\includegraphics[trim=0cm 0cm -1cm 0cm,clip,width=0.92\columnwidth]{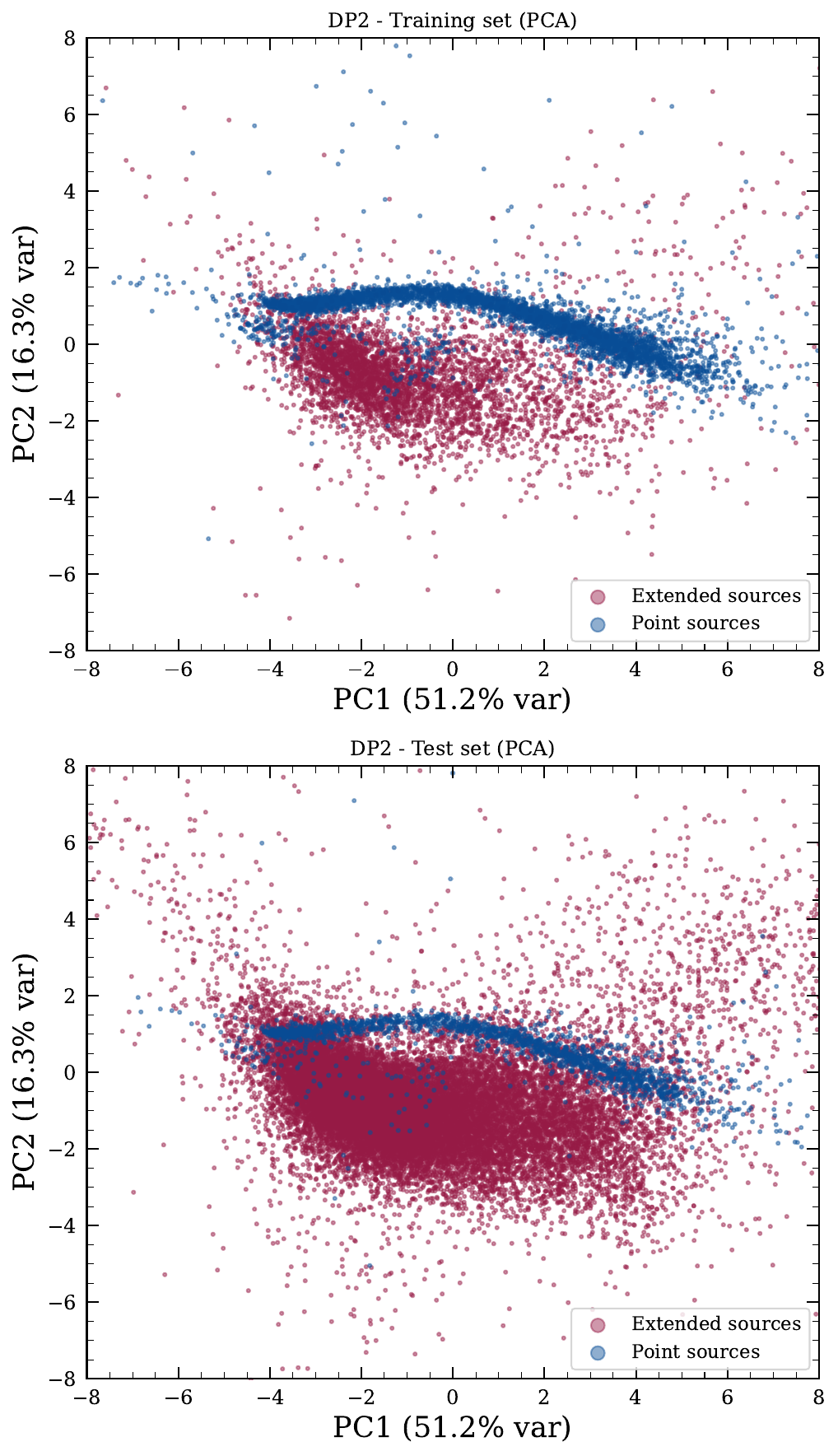}
\caption{PCA visualization of the training (top) and test (bottom) sets in the first two principal components. Left column shows the DP1 data, and the right column shows the DP2 data. Point sources (orange) and extended sources (blue) with a two-dimensional projection retains a significant fraction of the information contained in the original feature space.}
\label{fig:pca}
\end{figure*}

Figure~\ref{fig:pca} shows the PCA projection of the balanced training and test sets left column for DP1, and right columns for DP2. For DP1, the first two principal components capture 62.8\% and 13.9\% of the variance, respectively (76.7\% cumulative). For DP2, captures PC1 51.2\% and PC2 16.3\% of the variance and 67.4\% cumulative. The remaining variance ($\sim 23\%$ in DP1, $\sim 33\%$ in DP2) is spread smoothly across the higher components, with no single dominant component beyond PC2. The larger residual variance in DP2 is consistent with the substantial class overlap seen there.

The PCA transformation reveals that the morphological classes occupy distinct but overlapping regions of feature space. Point sources form a relatively narrow, curved locus, whereas extended sources occupy a substantially broader distribution. The first two principal components account for a high cumulative variance, indicating that the two-dimensional projection retains a significant fraction of the information contained in the original feature space. The similar qualitative structure observed in the training and test samples suggests that this organization is not restricted to the training data. However, the considerable overlap between the two classes indicates that PC1 and PC2 alone are insufficient for perfect morphological separation and that additional principal components and/or nonlinear classification methods may be required.

\begin{figure*}[t]
\centering
\includegraphics[trim=0cm 0cm -1cm 0cm,clip,width=0.92\columnwidth]{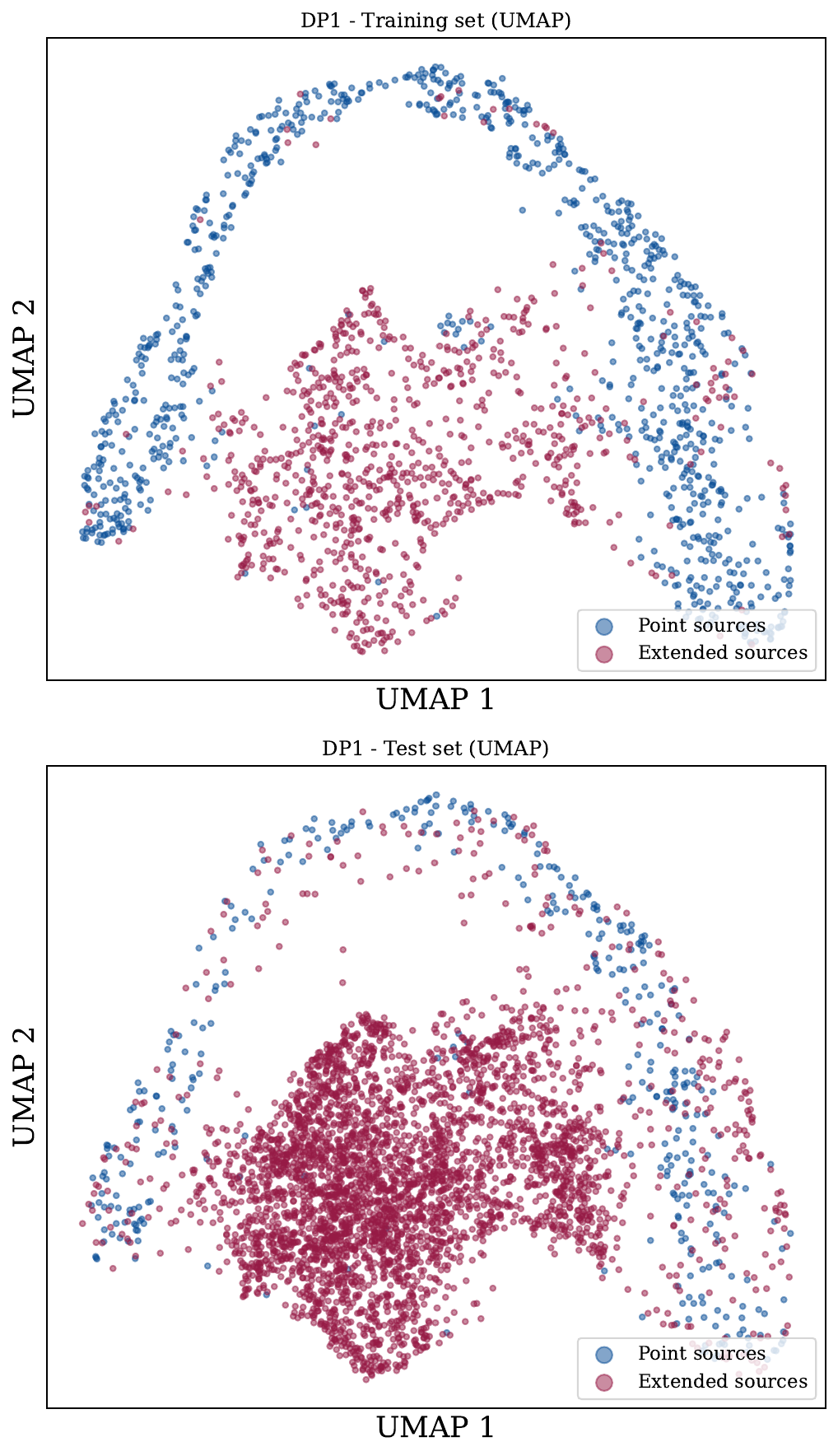}
\includegraphics[trim=0cm 0cm -1cm 0cm,clip,width=0.92\columnwidth]{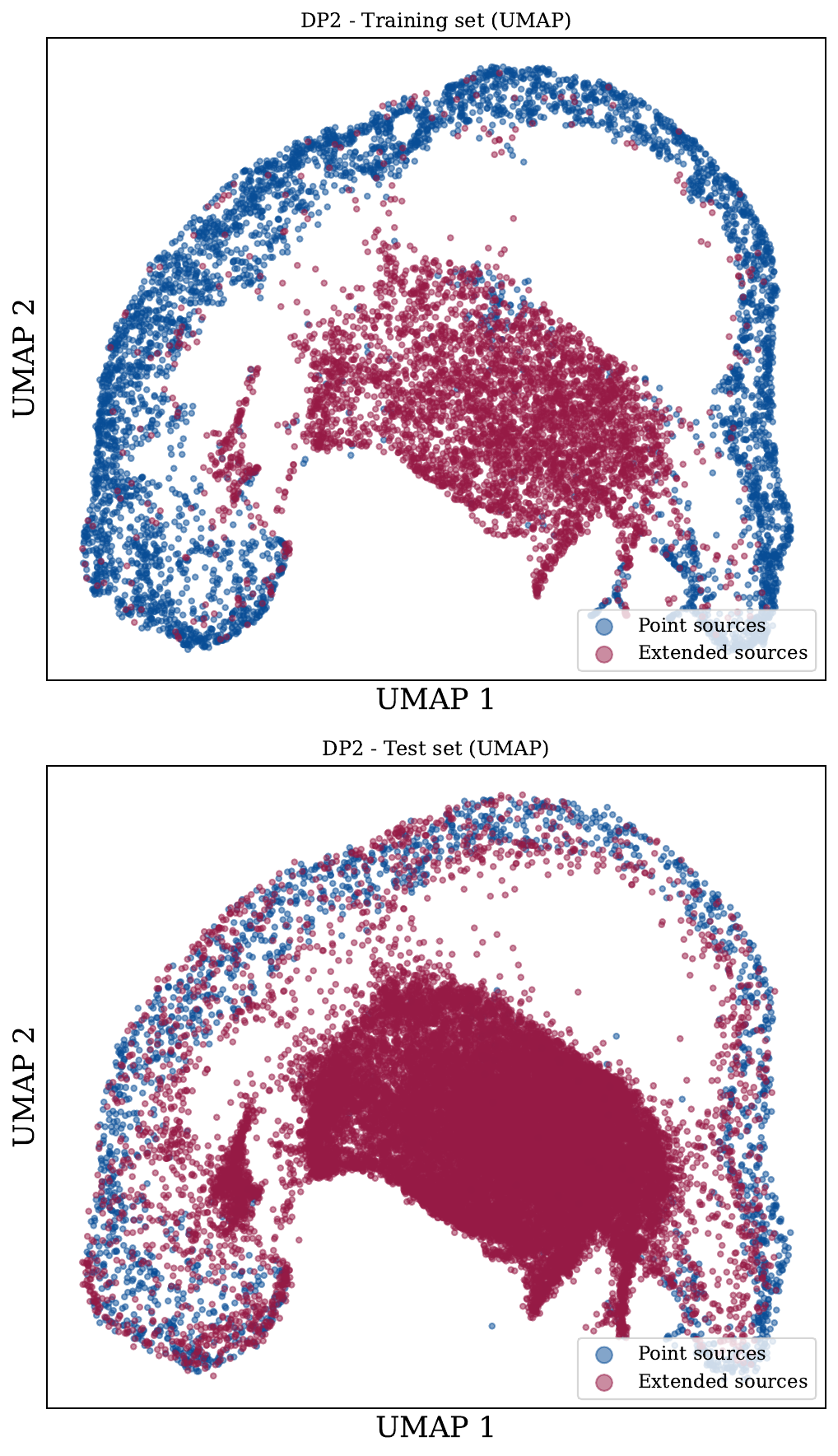}
\caption{UMAP visualization of the training (top) and test (bottom) sets. Left column shows the DP1 data, and the right column shows the DP2 data. Point sources (orange) and extended sources (blue) form distinct clusters, with overlap at the boundaries corresponding to faint, ambiguous sources.}
\label{fig:umap}
\end{figure*}

Figure~\ref{fig:umap} shows the UMAP embedding, which reveals more complex non-linear structure, in the left column DP1 data is shown, and in the right column the DP2. The two classes form clusters that overlap substantially,  especially in the DP2 test set, where the overlap is not confined to the cluster boundaries, with the overlap growing toward faint sources where photometric uncertainties increase.

\subsection{Hyperparameter Optimization} \label{sec:gridsearch}

For each classifier, we perform a grid search over hyperparameters using \texttt{GridSearchCV} with 5-fold stratified cross-validation, optimizing the macro-averaged F1-score. In $k$-fold cross-validation the training set is divided into $k$ equal parts of the same class proportions; each part serves once as a validation set while the model is trained on the other $k-1$ parts, and the $k$ validation scores are averaged. This gives a more stable performance estimate than a single split and guards against over-fitting to one particular partition. Table~\ref{tab:bestparams} reports the best parameters and cross-validation F1-scores for each method for DP1 and DP2. The best-fitting hyperparameters change for some methods between DP1 and DP2, but overall the cross-validation F1-scores range between 0.95 and 0.97 for both releases. The smaller ANN architecture selected for DP2 is simply the grid-search optimum on the larger DP2 training sample; neighboring architectures give indistinguishable cross-validation scores, so this difference should not be over-interpreted.

\begin{table*}[t]
\centering
\caption{Best hyperparameters and 5-fold cross-validation macro F1-scores.}
\label{tab:bestparams}
\begin{tabular}{lcccc}
\hline\hline
Method & DP1 CV F1 & Best parameters & DP2 CV F1 & Best parameters \\
\hline
SVM & 0.965 &$C\!=\!10$, $\gamma\!=`scale'$, RBF & 0.959 & $C\!=\!10$, $\gamma\!=\!0.1$, RBF \\
RF & 0.968 &$n_\mathrm{est}\!=\!100$, depth$=\!20$ & 0.960 & $n_\mathrm{est}\!=\!300$, depth$=\!10$ \\
XGBoost & 0.966 &$n_\mathrm{est}\!=\!200$, lr$=\!0.3$, depth$=\!5$ & 0.963 & $n_\mathrm{est}\!=\!300$, lr$=\!0.1$, depth$=\!6$\\
KNN & 0.952 &$K\!=\!3$, Euclidean, distance & 0.954 & $K\!=\!5$, Euclidean, distance \\
ANN & 0.952 &(100,100), ReLU, $\alpha\!=\!0.0001$ & 0.961 & (64,32), ReLU, $\alpha\!=\!0.0001$ \\
Voting & 0.966  &weights$=\![2,2,1]$, soft & 0.962 & weights$=\![2,2,1]$, soft \\
\hline
\end{tabular}
\end{table*}

\section{Results} \label{sec:results}

\subsection{Classification Performance} \label{sec:performance}

Table~\ref{tab:metrics} presents the per-class and macro-averaged classification metrics for all six methods on the held-out test set, see Section \ref{sec:split} for more details.

\begin{table*}[t]
\centering
\caption{Per-class and macro-averaged classification metrics for DP1 ECDFS (top table) and for DP2 ECDFS (bottom table) on the test set of 4{,}246 and 27{,}106 objects, respectively. Precision (the fraction of objects predicted to be a class that truly are) in the point-source column reflects the strong class imbalance (DP1 $\sim 10{:}1$, DP2 $\sim 13{:}1$) in the test set; for the imbalanced full-catalog setting, AUC is the more informative summary. The typical 68\% bootstrap confidence interval on the macro-$F_{1}$ is $\pm 0.007$ (DP1) and $\pm 0.004$ (DP2); see Section~\ref{sec:performance}.}
\label{tab:metrics}
\setlength{\tabcolsep}{5pt}
\begin{tabular}{l | cccc|  cccc | cccc}
\hline\hline
& \multicolumn{12}{c}{DP1} \\
 & \multicolumn{4}{c}{Point sources} & \multicolumn{4}{c}{Extended sources} & \multicolumn{4}{c}{Macro average} \\
Method \,\,\,\,\,\,\,\, & Prec. & Rec. & F1 & AUC & Prec. & Rec. & F1 & AUC & Prec. & Rec. & F1 & AUC \\
\hline
SVM     & 0.753 & 0.956 & 0.842 & 0.991 & 0.995 & 0.969 & 0.982 & 0.991 & 0.874 & 0.962 & 0.912 & 0.991 \\
RF      & 0.776 & 0.945 & 0.852 & 0.991 & 0.994 & 0.973 & 0.984 & 0.991 & 0.885 & 0.959 & 0.918 & 0.991 \\
XGBoost & 0.755 & 0.945 & 0.840 & 0.990 & 0.994 & 0.969 & 0.982 & 0.990 & 0.875 & 0.957 & 0.911 & 0.990 \\
KNN     & 0.638 & 0.966 & 0.769 & 0.972 & 0.996 & 0.945 & 0.970 & 0.972 & 0.817 & 0.956 & 0.869 & 0.972 \\
ANN     & 0.725 & 0.953 & 0.824 & 0.983 & 0.995 & 0.964 & 0.979 & 0.983 & 0.860 & 0.959 & 0.902 & 0.983 \\
Voting  & 0.777 & 0.943 & 0.852 & 0.993 & 0.994 & 0.973 & 0.984 & 0.993 & 0.886 & 0.958 & 0.918 & 0.993 \\
\hline\hline
& \multicolumn{12}{c}{DP2} \\
 & \multicolumn{4}{c}{Point sources} & \multicolumn{4}{c}{Extended sources} & \multicolumn{4}{c}{Macro average} \\
Method \,\,\,\,\,\,\,\, & Prec. & Rec. & F1 & AUC & Prec. & Rec. & F1 & AUC & Prec. & Rec. & F1 & AUC \\
\hline
 SVM     & 0.586 &0.971 &0.731 &0.985 & 0.998 &0.947 &0.971 &0.985 & 0.792 &0.959 &0.851 &0.985 \\
 RF      & 0.595 &0.965 &0.737 &0.990 & 0.997 &0.949 &0.972 &0.990 & 0.796 &0.957 &0.855 &0.990 \\
 XGBoost & 0.619 &0.966 &0.755 &0.991 & 0.997 &0.954 &0.975 &0.991 & 0.808 &0.960 &0.865 &0.991 \\
 KNN     & 0.547 &0.968 &0.699 &0.977 & 0.997 &0.938 &0.967 &0.977 & 0.772 &0.953 &0.833 &0.977 \\
 ANN     & 0.642 &0.957 &0.769 &0.991 & 0.997 &0.959 &0.977 &0.991 & 0.819 &0.958 &0.873 &0.991 \\
 Voting  & 0.611 &0.970 &0.750 &0.991 & 0.998 &0.952 &0.974 &0.991 & 0.804 &0.961 &0.862 &0.991 \\
\hline
\end{tabular}
\end{table*}

All classifiers achieve AUC\footnote{Area under the ROC curve: a threshold-independent measure of how well a classifier ranks a random point source above a random extended source; 1.0 is perfect, 0.5 random.} scores $\geq 0.97$ in both releases (the lowest is KNN, at 0.972 in DP1 and 0.977 in DP2) for both classes, demonstrating that the classes are well-separated in the 18-dimensional feature space.  Table \ref{tab:metrics} describes the different metrics for point sources, extended sources and macro average for DP1 dataset. The difference in macro-F1 per method is small, but the highest performance is for RF and Voting, followed by SVM, XGBoost, ANN, and KNN. Recall for both classes is high across all methods ($\geq 0.94$ for both point and extended sources). The spread in macro-F1 is driven almost entirely by point-source precision, which is dominated by false positives from the much more numerous extended class in the imbalanced test set.

Cross-validation macro-F1 scores (Table~\ref{tab:bestparams}) are uniformly higher than the test-set values because cross-validation is computed on the class-balanced training subset: DP1: 899 sources and DP2: 4{,}567 per class, whereas the held-out test set retains the $\sim 10:1$ and $\sim 13:1$ class imbalance, respectively. This emphasizes the importance of reporting both balanced and imbalanced metrics in this regime. The spread in test macro-$F_{1}$ among the top four methods ($\lesssim 0.01$) is comparable to the statistical uncertainty set by the small point-source test sample ($\approx 386$ objects in DP1), so no single method is significantly the best. We report bootstrap 68\% confidence intervals on the macro-$F_{1}$: in ECDFS DP1 the 68\% half-width is $\approx 0.007$, so Random Forest and Voting (both 0.918), SVM (0.912), and XGBoost (0.911) overlap within about $1\sigma$ and are statistically indistinguishable, with ANN (0.902) marginally lower and only KNN (0.869) clearly below. The much larger DP2 test set ($27{,}106$ objects) tightens the half-width to $\approx 0.004$; here the leading methods are ANN (0.873) and XGBoost (0.865), with Voting (0.862) and Random Forest (0.855) close behind. We therefore treat the leading methods as statistically comparable rather than identifying a single best model; the nominal best is field-dependent (ANN in ECDFS and COSMOS, XGBoost in EDFS and ELAIS-S1).

\subsection{Confusion Matrices and Performance Comparison} \label{sec:confusion_roc}

The normalized confusion matrices for all six classifiers are shown in Appendix~\ref{sec:CM_roc_fig}, Figure~\ref{fig:confusion}.
Point-source recall (the fraction of true point sources recovered) is uniformly high, from 0.945 (RF) to 0.966 (KNN) in DP1 and from 0.963 (RF) to 0.972 (ANN) in DP2; extended-source recall spans 0.945 (KNN) to 0.973 (Voting) in DP1 and 0.940 (KNN) to 0.955 (XGBoost) in DP2.
The dominant confusion pathway is the misclassification of true extended sources as point sources. In DP1 this rate ranges from 2.7\% (RF) to 5.5\% (KNN) of the extended-class support; in DP2 it ranges from 4.5\% (XGBoost) to 6.0\% (KNN).
Because the extended class is more numerous than the point-source class in the test set, even a small off-diagonal in this direction drives a substantial drop in point-source precision (see Table~\ref{tab:metrics}).

The ROC AUC for the extended-source class (Table~\ref{tab:metrics}) is high across all six classifiers: the tree-ensemble methods (RF, XGBoost, Voting) and SVM reach AUC $> 0.99$ while KNN and ANN sit at $\approx 0.98$; all methods reach average precision $\geq 0.998$.  The near-perfect separation reflects the strong distinction between the two classes in the 18-dimensional feature space, so the practical differences between methods appear in precision/recall trade-offs (see Table~\ref{tab:metrics}) rather than in AUC.

\subsection{Feature Importance} \label{sec:importance}

We analyze feature importance through two complementary approaches: cross-method permutation importance and feature ablation (Sect.\ref{sec:ablation}). Permutation importance measures how much a trained model's macro-$F_{1}$ drops when the values of a single feature are randomly shuffled across objects, which breaks that feature's association with the label while leaving all other features intact; a large drop means the model relied heavily on that feature. We also computed the built-in Gini importances for RF and XGBoost: they rank FWHM and Kron radius as the dominant features ahead of the color indices, qualitatively consistent with the permutation result below \citep{OrdenesBriceno2026}.  Because the model-agnostic permutation analysis covers all six classifiers and is robust to feature correlation, we report it as the headline diagnostic.

The permutation importance \citep{Breiman2001} across all six classifiers, with the colorbar capped at 0.05 to preserve contrast among non-dominant features, is shown in Appendix~\ref{sec:perm_fig}, Figure~\ref{fig:perm_importance}. FWHM is by far the most important feature across all six classifiers, with a six-method average $\Delta F_{1} = 0.043$ that is roughly $5\times$ the next-most-important feature (Kron radius, average $\Delta F_{1} = 0.008$) and dominates especially for the four strong classifiers (SVM, RF, XGBoost, Voting); for KNN and ANN the FWHM contribution is smaller but still the leading non-color feature. The remaining 17 features cluster near or below zero on the average axis, with $(g\!-\!r)$ as the strongest color contributor (average $\Delta F_{1} = 0.008$) and modest positive importance for KNN and ANN specifically. The consistency of FWHM dominance across methods confirms that morphology is the principal discriminator, with color information providing only marginal additional signal once FWHM is available.

The color indices are highly correlated with each other (particularly the $u$-band colors), while the morphological features (FWHM, ellipticity, Kron radius) are weakly correlated with the colors but strongly correlated among themselves; this redundancy among colors is why many individual color indices have low permutation importance, while morphological features remain individually load-bearing.

\subsection{Feature Ablation} \label{sec:ablation}

Following \citet{Zeraatgari2024} and \citet{OrdenesBriceno2026}, we test all classifiers with reduced feature sets (ablation) to assess their impact on model performance,

\begin{itemize}
    \item \textbf{All 18F}: The full feature set (15 colors + 3 morphological).
    \item \textbf{Top-8}: The 8 most important features by average permutation importance: FWHM, $g\!-\!r$, Kron radius, $z\!-\!y$, and others (see Fig.~\ref{fig:perm_importance}).
    \item \textbf{No $u$-band}: Removing all 5 $u$-band colors (13 features).
    \item \textbf{No $y$-band}: Removing all 5 $y$-band colors (13 features).
\end{itemize}

\begin{figure}[t]
\centering
\includegraphics[width=\columnwidth]{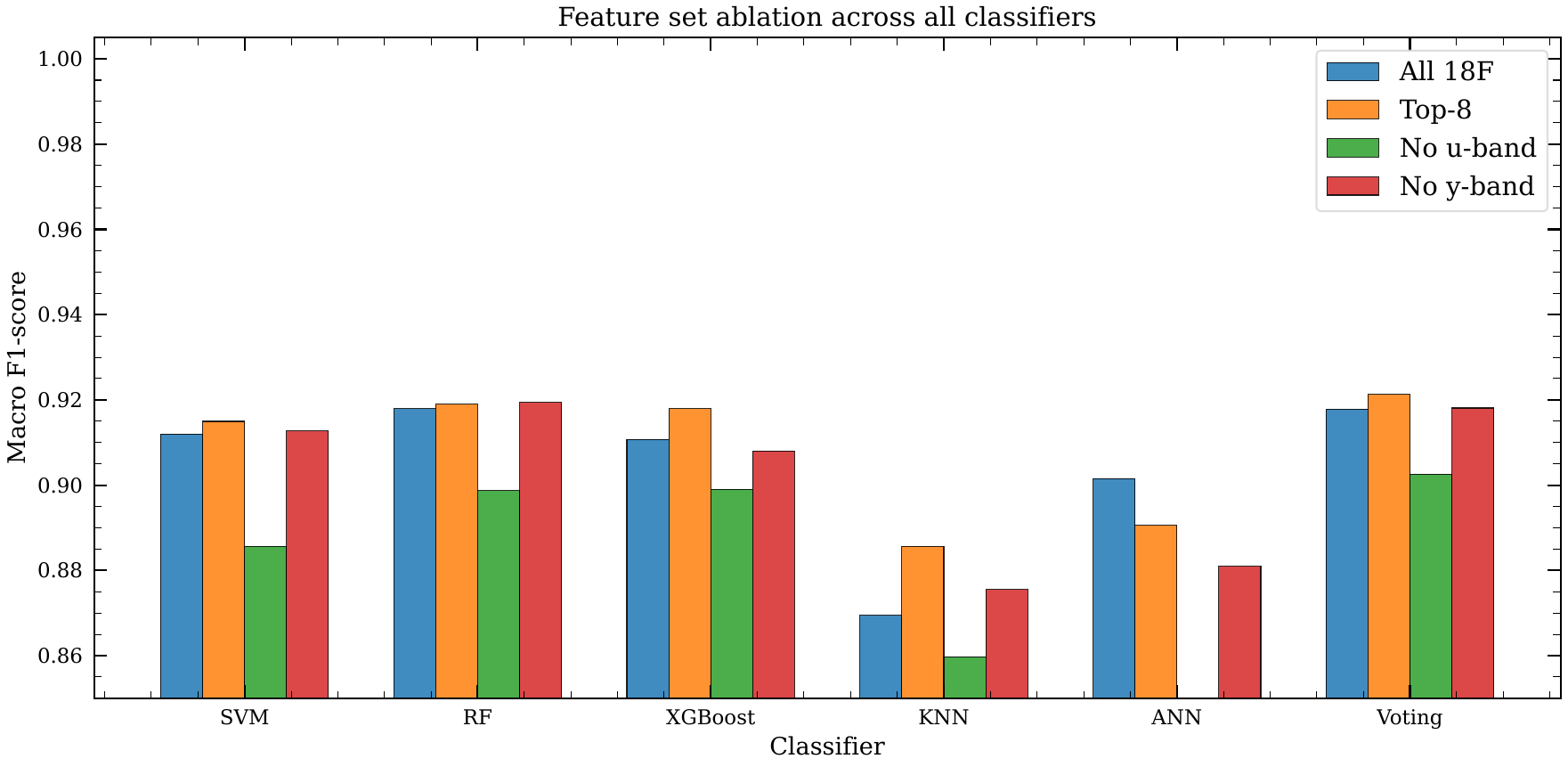}\\
\includegraphics[width=\columnwidth]{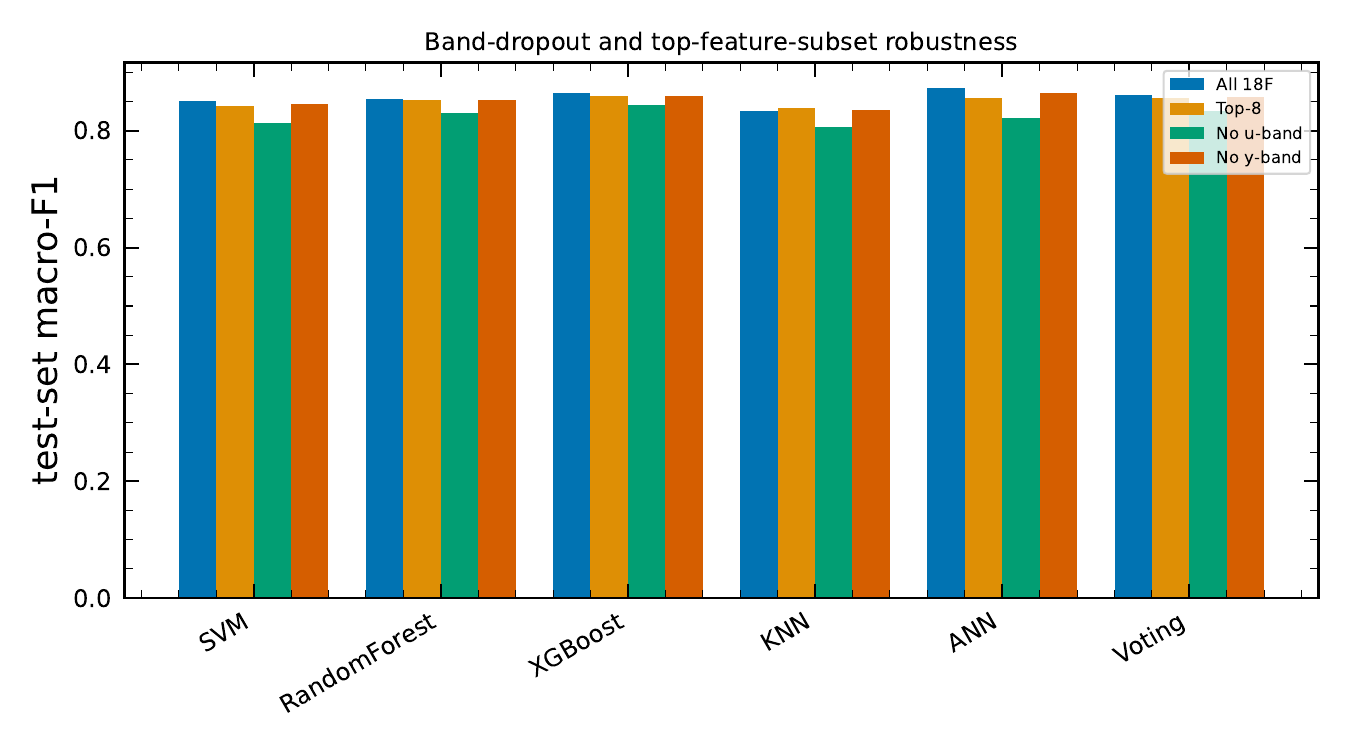}
\caption{Feature ablation results showing macro F1-scores for all classifiers under four feature set configurations. Removing the $u$-band produces the largest degradation, especially for KNN and ANN.}
\label{fig:f1_ablation}
\end{figure}

Table~\ref{tab:ablation} reports and Figure~\ref{fig:f1_ablation} illustrates the macro-$F_{1}$ for each classifier under the four feature sets for ECDFS DP1 and DP2.  The full 18F set and the Top-8 subset provide comparable best performance across methods, with Top-8 marginally exceeding 18F for the models that benefit most from feature selection (SVM, KNN, ANN, Voting) while the tree ensembles (RF, XGBoost) remain within $\Delta F_{1} \lesssim 0.002$ of their 18F values.  Removing $u$-band colors causes the largest degradation, most pronounced for the neural network and support vector methods (ANN $-0.057$, SVM $-0.026$) and smallest for the tree ensembles (RF $-0.019$, XGBoost $-0.012$), confirming the discriminating role of near-ultraviolet (NUV) photometry.  Removing $y$-band colors has a systematically smaller effect (largest drop ANN $-0.021$), consistent with the finding of \citet{OrdenesBriceno2026} that $y$ provides less discriminating power than $u$ for LSST \ugrizy{} data.

\begin{table}[t]
\centering
\caption{Macro $F_{1}$ under feature ablation for all classifiers for ECDFS DP1 (top) and DP2 (bottom).}
\label{tab:ablation}
\begin{tabular}{lcccccc}
\hline\hline
\multicolumn{7}{c}{DP1} \\
Feature set & SVM & RF & XGB & KNN & ANN & Voting \\
\hline
All 18F     & 0.912 & 0.918 & 0.911 & 0.869 & 0.902 & 0.918 \\
Top-8       & 0.915 & 0.919 & 0.918 & 0.886 & 0.891 & 0.921 \\
No $u$-band & 0.886 & 0.899 & 0.899 & 0.860 & 0.845 & 0.903 \\
No $y$-band & 0.913 & 0.919 & 0.908 & 0.876 & 0.881 & 0.918 \\
\hline\hline
\multicolumn{7}{c}{DP2} \\
Feature set & SVM & RF & XGB & KNN & ANN & Voting \\
\hline
All 18F     & 0.851 & 0.855 & 0.865 & 0.833 & 0.873 & 0.862 \\
Top-8       & 0.843 & 0.853 & 0.860 & 0.838 & 0.855 & 0.856 \\
No $u$-band & 0.812 & 0.831 & 0.844 & 0.807 & 0.822 & 0.834 \\
No $y$-band & 0.846 & 0.853 & 0.860 & 0.835 & 0.864 & 0.857 \\
\hline
\end{tabular}
\end{table}

\subsection{Feature Configuration Comparison} \label{sec:featconfig_results}

Table~\ref{tab:featconfig} shows the RF macro F1-score for the four feature configurations described in Section~\ref{sec:featconfig}.
The base 18F configuration (which excludes \texttt{refSizeExtendedness}) yields F1 $= 0.922$.
Adding \texttt{refSizeExtendedness} as a feature (+size\_extendedness, 19F) leaves F1 essentially unchanged at 0.922 (and in DP2 changes it only marginally, from 0.855 to 0.852, i.e.\ essentially unchanged; in both releases the base light-profile morphology already captures the size information that \texttt{refSizeExtendedness} would add, so the small DP2 shift is not significant), reflecting the fact that the base morphology features (FWHM, ellipticity, Kron radius) already carry most of the light-profile information that \texttt{refSizeExtendedness} would add.
Adding the six photometric magnitude errors (+Errors, 24F) improves F1 from 0.922 to 0.951, a gain of $\sim +0.030$ driven by improved point-source recall at faint magnitudes where the per-band flux uncertainty is the most informative axis.
The smaller gain that \citet{Gatto2026} report from the same feature addition on a spectroscopically labeled sample, and the likely reason for our larger amplitude, are discussed in Section~\ref{sec:gatto_comparison}.
Including both errors and \texttt{refSizeExtendedness} (+Full, 25F) gives F1 $= 0.958$, the highest of the four configurations, though the marginal gain over the +errors-only configuration is only $+0.007$. A feature-variant ablation confirms that the improvement comes from signal-to-noise information rather than the specific error parameterization: replacing the per-band magnitude errors with propagated \emph{color} errors, or with a per-band PSF signal-to-noise proxy ($\mathrm{S/N}\simeq 1.0857/\sigma_m$), gives essentially the same Random Forest macro-$F_{1}$ (within $\sim 0.001$ of the +errors value in both ECDFS and COSMOS DP2). The color-error variant is the most consistent with our colors-only base feature set.
We nevertheless retain the base 18F configuration (colors + light-profile morphology) as the headline throughout this paper because it is the most transferable choice and does not depend on the per-band \texttt{psfMagErr} columns, which are underestimated for the $u$ and $y$ bands in this commissioning-era release.

The magnitude-dependent behavior of the four configurations is very similar over the $i \lesssim 24$ range where the Gaia-anchored test set has support: stellar recovery remains $\approx 1.0$ in every bin, and the per-bin differences in galaxy misclassification are small ($\Delta F_{1} \lesssim 0.01$ around the corresponding integrated values in Table~\ref{tab:featconfig}).

\begin{table}[t]
\centering
\caption{DP1 and DP2 Random Forest macro F1-score for different feature configurations. The Random Forest is re-trained on each configuration; for DP1 this uses a fixed representative RF, so its base value (0.922) differs slightly from the tuned headline RF of Table~\ref{tab:metrics} (0.918), while for DP2 the tuned RF is used throughout (base 0.855).}
\label{tab:featconfig}
\begin{tabular}{lccc}
\hline\hline
Configuration & N$_F$ & DP1 & DP2\\
 &  &  F1$_{macro}$ & F1$_{macro}$\\
\hline
Base {\footnotesize (colors + morpho)}           & 18 & 0.922 & 0.855\\
+ {\footnotesize Photometric errors}             & 24 & 0.951 & 0.968\\
+ {\footnotesize \texttt{size\_extendedness}} & 19 & 0.922 & 0.852\\
+ {\footnotesize Full (errors + size\_ext.)}     & 25 & 0.958 & 0.968\\
\hline
\end{tabular}
\end{table}

\begin{figure}[t]
\centering
\includegraphics[width=\columnwidth]{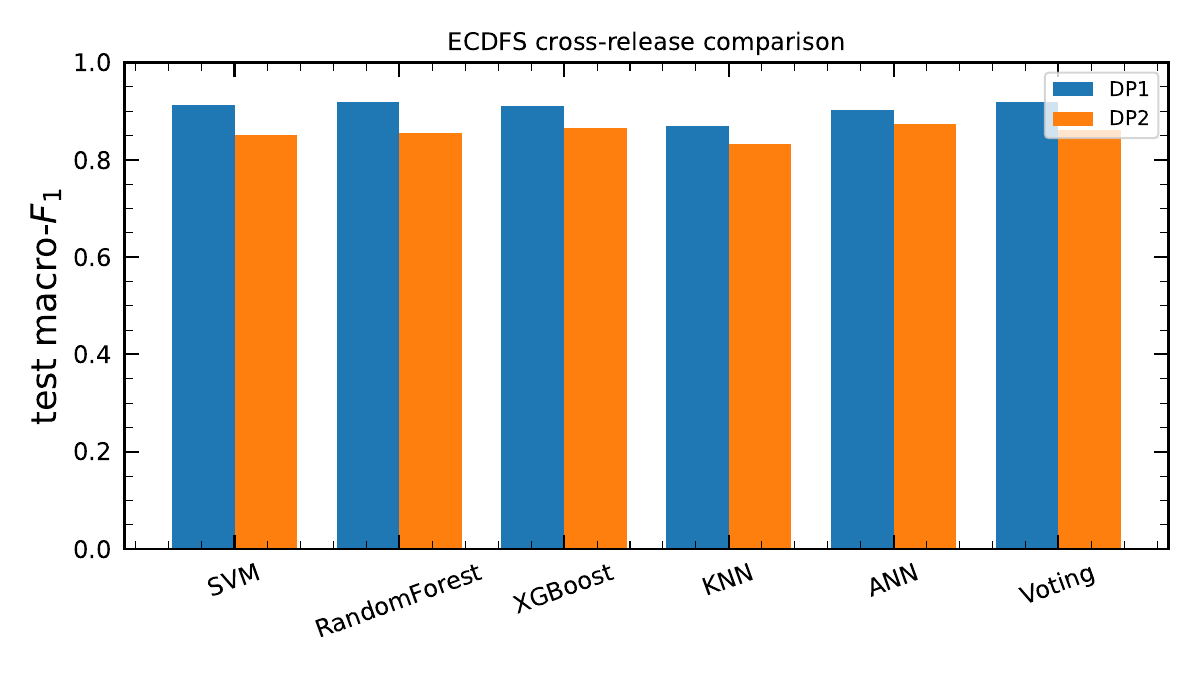}
\caption{Per-method test macro-$F_{1}$ in ECDFS for DP1 and DP2, using the identical pipeline on the same field. The DP1$\rightarrow$DP2 decrease and its interpretation are discussed in Section~\ref{sec:dp2}.}
\label{fig:cross_release}
\end{figure}

\subsection{Label Threshold Sensitivity} \label{sec:threshold_sensitivity}

As an additional robustness check, we test the sensitivity of the headline F1 to the \texttt{refSizeExtendedness} threshold used as an alternative morphology-only galaxy-label proxy (the headline labels used elsewhere in this paper are photo-$z$-anchored).
The DP1 tutorial 309.3 \citep{LSSTTutorials} sweeps several diagnostic thresholds (0.05, 0.25, 0.5, 0.75) without prescribing a single value; we adopt the natural midpoint 0.5 and verify robustness against the same threshold grid.
As a sensitivity check, we re-run the base-feature RF pipeline using morphology-only galaxy labels with the threshold swept across [0.05, 0.25, 0.5, 0.75], with the Monster star labels held fixed.  We obtain macro F1 = 0.861, 0.887, 0.943, and 0.947 respectively; the narrow spread above the 0.5 midpoint confirms that the natural \texttt{refSizeExtendedness} threshold is not load-bearing.  When the stricter $\geq 0.75$ morphology cut is applied as an additional filter on top of the headline photo-z label set (i.e., \texttt{refSizeExtendedness} $\geq 0.75$ \emph{and} photo-z reliable), the retained galaxy training set drops from 12{,}867 to 10{,}215 (a $\sim 21\%$ reduction) and the base-feature RF macro F1 rises to 0.986; the improvement reflects the fact that the double cut selects galaxies on which the SED-based and morphology-based classifiers already agree, making the test set easier by construction. We keep the headline labels purely photo-z-anchored to preserve the larger training sample and to avoid conflating the photo-z-only performance with the double-cut result.
The macro F1 spread of 0.087 reflects the expected monotonic improvement with stricter consensus (cleaner labels at higher threshold), and the 0.5 choice yields 99\% of the maximum F1 with $\sim 19\%$ more galaxy training examples than threshold 0.75.

\subsection{Baseline Comparisons} \label{sec:baselines}

To contextualize the ML classifier performance, we compare against simple baselines:
\begin{itemize}
    \item \textbf{PSF$-$cModel threshold}: optimizing the $i$-band PSF$-$cModel magnitude difference threshold yields macro F1 $= 0.933$ on the test set, dominated by low point-source precision.
    \item \textbf{\texttt{refSizeExtendedness} $\geq 0.5$ (morphology-only proxy)}: applying the DP1 flux-ratio morphology classifier directly as a star/galaxy rule against our photo-$z$-anchored labels achieves macro F1 $= 0.782$; it is reported as a baseline that quantifies how much of the classification task is solvable from morphology alone, without any color or feature-space combination.
    \item \textbf{Morphology-only RF}: training RF using only the 3 light-profile morphological features (FWHM, ellipticity, Kron radius) achieves F1 $= 0.883$.
    \item \textbf{Photometry-only RF}: training RF using only the 15 color indices achieves F1 $= 0.887$.
\end{itemize}
The full 18F RF (F1 $= 0.918$) outperforms the morphology-only baseline (F1 $= 0.883$) and the photometry-only baseline (F1 $= 0.887$), confirming that morphological and color information are complementary.  We also report the \texttt{refSizeExtendedness}~$\geq 0.5$ baseline (F1 $= 0.782$); because galaxy labels in this paper come from the photo-z catalog and not from \texttt{refSizeExtendedness}, this baseline is no longer a self-consistency check and the $\sim$0.14 gap to the ML methods quantifies the genuine information added by the SED-based labels and the multi-feature classifier.
In this deep, high-Galactic-latitude field the optimized PSF$-$cModel threshold is itself strong (F1 $= 0.933$), slightly above the ML best; a single flux-ratio cut works well where the sources are well resolved and the stellar sample is sparse. This ordering reverses in harder regimes: in the Low Galactic Latitude field the same PSF$-$cModel baseline falls to $0.853$ while the ML best rises to $0.948$ (Section~\ref{sec:xfield}), so the ML advantage is largest exactly where a single morphological parameter struggles. For DP2 ECDFS the corresponding baselines are PSF$-$cModel $0.820$, \texttt{refSizeExtendedness} $0.860$, morphology-only RF $0.824$, and photometry-only RF $0.845$, all at or below the ML best of $0.873$ (Table~\ref{tab:metrics}).

\subsection{Probability Calibration and Robustness}
\label{sec:calibration}
\begin{figure*}[t]
\centering
\includegraphics[width=0.95\textwidth]{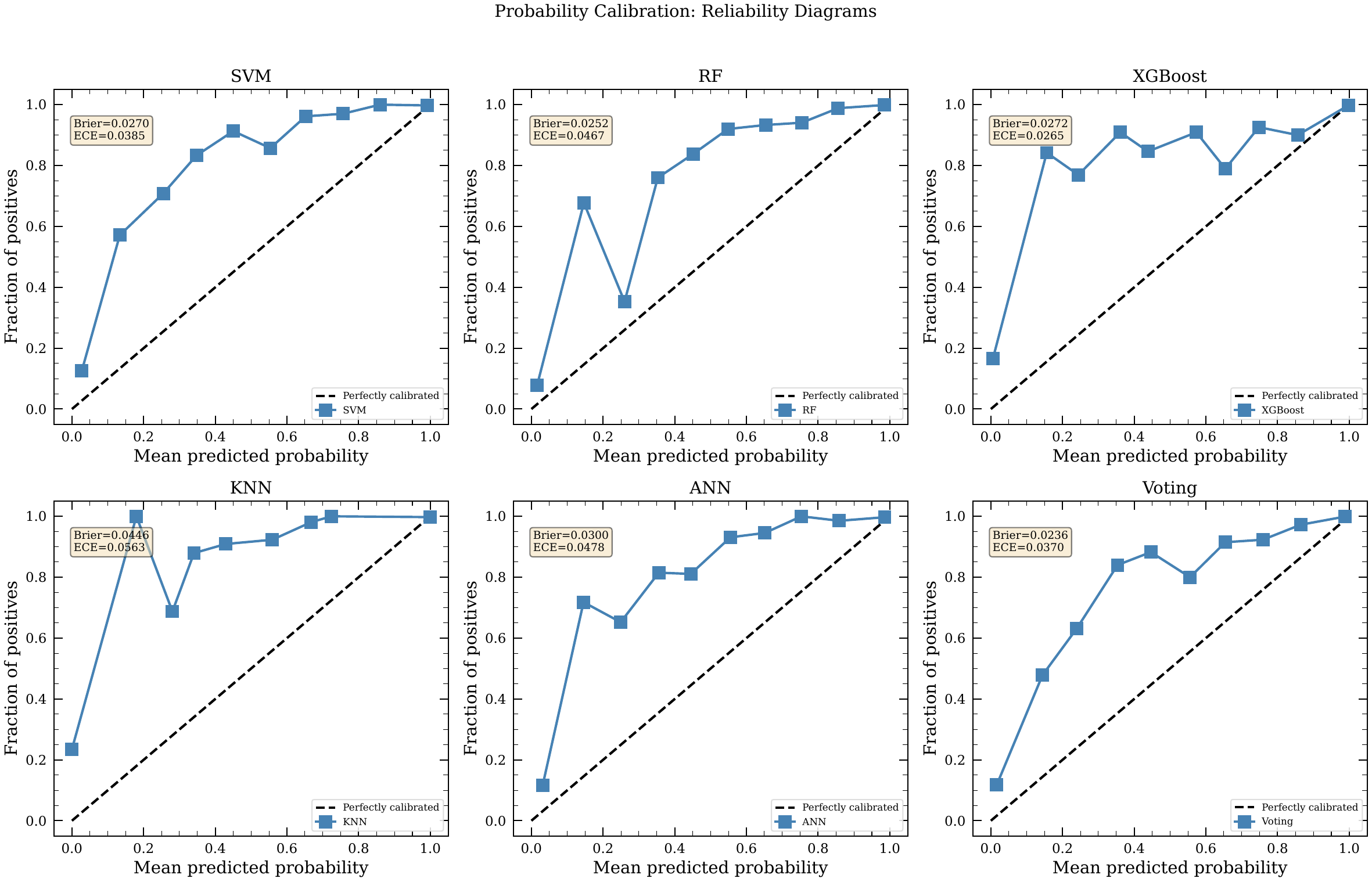}\\
\caption{ECDFS DP1 reliability diagrams for all six
classifiers, showing the fraction of positives vs.\ mean predicted
probability. A perfectly calibrated classifier falls on the diagonal;
points above it are under-confident and points below are
over-confident. Brier scores and ECE values are annotated.
XGBoost has the lowest ECE and Voting the lowest Brier
score.}
\label{fig:calibration}
\end{figure*}

Well-calibrated posterior probabilities are important for downstream
applications that rely on probability thresholds, e.g.\ selecting
high-confidence samples. We assess calibration with reliability diagrams and two scores: the Brier
score (BS; mean squared difference between predicted probabilities and true
labels, $0$ for perfect predictions) and the Expected Calibration Error (ECE;
the bin-count-weighted average gap, over probability bins, between the mean
predicted probability and the observed fraction of positives). The reliability
diagrams for the six ECDFS DP1 models (Figure~\ref{fig:calibration}, BS and ECE
annotated) plot the fraction of positives against the mean predicted
probability: a perfectly calibrated classifier lies on the diagonal, points
above it are under-confident and points below over-confident. All six
classifiers lie above the diagonal over essentially the full probability range,
so the probabilities are systematically under-confident, and a cut at a given
probability returns a purer sample than its nominal value implies.
The Voting method achieves the lowest Brier Score (0.023), indicating the most
accurate probabilistic predictions overall. In contrast, XGBoost has the
lowest ECE (0.027), suggesting that its predicted probabilities are the most
faithfully calibrated. The SVM also performs very well on both metrics
(BS = 0.027; ECE = 0.039). Although RF attains one of the lowest Brier Scores
(0.025), its slightly higher ECE (0.046) indicates that its confidence
estimates are somewhat less well calibrated. KNN consistently yields the
highest Brier Score (0.045) and ECE (0.056), reflecting lower-quality
probability estimates, while the ANN shows intermediate performance
(BS = 0.03; ECE = 0.048). Compared to ECDFS DP2 the lowest Brier score is the
Voting method with 0.022, and the highest BS is KNN method with 0.0382,
consistent with DP1 estimates.

Post-hoc isotonic regression \citep{Pedregosa2011} via
\texttt{CalibratedClassifierCV} substantially reduces the Brier score for the
two worst-calibrated methods, KNN (DP1: $0.045 \to 0.037$, DP2:
$0.038 \to 0.031$) and ANN (DP1: $0.030 \to 0.027$, DP2: $0.032 \to 0.027$),
while the already well-calibrated tree ensembles (RF, XGBoost, Voting) change
by $|\Delta \mathrm{BS}| \lesssim 0.002$. In DP1 SVM improves marginally by
$\Delta \mathrm{BS} = 0.002$, but in DP2 SVM improves considerably
$0.030 \to 0.026$.

To assess the robustness of the classifiers during training, we performed
stratified 10-fold cross-validation (Section~\ref{sec:gridsearch}) on the
balanced training set. All models achieved consistently
high mean F1 scores with small fold-to-fold variations, indicating stable
performance across different training subsets. The Voting ensemble obtained
the highest mean score ($0.969 \pm 0.008$), followed closely by RF
($0.968 \pm 0.010$) and XGBoost ($0.968 \pm 0.007$), then SVM
($0.964 \pm 0.009$), while KNN ($0.951 \pm 0.012$) and ANN
($0.951 \pm 0.013$) sat about $0.015$ lower with slightly larger variability.
As noted in Section~\ref{sec:performance}, these balanced-set scores exceed the
imbalanced test-set values, which remain the operationally relevant number.

\section{Application to the Full Catalog} \label{sec:catalog}

We apply all six trained classifiers to the full catalog of 97{,}383 objects, producing predicted class labels and posterior probability estimates for each source. Table~\ref{tab:fullcat} summarizes the predicted class counts.

\begin{table}[t]
\centering
\caption{Full catalog classification summary for all methods for DP1 and DP2.}
\label{tab:fullcat}
\begin{tabular}{lcc}
\hline\hline
\multicolumn{3}{c}{DP1} \\
Method & Point sources & Extended sources \\
\hline
SVM     &  7{,}433 & 89{,}950 \\
RF      & 14{,}262 & 83{,}121 \\
XGBoost & 14{,}225 & 83{,}158 \\
KNN     &  7{,}843 & 89{,}540 \\
ANN     &  8{,}216 & 89{,}167 \\
Voting  & 11{,}433 & 85{,}950 \\
\hline\hline
\multicolumn{3}{c}{DP2} \\
Method & Point sources & Extended sources \\
\hline
SVM     &  32{,}059 & 457{,}444 \\
RF      & 28{,}860 & 460{,}643 \\
XGBoost & 27{,}012 & 462{,}491 \\
KNN     &  33{,}807 & 455{,}696 \\
ANN     &  46{,}665 & 442{,}838 \\
Voting  &  23{,}660 & 465{,}843 \\
\hline
\end{tabular}
\end{table}

\subsection{Classifier Agreement} \label{sec:agreement}

\begin{figure}[t]
\centering
\includegraphics[width=\columnwidth]{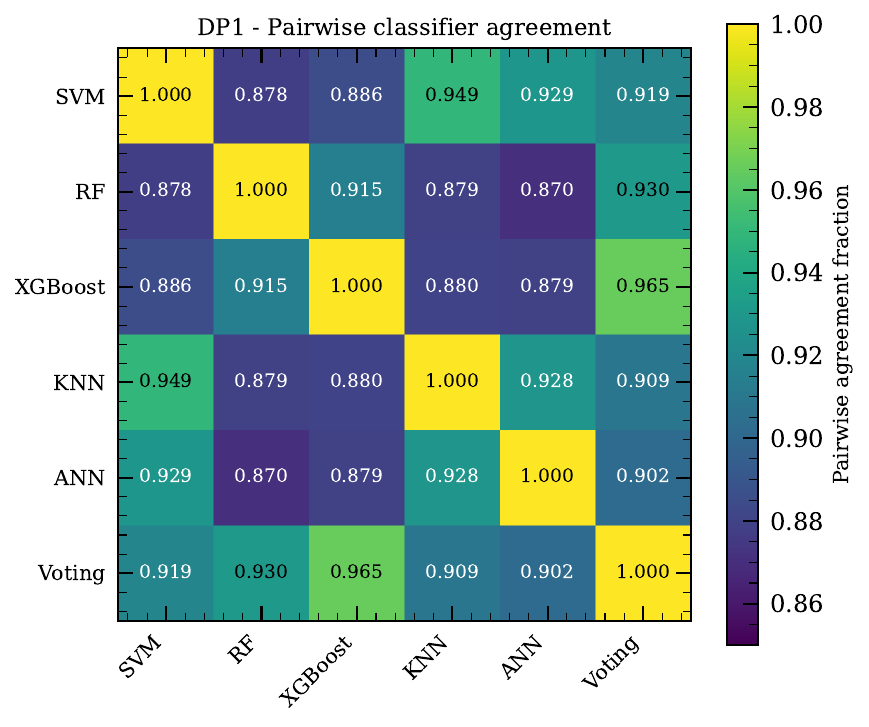}\\
\includegraphics[width=\columnwidth]{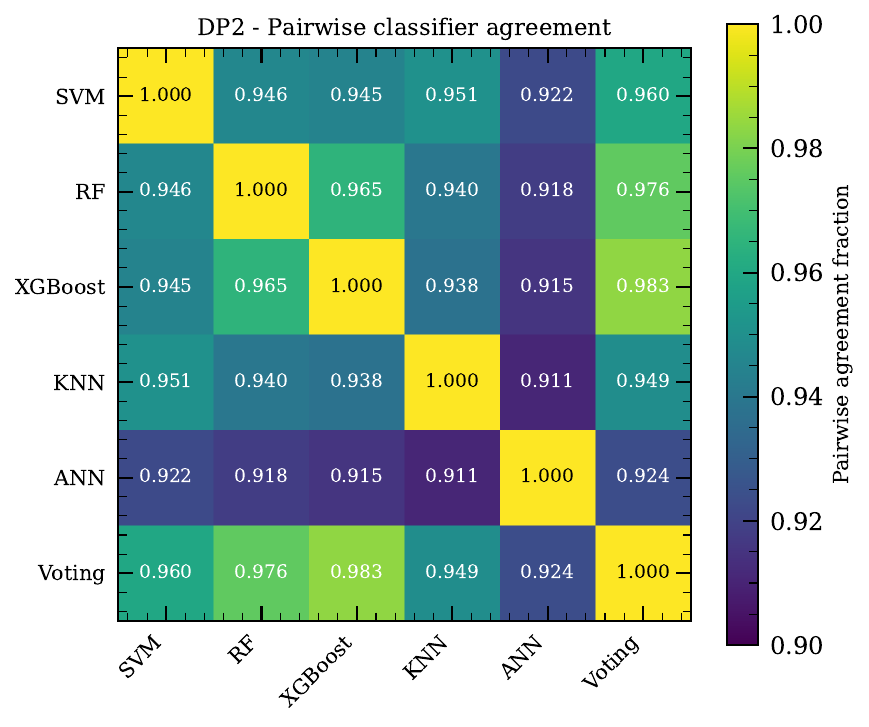}
\caption{Pairwise agreement fraction between the six classifiers on the full ECDFS catalog, DP1: 97{,}383 objects (top panel) and DP2: 489{,}503 objects (bottom panel).  Diagonal entries are unity by construction; off-diagonal entries are color-coded by the fraction of catalog objects on which the two classifiers assign the same label.  The all-six-agree fraction is annotated in the title.}
\label{fig:agreement_heatmap}
\end{figure}

We compute the pairwise agreement between all six classifiers on the full ECDFS catalog; the resulting $6 \times 6$ matrix is shown as Figure~\ref{fig:agreement_heatmap}. Pairwise agreement values range from $0.870$ to $0.965$. The highest agreement is between the XGBoost and Voting classifiers ($0.965$), reflecting the direct contribution of XGBoost (weight $2$) to the soft-Voting ensemble; the lowest is between RF and ANN ($0.870$), reflecting the different decision boundaries of tree- and neural-network-based methods.  All six classifiers assign the same label to $79.5\%$ of the $97{,}383$ objects ($77{,}438$ sources), indicating substantial overall consistency.

\subsection{Probability Maps and High-Confidence Sources} \label{sec:bestmodel}

\begin{figure*}[t]
\centering
\includegraphics[width=\textwidth]{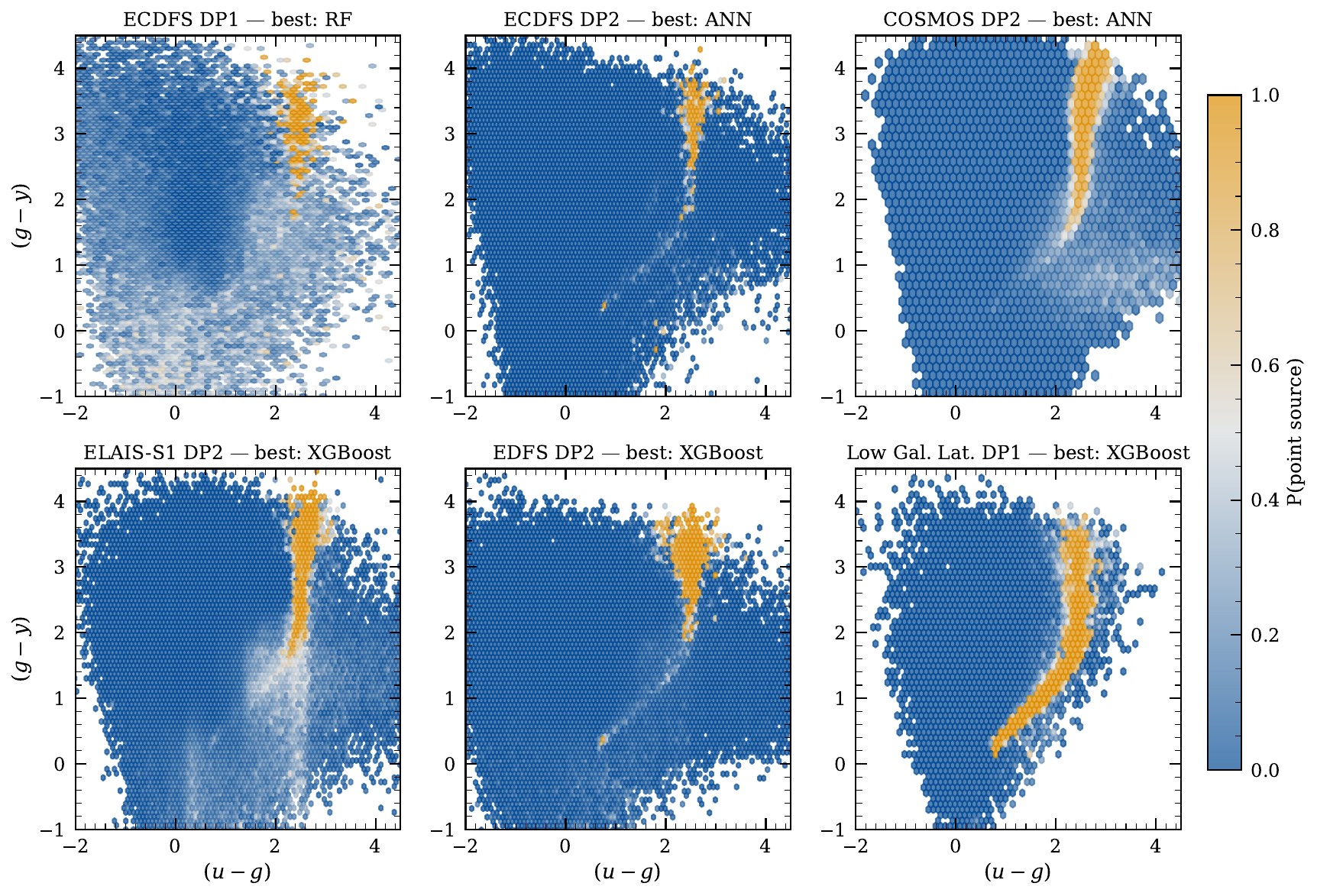}
\caption{Point-source classification probability maps for the best method in the $(u\!-\!g)$ vs.\ $(g\!-\!y)$ color-color space for the different fields.  All methods show consistent classification patterns, with high-confidence regions for both point sources and extended sources.}
\label{fig:prob_maps_all}
\end{figure*}

Figure~\ref{fig:prob_maps_all} shows the point-source probability map of the best model in each of the six fields, in the $(u\!-\!g)$ vs.\ $(g\!-\!y)$ color-color plane, with each hex cell colored by the median predicted point-source probability (see the color bar). In every field the stellar locus forms a compact region of uniformly high point-source probability, while the broader, bluer galaxy distribution is assigned low probability; the confidence drops only in the narrow transition at the lower, blue end of the locus (around $(u\!-\!g)\lesssim 1.5$ and $(g\!-\!y)\lesssim 1.0$), where the two populations overlap. The same structure recurs across DP1 and DP2 and across the full range of depth, extinction, and stellar density probed here, showing that the classifier places the high-confidence stellar locus consistently from field to field.

Applying a conservative probability threshold of $P \geq 0.80$ to the best model (RF), we identify 2{,}306 high-confidence point sources and 56{,}468 high-confidence extended sources, together comprising 60.4\% of the catalog; the point sources span $15.4 \leq i_\mathrm{PSF} \leq 27.2$~mag.

\section{Other Applications} \label{sec:other_app}

\subsection{Additional Rubin Fields} \label{sec:xfield}

To assess how the spectroscopy-free, photo-z-based pipeline of Sections~\ref{sec:training}--\ref{sec:catalog} transfers across observing conditions, we apply the identical workflow without retuning the settings to two other DP1 fields and three other DP2 fields that combine full \ugrizy{} coverage with the \citet{Charles2025} ancillary photo-z catalog: the Euclid Deep Field South (EDFS) at the south Galactic cap available in DP1 and DP2, and the Rubin SV Low Galactic Latitude Field at $b \approx -10^\circ$ available in DP1. In DP2 we use the DDF COSMOS and ELAIS-S1.

Table~\ref{tab:crossfield_t1} and Table~\ref{tab:crossfield_t2} summarize the field properties, training composition, and headline classification metrics: Table~\ref{tab:crossfield_t1} covers ECDFS and EDFS, which have both DP1 and DP2 data, and Table~\ref{tab:crossfield_t2} covers COSMOS, ELAIS-S1, and the Low Galactic Latitude Field. Figure~\ref{fig:crossfield_f1} shows the per-method test macro-$F_{1}$ for the six classifiers across the fields. Three findings follow from the cross-field comparison.

First, the photo-z based, six-method pipeline transfers across fields without retuning. The headline macro-$F_{1}$ stays above 0.87 in every field and the best model is always a tree ensemble (RF in ECDFS and in EDFS, XGBoost in the Low Galactic Latitude field), with the other methods clustering within $\Delta F_{1} \leq 0.06$ of the best. Six-method full-catalog agreement also rises with stellar abundance, from 79.5\% in ECDFS to 89.2\% in the Low Galactic Latitude field, where the more balanced training sample reduces the contribution of point-source-precision dispersion to inter-method disagreement.

Second, the relative ranking of the non-ML baselines is field-dependent and tracks Galactic latitude.  The PSF$-$cModel threshold baseline drops from $F_{1} = 0.933$ in ECDFS to $F_{1} = 0.853$ in the Low Galactic Latitude field, while the photometry-only RF baseline improves in the opposite direction, from $F_{1} = 0.887$ to $F_{1} = 0.936$.  This crossover reflects two effects.  At low Galactic latitude the much higher stellar density tightens the stellar locus in color-color space (locus widths of $\sigma \approx 0.02$ in the $(g\!-\!i)$--$(r\!-\!i)$ plane in the Low Galactic Latitude field versus $\sigma \approx 0.03$ in EDFS in the same plane, and up to a factor of $\sim 8\times$ narrower for the $(g\!-\!r)$-anchored planes), so colors carry more discriminating power than morphology; at high Galactic latitude with sparse stellar samples the situation is reversed.  The cross-field consistency of the morphology-only RF baseline ($F_{1} = 0.858$--$0.888$) further confirms that the dominance of FWHM in the permutation importance ranking of Section~\ref{sec:importance} is a property of the feature space rather than of the high-latitude ECDFS field specifically.

Third, the photo-z and morphology label sources disagree most where the morphology classifier is most stressed.  In ECDFS and EDFS, 78--80\% of photo-z galaxies are also \texttt{refSizeExtendedness} galaxies; in the Low Galactic Latitude field this fraction drops to 64\%, even though the field has the highest individual classifier $F_{1}$.  This is the strongest single piece of cross-field evidence that the photo-z-anchored labels test a physically orthogonal classifier from the moments-based morphology label, and that the divergence is largest where photo-z information is most valuable (high stellar density, larger extinction).  We retain ECDFS as the primary deepest validation field; the EDFS and Low Galactic Latitude results show that the pipeline generalizes beyond the deepest, high-latitude, extragalactic baseline assumed by single-field analyses.

\begin{table*}[t]
\centering
\caption{Cross-field summary of properties, training composition, and headline classification metrics for the two fields with DP1 and DP2 data, photo-z coverage and full \ugrizy{} bands. Field properties are derived from \citet{RubinDP1,RTN115} Tables 1--3 and our own catalog reductions. The headline classification number is the test-set macro-$F_{1}$ of the best model in each field; per-method values appear in Figure~\ref{fig:crossfield_f1}. `Leading methods' lists the classifiers within $\Delta F_1 \le 0.01$ of the best (a practical comparability threshold; the bootstrap 68\% CI is $\approx 0.007$ for DP1 and $0.004$ for DP2, Section~\ref{sec:performance}).}
\label{tab:crossfield_t1}
\begin{tabular}{lcccc}
\hline\hline
Property & ECDFS DP1 & ECDFS DP2 & EDFS (DP1) & EDFS b (DP2) \\
\hline
RA, Dec (deg)               & (53.13, -28.10) & (53.00, -28.10) & (59.10, -48.73) & (63.2, -47.8) \\
Median $E(B-V)$ (mag)       & 0.008           &  0.009  & 0.006 & 0.014 \\
Total visits                & 855             &  289  & 272  & 178 \\
$u$-band depth (coadd $5\sigma$) & 24.55 & 24.35 & 23.42 & 23.69 \\
$g$-band depth (coadd $5\sigma$) & 26.18 & 25.84 & 25.77 & 25.75 \\
$i$-band depth (coadd $5\sigma$) & 25.71 & 25.87 & 25.17 & 25.71 \\
\hline
Training: stars             & 1{,}284         &  6{,}525  & 1{,}881 & 8{,}594  \\
Training: galaxies          & 12{,}867        &  83{,}827 & 13{,}897 & 84{,}135 \\
Training balance            &  1{,}798        &  9{,}134   & 2{,}634 & 12{,}032\\
Galaxy:star imbalance       & 10:1            & 13:1 & 7.4:1 & 10:1 \\
Full catalog ($N$)          & 97{,}383        & 489{,}503  & 89{,}188 & 449{,}341 \\
\hline
Leading methods             & RF, Vot, SVM, XGB & ANN, XGB & RF, Vot, XGB & XGB, RF, Vot, ANN \\
Best macro $F_{1}$          & 0.918 & 0.873 & 0.918 & 0.880 \\
$griz$-only (9F)            & ---           &  0.830      & ---  & 0.835 \\
Base + Errors (RF, 24F)     & 0.951 & 0.968 & 0.971 & 0.982\\
PSF$-$cModel baseline       & 0.933 & 0.820 & 0.899 & 0.815\\
Photometry-only RF          & 0.887 & 0.845 & 0.894 & 0.868 \\
Morphology-only RF          & 0.883 & 0.824 & 0.858 & 0.822 \\
Six-method agreement (\%)   & 79.5 & 86.1 & 86.5 & 85.6 \\
HC point sources ($P \geq 0.8$) & 2{,}306 & 34{,}028 & 3{,}028 & 20{,}621 \\
HC extended sources ($P \geq 0.8$) & 56{,}468 & 425{,}571 & 55{,}929 & 381{,}142\\
\hline
\end{tabular}
\end{table*}

\begin{table*}[t]
\centering
\caption{Cross-field summary of properties, training composition, and headline classification metrics for the remaining fields tested, with photo-z coverage and full \ugrizy{} bands. Field properties are derived from \citet{RubinDP1,RTN115} Tables 1--3 and our own catalog reductions. The headline classification number is the test-set macro-$F_{1}$ of the best model in each field; per-method values appear in Figure~\ref{fig:crossfield_f1}. `Leading methods' as in Table~\ref{tab:crossfield_t1}.}
\label{tab:crossfield_t2}
\begin{tabular}{lccc}
\hline\hline
Property & COSMOS (DP2) & ELAIS\_S1 (DP2) & Low Gal.\ Lat. (DP1) \\
\hline
RA, Dec (deg) & (150.1,2.1) & (9.5, -44.0) & (95.00, -25.00) \\
Median $E(B-V)$ (mag) & 0.020 & 0.007 & 0.044 \\
Total visits & 714 & 799 & 292 \\
$u$-band depth (coadd $5\sigma$) & 25.77 & 25.18 & 24.29 \\
$g$-band depth (coadd $5\sigma$) & 26.73 & 26.57 & 25.46 \\
$i$-band depth (coadd $5\sigma$) & 25.94 & 26.07 & 24.86 \\
\hline
Training: stars    & 7{,}473 & 7{,}238 & 10{,}929 \\
Training: galaxies & 106{,}432 & 96{,}665 & 17{,}464 \\
Training balance   & 10{,}462  &  10{,}134  & 15{,}300    \\
Galaxy:star imbalance & 14:1 & 13:1 & 1.6:1 \\
Full catalog ($N$) & 778{,}313 & 686{,}425 & 112{,}690 \\
\hline
Leading methods (Base -18F) & ANN, XGB & XGB, RF, Vot, ANN & XGB, Vot, ANN, RF \\
Best macro $F_{1}$ (Base -18F) & 0.833 & 0.857 & 0.948 \\
$griz$-only (9F) & 0.791 &  0.811 & --- \\
Base + Errors (RF) & 0.974 & 0.968 & 0.980 \\
PSF$-$cModel baseline & 0.760 & 0.808 & 0.853 \\
Photometry-only RF & 0.825 & 0.847 & 0.936 \\
Morphology-only RF & 0.777 & 0.811 & 0.888 \\
Six-method agreement (\%) & 76.7 & 78.8 & 89.2 \\
HC point sources ($P \geq 0.8$) & 43{,}025 & 24{,}970 & 14{,}212 \\
HC extended sources ($P \geq 0.8$) & 659{,}834 & 551{,}400 & 91{,}101 \\
\hline
\end{tabular}
\end{table*}

We also re-ran the identical pipeline in the four DP2 fields with a griz-only feature set (six colors plus the three morphological features, nine features total), training both feature sets on the same objects and the same train/test split so that only the feature columns differ. Figure~\ref{fig:ugrizy_vs_grizonly_f1} reports the outcome. Dropping the $u$ and $y$ bands lowers the best-model macro-$F_{1}$ by a nearly uniform $\Delta F_{1} \approx 0.04$ in every field (ECDFS $+0.043$, COSMOS $+0.042$, EDFS $+0.045$, ELAIS-S1 $+0.046$), while the cross-field ordering (EDFS $>$ ECDFS $>$ ELAIS-S1 $>$ COSMOS) is preserved in both configurations. The field-to-field differences are population- and depth-driven rather than a consequence of $u$/$y$ feature richness.

The per-field analogues of the ECDFS diagnostic plots in Sections~\ref{sec:performance}--\ref{sec:stellarlocus} (field characteristics, confusion matrices, completeness/contamination versus magnitude) will be released alongside the cross-field comparison notebook described in the Data Availability statement, and the field-level visit/depth numbers tabulated in \newtext{Table~\ref{tab:crossfield_t1}  and Table~\ref{tab:crossfield_t2} are taken from \citet{RubinDP1,RTN115}.}  The qualitative cross-field behavior is consistent with the ECDFS case: the dominant misclassification pathway is true extended sources mistakenly classified as point sources (rate $\lesssim 3\%$ in the tree ensembles and $\lesssim 6\%$ in KNN, as fractions of the extended-class support), and the methods are nearly indistinguishable for $i \lesssim 23$ in every field, with KNN and ANN diverging first at the faint end.

\subsection{Additional DP2 Fields} \label{sec:dp2}

We apply the same pipeline to DP2 with the identical label recipe: point-source labels from the Gaia-anchored Monster cross-match, and extended-source labels from the DP2 photometric-redshift catalog produced with the RAIL package \citep{RTN124} and accessed through LSDB, keeping the FlexZBoost estimate with $\sigma_{p(z)} < 0.15$, $0.02 < z_\mathrm{mode} < 1.5$, and $i_\mathrm{cModel} < 23$, the same galaxy-label cuts used for DP1 so that the two releases are directly comparable. Because the DP2 depth varies within each field, we first apply a coverage cut that keeps objects whose $i$-band input-visit count is at least 0.85 times the field median, selecting the region of near-uniform depth. We analyze four DP2 fields: ECDFS (also a DP1 field, which enables the cross-release comparison), COSMOS (the deepest Deep Drilling Field), EDFS, and ELAIS-S1. COSMOS has 7{,}473 point-source and 106{,}432 extended-source labels (a class imbalance near $14{:}1$) out of 778{,}313 classified objects.

On the COSMOS base feature set (colors + morphology) the six methods reach test macro-$F_{1}$ from 0.791 (SVM) to 0.833 (ANN), and agree on 76.7\% of the full catalog (Table~\ref{tab:dp2cosmos}). Adding the six per-band magnitude uncertainties raises the best macro-$F_{1}$ to about 0.98 (RF 0.974), the same direction as in DP1 and larger in size because a larger fraction of the COSMOS sample lies near the detection limit. Permutation importance again ranks $i$-band FWHM first. The non-ML baselines are PSF$-$cModel $0.760$, \texttt{griz\_model\_extendedness} $0.737$, morphology-only RF $0.777$, and photometry-only RF $0.825$, so the ML best exceeds every single-parameter baseline. We tested the two DP2-only morphology features as additions to the base set: replacing the Kron radius with the S\'ersic effective radius raises the base Random Forest by $0.039$ (from 0.889 to 0.927 on the common finite sample), and the S\'ersic radius is defined for essentially all labeled objects; adding the six per-band \texttt{extendedness} values instead gives a negligible improvement ($+0.004$ at fixed hyperparameters) and is available for only 59\% of labeled objects, so we do not adopt it. Removing the $i_\mathrm{cModel} < 23$ cut and keeping all reliable photo-z galaxies adds fainter sources and lowers the COSMOS best base macro-$F_{1}$ to about 0.80; we retain the $i < 23$ cut for consistency with the DP1 analysis.

Because ECDFS is present in both releases, we run the identical pipeline on the same field in DP1 and DP2. The test macro-$F_{1}$ decreases from DP1 to DP2 for every method, by 0.03 to 0.06 (Figure~\ref{fig:cross_release}), consistent with the shallower and less uniform commissioning-era DP2 depth in this field; the photo-z and morphology label definitions remain mutually consistent across the two releases (Section~\ref{sec:xfield}). Because the galaxy training labels use the same $i_\mathrm{cModel}<23$ cut in both releases, this drop is not caused by fainter galaxies entering the DP2 sample; it instead reflects the shallower, less uniform commissioning-era coadds and the poorer, more variable PSF over the much larger LSSTCam area (outer-detector systematics), rather than the per-visit seeing tail of Figure~\ref{fig:field_characteristics}, whose worst-seeing visits largely do not enter the coadds. Across the four DP2 fields the best base macro-$F_{1}$ ranges from 0.833 (COSMOS) to 0.880 (EDFS), with ECDFS at 0.873 and ELAIS-S1 at 0.857; the leading method is field-dependent --- ANN in ECDFS and COSMOS, XGBoost in EDFS and ELAIS-S1 --- with the tree ensembles and the ANN statistically comparable within each field (Section~\ref{sec:performance}). EDFS has 8{,}594 point-source and 84{,}135 extended-source labels (449{,}341 objects), and ELAIS-S1 has 7{,}238 and 96{,}665 (686{,}425 objects). Full field-by-field properties, baselines, and six-method agreement fractions are collected in Tables~\ref{tab:crossfield_t1}--\ref{tab:crossfield_t2}.

\begin{table}[t]
\centering
\caption{Per Field Test macro-$F_{1}$ by method, with the Base features (18F) and the Base+Errors (24F) parameters.}
\label{tab:dp2cosmos}
\begin{tabular}{lcccccc}
\hline
\hline
\multicolumn{7}{c}{ECDFS DP2}\\
& SVM & RF & XGBoost & KNN & ANN &Voting \\
\hline
Base      & 0.851 & 0.855 & 0.865 & 0.833 & 0.873 & 0.862\\
+Errors   & 0.936 & 0.968 & 0.969 & 0.882 & 0.956 & 0.970\\
\hline
\hline
\multicolumn{7}{c}{COSMOS DP2}\\
     & SVM & RF & XGBoost & KNN & ANN &Voting \\
\hline
Base      & 0.791 & 0.810 & 0.827 & 0.800 & 0.833 & 0.817\\
+Errors   & 0.939 & 0.974 & 0.974 & 0.893 & 0.954 & 0.974\\
\hline
\hline
\multicolumn{7}{c}{EDFS DP2}\\
     & SVM & RF & XGBoost & KNN & ANN &Voting \\
\hline
Base      & 0.858 & 0.879 & 0.880 & 0.851 & 0.875 & 0.879\\
+Errors   & 0.953 & 0.982 & 0.982 & 0.909 & 0.976 & 0.982\\
\hline
\hline
\multicolumn{7}{c}{ELAIS-S1 DP2}\\
     & SVM & RF & XGBoost & KNN & ANN &Voting \\
\hline
Base      & 0.833 & 0.856 & 0.857 & 0.831 & 0.854 & 0.856\\
+Errors   & 0.947 & 0.968 & 0.970 & 0.918 & 0.956 & 0.969\\
\hline
\end{tabular}
\end{table}

\subsection{Stellar-locus Cross-check}
\label{sec:stellarlocus}

\begin{figure*}[t]
\centering
\includegraphics[trim=0cm 0cm 0cm 0cm,clip,width=\textwidth]{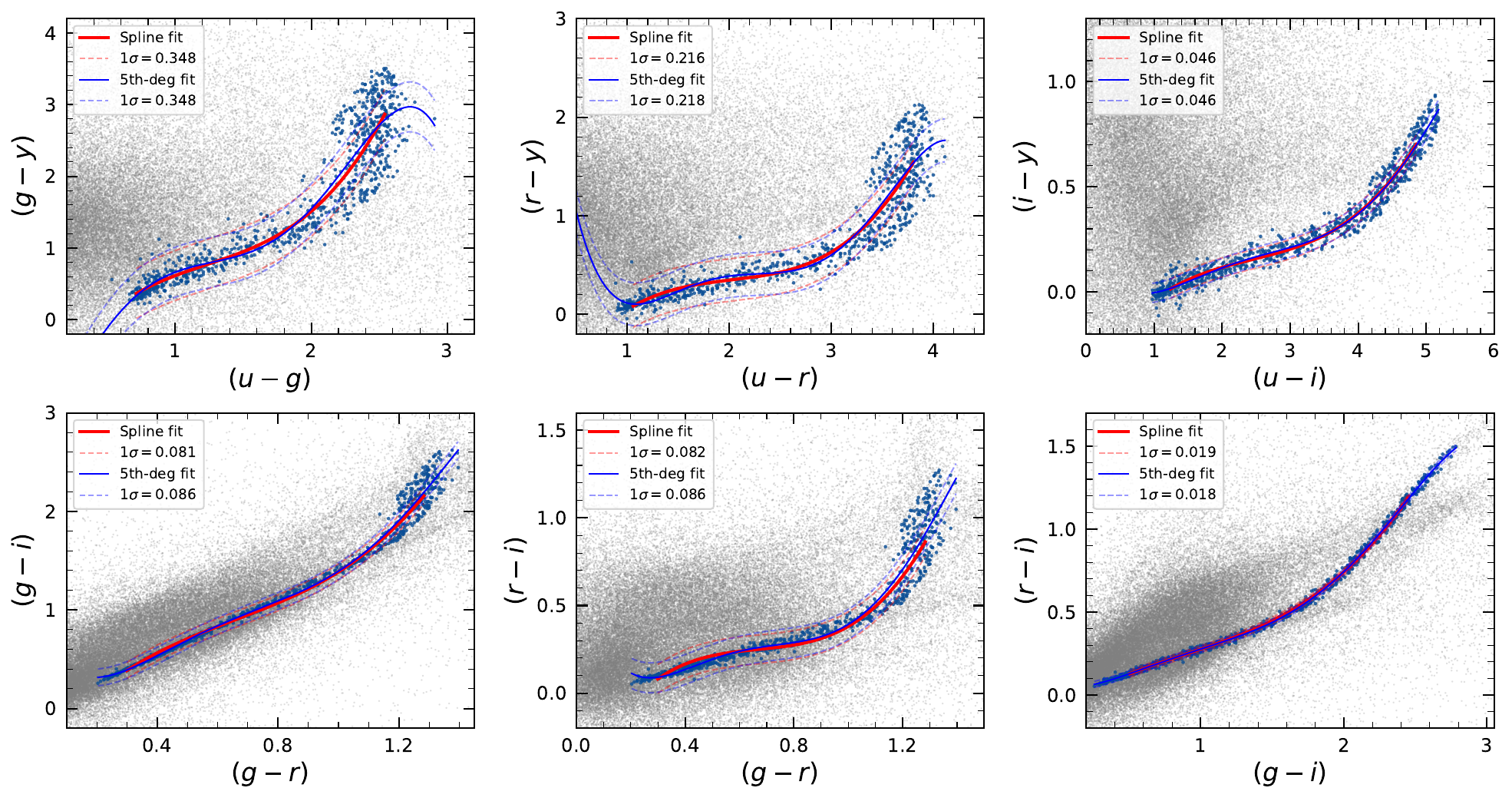}
\caption{Stellar-locus fits in the six color-color planes for ECDFS DP1.  Gray dots show all sources, orange dots the stars selected for the fit. The blue curve is the fifth-degree polynomial fit and the red curve is the binned spline fit; dotted blue and red curves show the corresponding $\pm 1\sigma_\mathrm{locus}$ envelopes, with the $\sigma_\mathrm{locus}$ values annotated in each panel legend.}
\label{fig:locus_fits}
\end{figure*}

We test a non-ML, physically motivated star/galaxy classifier based on residuals from the stellar locus in color-color space, following \citet{Choi2025}, where they use multiple stellar loci to further refine their 47~Tuc membership selection. For 47~Tuc, which lacks $u$-band coverage and has very shallow $y$-band, we use their three $gri$ planes [$(g-r)$ vs $(g-i)$, $(g-r)$ vs $(r-i)$, $(g-i)$ vs $(r-i)$]; for the ECDFS we add three $u$/$y$ extensions [$(u-g)$ vs $(g-y)$, $(u-r)$ vs $(r-y)$, $(u-i)$ vs $(i-y)$]. Whereas \citet{Choi2025} fit a fifth-degree polynomial to an isochrone-selected locus, we select the ECDFS locus with $\mathrm{SNR} > 5$ and \texttt{refSizeExtendedness} $\leq 0.001$ (1{,}002 stars; orange circles in Figure~\ref{fig:locus_fits}) and fit a fifth-degree polynomial after one round of median-absolute-deviation (MAD) $\sigma$-clipping. Because some loci are poorly fit by a polynomial, we also compute a binned spline (red curve in Figure~\ref{fig:locus_fits}); the blue solid and dashed curves show the polynomial fit and its $1\sigma$ envelope.

These fits show that the stellar locus is a well-defined sequence in all six color-color planes---tightest in the red adjacent-band planes and broader in the $u$-anchored planes---providing an independent, physically motivated view of the point-source population that is consistent with the ML stellar locus. A fully quantitative residual-based star/galaxy classifier along these lines is sensitive to the per-plane $\sigma$ threshold adopted and to the sparse, galaxy-dominated nature of this high-latitude field; we defer it to future work and adopt the space-based Euclid Q1 cross-match (Section~\ref{sec:limitations}) as our quantitative independent validation.

\subsection{Transfer to Crowded Stellar Fields} \label{sec:crowded}

We additionally study whether a classifier trained on a sparse, high-Galactic-latitude, extragalactic field transfers to a dense stellar field such as a globular cluster or a dwarf galaxy. We test this directly by applying the ECDFS-trained classifier, without any retraining, to two crowded DP1 fields: the globular cluster 47~Tucanae and the Fornax dwarf spheroidal. These fields have no reliable photo-z DP1 catalog and their cores are blend-dominated, so we cannot build training labels or measure $F_{1}$ there; instead we restrict to an annulus (excluding the inner $0.10^\circ$ for 47~Tuc and $0.05^\circ$ for Fornax) and compare, object by object, the ML prediction against the DP1 \texttt{refExtendedness} classifier. Because these fields lack useful $u$- and $y$-band coverage, we use the $gri$ colors plus the three morphology descriptors and the ECDFS-trained Random Forest.

\begin{figure*}[t]
\centering
\includegraphics[width=0.49\textwidth]{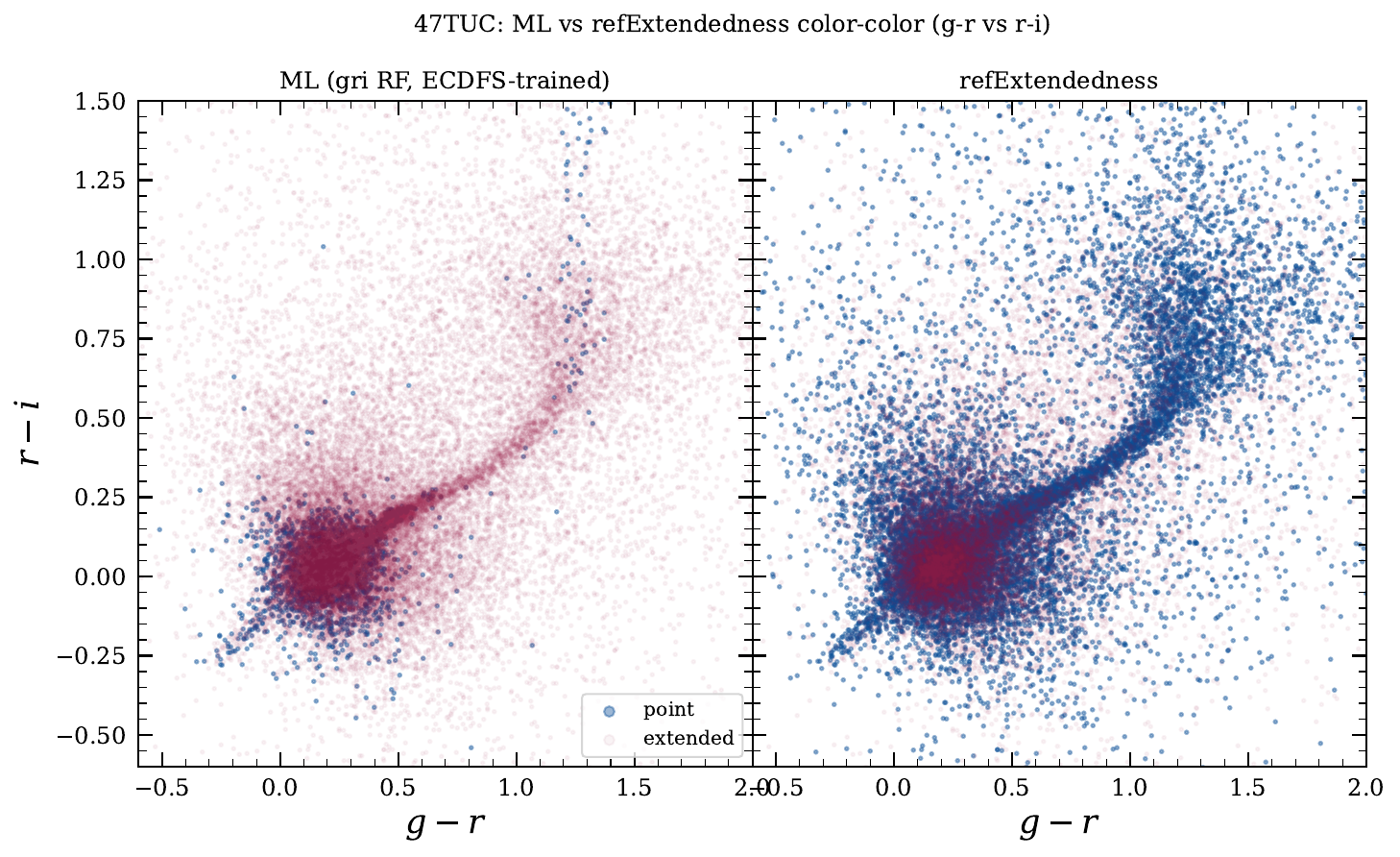}
\includegraphics[width=0.49\textwidth]{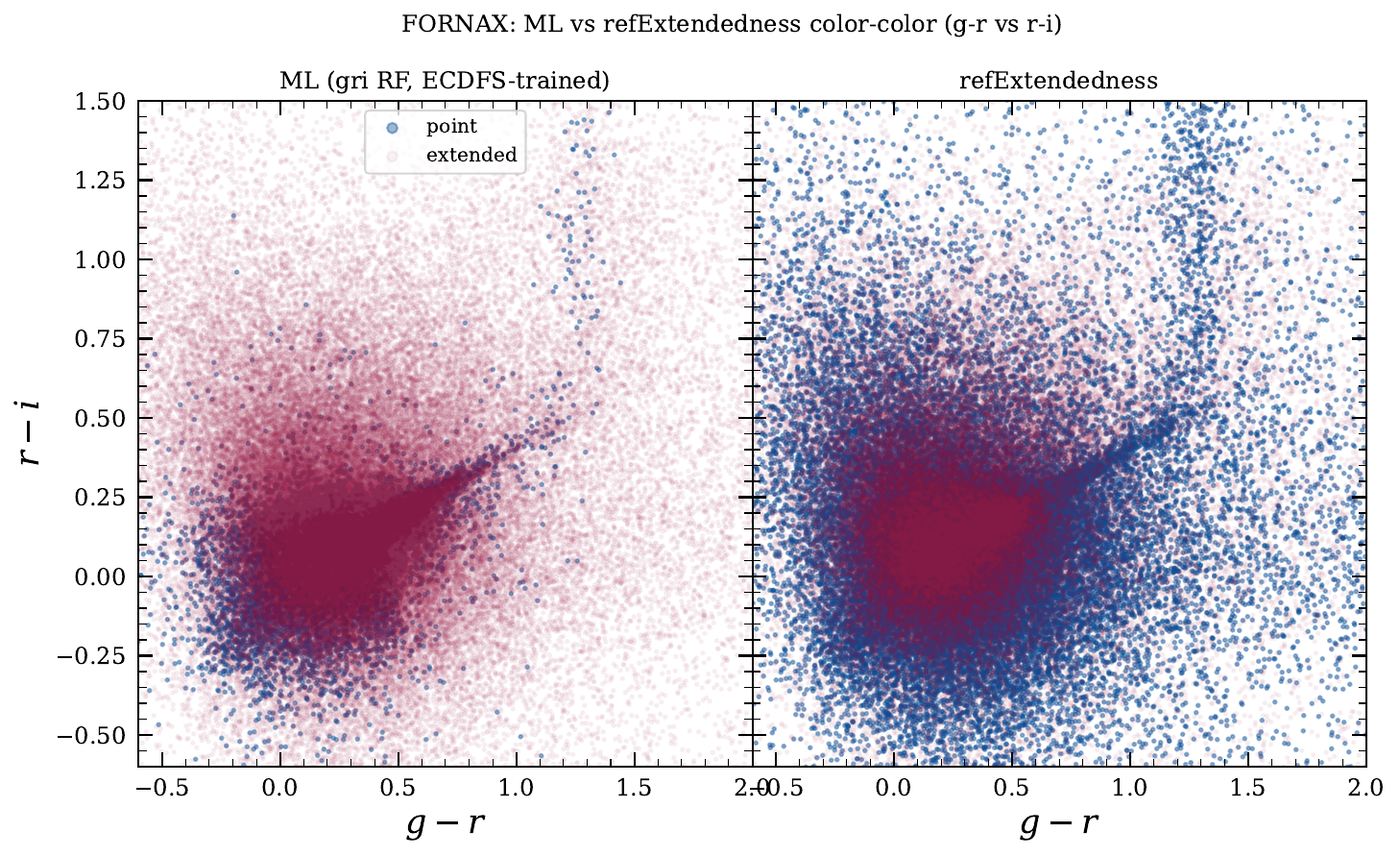}
\caption{Color--color diagrams ($r-i$ vs $g-r$) for the crowded DP1 fields 47~Tucanae (left block) and Fornax (right block), in an annulus that excludes the blend-dominated core. Within each block, the left panel colors objects by the ECDFS-trained $gri$ Random Forest prediction and the right panel by the DP1 \texttt{refExtendedness} classifier (orange = point source, blue = extended). The ML model, trained on a sparse high-latitude field, assigns the tight cluster stellar locus to the extended class, whereas \texttt{refExtendedness} recovers it as point sources: the transfer failure discussed in Section~\ref{sec:crowded}.}
\label{fig:crowded_cmd}
\end{figure*}

Figure~\ref{fig:crowded_cmd} shows the result as color--color diagrams ($g-r$ vs $r-i$). The ECDFS-trained model labels only 6.2\% (47~Tuc) and 10.8\% (Fornax) of objects as point sources, and it assigns the tight cluster stellar locus (clearly visible as a narrow sequence in the $g-r$ vs $r-i$ plane) to the \emph{extended} class. The moments-based \texttt{refExtendedness} classifier, by contrast, recovers that locus as point sources (43--47\% of objects labeled point-like). In these crowded fields the high-latitude-trained ML model does not produce a cleaner stellar sample than the pipeline morphology classifier; it fails in the opposite direction, misclassifying resolved cluster stars as galaxies.

These results represent a transfer, or domain-shift, limitation. The model was trained where stars are sparse and relatively faint, and it has never seen the crowding, blending, and bright red-giant-branch stars that dominate these fields, whose color and magnitude distributions differ substantially from the training field. It is the honest answer to whether a spectroscopy-free, high-latitude-trained classifier can be deployed unchanged on crowded stellar fields: it cannot, and either field-matched training labels or a dedicated morphology-based classifier is required there (Section~\ref{sec:limitations}). This is consistent with ultra-faint dwarf studies on deep DP2 imaging, where the pipeline \texttt{refExtendedness} classifier is found to give the best faint-end star completeness \citep{2026arXiv260725044P}. The result complements the qualitative stellar-locus check of Section~\ref{sec:stellarlocus}.

\section{Discussion \& Summary} \label{sec:discussion}

We have presented a cloud-native, spectroscopy-free pipeline for the binary classification of point-like and extended sources in Rubin DP1 and DP2, anchoring stellar labels on the Gaia-aligned Monster reference catalog \citep{DMTN277} and extended-source labels on the ancillary photometric-redshift catalogs \citep{Charles2025, RTN124} rather than on external spectroscopy.  This is a deliberate design choice: photo-$z$'s are, in principle, available wherever LSST observes, whereas homogeneous, high-purity spectroscopic training sets are not, since spectroscopic depth, completeness, and selection vary strongly across the sky. To summarize the main outputs: six classifiers reach a test macro-$F_{1}$ of $0.87$--$0.92$ on the imbalanced ECDFS DP1 held-out set (Random Forest best at $0.918$; balanced 10-fold cross-validation $F_{1} = 0.95$--$0.97$), with $i$-band FWHM the dominant feature. Applied without retuning, the identical pipeline keeps macro-$F_{1} > 0.87$ across DP1 fields spanning a factor of $\sim 7$ in $E(B-V)$ and $\sim 10$ in stellar density, and extends to four DP2 fields (best base macro-$F_{1} = 0.83$--$0.88$); the same-field ECDFS score falls by $\sim 0.05$ from DP1 to DP2, consistent with the shallower commissioning-era depth, and adding the per-band photometric errors raises the best macro-$F_{1}$ to $\sim 0.95$ (DP1) and $\sim 0.97$ (DP2) in ECDFS. The one clear failure mode is transfer to crowded stellar fields (47~Tuc and Fornax), where the high-latitude-trained model mislabels resolved cluster stars as extended (Section~\ref{sec:crowded}), an honest limitation of spectroscopy-free, high-latitude training rather than of the method itself.

\subsection{Comparison with Literature}

Our results are broadly consistent with previous multi-method ML classification studies, while highlighting the specific consequences of working without external spectroscopy.
\citet{Zeraatgari2024} achieved F1 $\sim 0.99$ for three-class star/galaxy/QSO classification with SDSS+ALLWISE optical+infrared data, using a spectroscopic training set.
Our binary classification with DP1 \ugrizy{}-only data and RSP-only labels achieves a test macro F1 in the range 0.87--0.92, with the residual spread between methods (rather than the absolute level) reflecting algorithmic robustness to the $\sim$10:1 class imbalance retained in the held-out test set and the absence of infrared data in DP1.

Consistent with \citet{Zeraatgari2024} and \citet{Feng2025}, we find that ensemble methods (RF, XGBoost, Voting) tend to match or outperform single-model approaches (SVM, KNN, ANN), with RF achieving the best test-set performance.
This pattern is also reported by \citet{vonMarttens2024} for J-PLUS DR3.

The importance of morphological features (especially FWHM) is consistent with the canonical view that compact size is the primary discriminant for point-vs-extended separation \citep{OrdenesBriceno2026, Angora2019, Barbisan2022}. In deep DP2 imaging, \citet{2026arXiv260725044P} similarly find that a morphology-based classifier (\texttt{refExtendedness}) provides the most complete faint-end star selection for ultra-faint dwarf detection, complementing our result that morphology remains competitive with, and in the crowded and faint regimes more robust than, the color-plus-morphology ML classifier (Section~\ref{sec:crowded}).
In our setup FWHM dominates the marginal feature ranking while the color indices rank lower.  The colors nonetheless carry SED information that is genuinely complementary to morphology, as shown by the $\Delta F_1 = +0.035$ improvement of the full RF over the morphology-only baseline in Section~\ref{sec:baselines}; their lower marginal importance reflects the strong point/extended discrimination already provided by the light-profile morphology in this depth regime. Our finding that morphology dominates, with a strength that depends on depth and seeing, is consistent with the statistical framework of \citet{Slater2020}, who show that the widely used morphological separators (e.g.\ SDSS and SExtractor \texttt{spread\_model}) are closely related to one another and that their performance is governed primarily by signal-to-noise (limiting depth) and seeing.

\subsection{Comparison with Gatto et al.\ (2026)} \label{sec:gatto_comparison}

\citet{Gatto2026} performed a star-galaxy classification study on the same ECDFS field using DP1 data.
Their approach differs from ours in three key respects:
\begin{itemize}
    \item \textbf{Training labels}: \citet{Gatto2026} used spectroscopically confirmed sources from overlapping surveys; we use only RSP-resident products: Gaia/Monster \citep{DMTN277} for stars, and the DP1 ancillary photometric-redshift catalog of \citet{Charles2025} (with the $\sigma_{p(z)} < 0.15$ quality cut and $0.02 < z_\mathrm{mode} < 1.5$) for galaxies.
    \item \textbf{Method}: \citet{Gatto2026} focused on Random Forest with extensive feature engineering; we compare six methods using a common feature set.
    \item \textbf{Generalizability}: their spectroscopic labels avoid the label-generalization caveat but are tied to fields with overlapping spectroscopic surveys; our photo-$z$-anchored labels generalize directly to any LSST field where the \citet{Charles2025} photo-$z$ catalog (or its DP2 successor) is available, as demonstrated by the cross-field transfer of Section~\ref{sec:xfield}.
\end{itemize}
The two studies are complementary.
The Gatto et al.\ analysis provides the more rigorous ground-truth benchmark for the bright end of the ECDFS field; our analysis provides a multi-method comparison and a reproducible RSP workflow, that could be applied to fields for which no external spectroscopy exists.

The two approaches yield consistent qualitative conclusions about the relative importance of color and morphology features, and about the value of including photometric uncertainty. Concretely, our feature-configuration ablation of Section~\ref{sec:featconfig_results} adds the six per-band PSF magnitude uncertainties on top of the base 18-feature set and finds that the RF macro-$F_{1}$ improves from $0.922$ to $0.951$ under our balanced, photo-$z$-anchored training set (Table~\ref{tab:featconfig}), a gain of $+0.030$ that is qualitatively consistent with, and quantitatively larger than, the $\sim +0.002$ improvement \citet{Gatto2026} report from the same feature addition on their spectroscopically labeled sample. That the direction is the same across the two label regimes supports the more general point that per-band photometric-uncertainty features carry independent signal-to-noise information not fully captured by the flux measurements themselves; the larger amplitude in our case likely reflects the higher fraction of faint sources near the DP1 detection limit in the photo-$z$-anchored galaxy sample. We nevertheless quote the headline metrics on the base 18F configuration to keep the primary results transparent and to preserve strict independence between the feature set and the label-defining photo-$z$ pipeline.

\subsection{Limitations} \label{sec:limitations}

\textit{Photo-z labels.} We do not independently validate the FlexZBoost photo-z's against an external spec-z catalog and adopt the \citet{Charles2025} validation numbers ($\sigma = 0.026$, 19.2\% 3-$\sigma$ outlier fraction at six bands, $n = 2{,}905$); the $\sigma_{p(z)} < 0.15$ cut is expected to reject the catastrophic outliers.  Spectroscopic training samples from 4MOST, DESI, or PFS, where they exist, would provide an independent cross-validation that is not available in the spectroscopy-free setting demonstrated here.

\begin{figure*}[t]
\centering
\includegraphics[width=\textwidth]{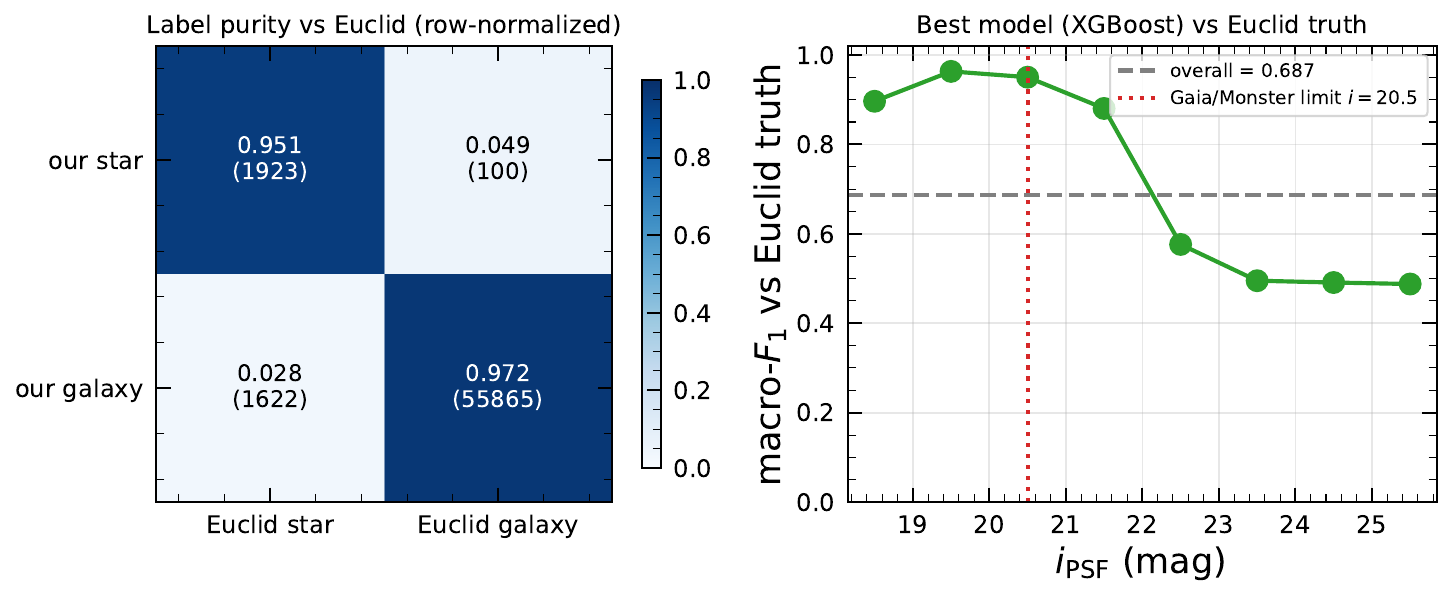}
\caption{Independent validation against the space-based Euclid Q1 MER classification in DP2 ECDFS (EDFS gives a consistent picture). \emph{Left:} row-normalized confusion between our spectroscopy-free labels and the Euclid classification, for labeled objects that Euclid classifies unambiguously; the diagonal purities are $0.95$--$0.97$ in both classes (stellar contamination of the galaxy labels, $\sim 3\%$, is confined to the bright end, $i \lesssim 22$). \emph{Right:} macro-$F_{1}$ of the best model (XGBoost) against the Euclid truth as a function of $i$-band magnitude, with the overall value (dashed) and the Gaia/Monster bright limit (dotted); agreement is excellent within the $i<23$ labeled regime and degrades beyond it, where the classifier extrapolates out of the training range and the Euclid morphological truth itself becomes less reliable.}
\label{fig:euclid_validation}
\end{figure*}

\textit{Label-magnitude asymmetry and transfer to faint sources.} The two label classes are drawn from different magnitude regimes. Point-source labels come from the Monster catalog, which is bright-limited (roughly $r \lesssim 19$--$21$), whereas galaxy labels extend to $i_{\rm cModel} < 23$; objects fainter than the Monster limit that also fail the photo-z cut are left unlabeled and used only for prediction, never for training. The training problem is not symmetric in magnitude, and a classifier can in principle learn a partial ``faint $\Rightarrow$ galaxy'' association from the availability of labels rather than from the photometry itself. We checked this directly by measuring per-magnitude completeness and contamination for all six classifiers in ECDFS DP1 and DP2: point-source completeness stays high ($\sim0.94$--$0.98$) and extended-source completeness above $0.92$ across the sampled range, and the threshold-independent AUC ($\sim0.99$) confirms the ranking is robust. Point-source \emph{purity} in the faintest bins is the main casualty of the asymmetry, but the sharp contamination rise in the faintest point-source bin is dominated by small-number statistics in the bright-limited Monster test sample rather than by genuine classifier degradation. Transfer to the faint, crowded regime is limited, as the 47~Tuc and Fornax test of Section~\ref{sec:crowded} makes explicit. A related concern is galaxy-label \emph{purity}: because the photo-$z$ galaxy criterion carries no explicit point-source veto beyond the bright Monster cross-match, a small number of faint stars or QSOs with confident but incorrect low-$z$ photo-$z$ estimates can enter the galaxy training set. We bound this with an independent cross-match to the space-based Euclid Q1 MER morphological classification\footnote{The Euclid Q1 MER (\emph{MERge}) catalog is the merged, multi-wavelength object catalog produced by the Euclid MERge processing function \citep{2026A&A...711A...4E}. We accessed it on the Rubin Science Platform with the \texttt{lsdb} package as a HATS catalog served from the IRSA cloud archive and cross-matched it to \texttt{dp2.Object} within $0.5''$. The star/galaxy separation follows the Euclid Q1 recommendations, the morphological \texttt{point\_like\_flag} refined by the PHZ SED-based galaxy probability \texttt{galaxysed\_phz\_gal\_prob},  as documented in the IRSA Euclid Q1 MER catalog tutorial (\url{https://caltech-ipac.github.io/irsa-tutorials/euclid-intro-mer-catalog/}) and the MER Morphology Cookbook (\url{https://euclid.esac.esa.int/dr/q1/dpdd/merdpd/mermorphologycookbook.html}).} (\texttt{point\_like\_flag}, refined by the SED galaxy probability \texttt{galaxysed\_phz\_gal\_prob}) in the ECDFS and EDFS overlap (Figure~\ref{fig:euclid_validation}). The photo-$z$ labels are clean: $96$--$97\%$ of our galaxy-labeled objects that Euclid classifies unambiguously are Euclid galaxies (stellar contamination $3$--$4\%$), and $95$--$97\%$ of our Monster star labels are Euclid point sources. The training galaxy set is, however, only a bright, high-reliability \emph{subset} of the true galaxy population: of all Euclid galaxies in the field, only $\sim 19\%$ pass the photo-$z$ reliability and $i_{\rm cModel}<23$ cuts and enter the labeled set (the remainder are fainter than the label cap or lack a reliable photo-$z$, and are used for prediction only). Comparing the best-model predictions to the Euclid truth as a function of magnitude, agreement is excellent within the labeled regime (macro-$F_{1}\approx0.95$--$0.98$ at $i\lesssim21$) but degrades to $\approx0.5$ beyond $i\approx23$; the high-latitude, $i<23$-trained classifier does not extrapolate reliably to the much fainter Euclid-depth population, consistent with the faint-regime and crowded-field limitations above. (Euclid's own morphological truth also degrades at the faint end, so this is a conservative bound rather than a pure classifier error.)

\textit{Independent truth.} Our completeness and purity are measured against Gaia/Monster stars and photo-z galaxies, both of which are themselves imperfect and, for the stars, bright. An external, deeper truth catalog would tighten these estimates and directly quantify how many true extended sources are missed by the photo-z-anchored labels. The Euclid Q1 release, which overlaps ECDFS and EDF-S and is accessible through LSDB on the Rubin Science Platform, is a promising independent-truth cross-match for a future iteration; we did not use it here in order to keep the workflow to products available directly on the public RSP.

\textit{Metric interpretation.}  The held-out test set retains the natural $\sim 10{:}1$ class imbalance, so point-source precision is sensitive to a small number of false positives drawn from the dominant extended class and the macro-$F_{1}$ conflates this precision penalty with genuine recall.  We therefore report both the balanced 10-fold cross-validation $F_{1}$ (directly comparable to metrics from spectroscopically labeled studies) and the imbalanced test $F_{1}$ (the operationally relevant number for the full catalog); the threshold-sensitivity sweep of Section~\ref{sec:threshold_sensitivity} bounds the dependence on the morphology-label threshold at $\Delta F_{1} \leq 0.09$.

\subsection{Future Work} \label{sec:future}

Several developments will extend this analysis as Rubin data releases mature.
The most immediate extension is to multi-class classification, separating stars, galaxies, and QSOs and ultimately resolving stellar and galaxy subtypes such as emission-line galaxies and AGN, following the approach of \citet{Zeraatgari2024} and \citet{Feng2025} on larger photometric samples.
A related direction is to expand the photometric feature space through cross-matching with WISE, Euclid, or Roman infrared data, both to improve QSO/star separation through the characteristic mid-infrared excess of obscured active galactic nuclei (AGN) and to enable classification of dust-obscured sources that are missed in the optical.
As spectroscopically confirmed labels from DESI, 4MOST, and MUSE become available across the LSST footprint, they will complement the spectroscopy-free workflow demonstrated here, allowing direct cross-validation of the labels in the regions where spectroscopy is available and providing an empirical bound on the residual label bias of the moments-based criterion elsewhere.
Beyond the tabular ML methods explored in this paper, convolutional neural networks trained on multi-band image cutouts may improve classification at the faintest magnitudes by leveraging spatial information beyond photometry and morphological summary statistics.

Finally, the incorporation of variability statistics from future Rubin time-domain data products will provide an essentially independent classification axis that should improve QSO and variable-star identification beyond what is achievable with single-epoch photometry.

\section*{Acknowledgments}
This work is dedicated to the memory of R\'egulo Plazas.

We thank the Rubin Observatory commissioning team and the LSST Science
Pipelines developers for producing the DP1 data products that this work
is built on, and we thank colleagues in the Rubin/LSST DESC and SAG
working groups for useful discussions. The work of AAPM was supported by the U.S.\ Department of Energy under contract number DE-AC02-76SF00515. AAPM thanks the Department of Physics of Harvard University and the Laboratory of Particle Astrophysics and Cosmology, the Cosmology Group at Boston University, and the Department of Physics at Washington University in St.\ Louis for their hospitality during the preparation of this paper. Y. Ordenes-Brice\~no acknowledges the support from the Comit\'e Mixto ESO-Chile.
This material or work is supported in part by the National Science
Foundation through Cooperative Agreement AST-1258333 and Cooperative
Support Agreement AST-1836783 managed by the Association of Universities
for Research in Astronomy (AURA), and by the Department of Energy under
Contract No.\ DE-AC02-76SF00515 with the SLAC National Accelerator
Laboratory managed by Stanford University.

This paper makes use of software developed for Vera C.\ Rubin Observatory. We thank the Rubin Observatory for making their code available as free software at \url{https://pipelines.lsst.io}.
The analysis was carried out on the Rubin Science Platform (RSP) at
\url{https://data.lsst.cloud}.

This work has made use of data from the European Space Agency (ESA)
mission \emph{Gaia} (\url{https://www.cosmos.esa.int/gaia}), processed
by the \emph{Gaia} Data Processing and Analysis Consortium
(DPAC, \url{https://www.cosmos.esa.int/web/gaia/dpac/consortium}).

\vskip1ex\noindent\textbf{Facilities:} \facility{Rubin:Simonyi (LSSTComCam)}, \facility{Rubin:USDAC}

\software{Astropy \citep{AstropyCollab2022}, NumPy \citep{Harris2020}, pandas \citep{McKinney2010}, matplotlib \citep{Hunter2007}, scikit-learn \citep{Pedregosa2011}, XGBoost \citep{ChenGuestrin2016}, imbalanced-learn \citep{Lemaitre2017}, UMAP \citep{McInnes2018}}

During the preparation of this work, Claude Code (Anthropic; Claude Opus and Sonnet 4-series models) and ChatGPT (OpenAI; GPT-4-series models) were used both to
help overcome language barriers associated with the use of English as an
additional language by improving grammar and writing, and to assist with development and debugging of the analysis code; all code and results were reviewed and validated by the authors. After using these tools, the authors reviewed and edited the content as needed and take full responsibility for the final publication.

\section*{Data Availability}
This paper uses Rubin Observatory DP1 and DP2 data products, which are available to LSST Data Rights Holders via the Rubin Science Platform at \url{https://data.lsst.cloud}. The TAP and LSDB queries used to retrieve the \texttt{dp1.Object} and \texttt{dp2.Object} catalogs and the photometric-redshift catalogs of \citet{Charles2025} (DP1) and the RAIL product (DP2) are reproduced in Appendix~\ref{sec:adql}. The Jupyter notebooks implementing the full classification pipeline (one per field for DP1 and DP2, plus the cross-field comparison), the figure-generation scripts, and the classified source catalogs will be archived together with the published version of this paper on Zenodo.

\begin{appendix}

\section{ADQL Query} \label{sec:adql}
The following ADQL query was used to retrieve sources from the \texttt{dp1.Object} table:

\begin{verbatim}
SELECT objectId, coord_ra, coord_dec,
  ebv, refBand,
  refExtendedness, refSizeExtendedness,
  u_psfMag, g_psfMag, r_psfMag,
  i_psfMag, z_psfMag, y_psfMag,
  u_psfMagErr, g_psfMagErr, r_psfMagErr,
  i_psfMagErr, z_psfMagErr, y_psfMagErr,
  u_cModelMag, g_cModelMag, r_cModelMag,
  i_cModelMag, z_cModelMag, y_cModelMag,
  i_ixx, i_iyy, i_ixy,
  i_kronRad, i_blendedness,
  u_psfFlux, u_psfFluxErr,
  g_psfFlux, g_psfFluxErr,
  r_psfFlux, r_psfFluxErr,
  i_psfFlux, i_psfFluxErr,
  z_psfFlux, z_psfFluxErr,
  y_psfFlux, y_psfFluxErr
FROM dp1.Object
WHERE CONTAINS(
  POINT('ICRS', coord_ra, coord_dec),
  CIRCLE('ICRS', 53.13, -28.10, 1.0)
) = 1
  AND u_psfFlux_flag = 0
  AND g_psfFlux_flag = 0
  AND r_psfFlux_flag = 0
  AND i_psfFlux_flag = 0
  AND z_psfFlux_flag = 0
  AND y_psfFlux_flag = 0
  AND i_cModel_flag = 0
  AND i_pixelFlags_saturatedCenter = 0
  AND i_pixelFlags_crCenter = 0
  AND i_pixelFlags_interpolatedCenter = 0
  AND i_pixelFlags_sensor_edgeCenter = 0
  AND i_pixelFlags_clippedCenter = 0
  AND (
    (refBand='u' AND u_sizeExtendedness_flag=0) OR
    (refBand='g' AND g_sizeExtendedness_flag=0) OR
    (refBand='r' AND r_sizeExtendedness_flag=0) OR
    (refBand='i' AND i_sizeExtendedness_flag=0) OR
    (refBand='z' AND z_sizeExtendedness_flag=0) OR
    (refBand='y' AND y_sizeExtendedness_flag=0)
  )
ORDER BY objectId ASC
\end{verbatim}

For DP2 we run the analogous query on the \texttt{dp2.Object} table. The main differences from the DP1 query above are: (i) the table name and the field center; (ii) DP2 uses the per-band \texttt{\{band\}\_psfMagErr} and \texttt{\{band\}\_cModelMagErr} columns directly rather than deriving errors from PSF fluxes; (iii) DP2 selects the per-object visit count \texttt{\{band\}\_inputCount}, used for the survey-homogeneity cut of Section~\ref{sec:dp2} (DP1 instead builds a visit-count map from a separate \texttt{dp1.CcdVisit} query); (iv) DP2 adds the model-based classifiers \texttt{griz\_model\_extendedness} and \texttt{g\_sizeExtendedness}, and, for the S\'ersic ablation of Section~\ref{sec:dp2}, \texttt{sersic\_reff\_major} and the per-band \texttt{\{band\}\_extendedness}; and (v) the reference-band \texttt{sizeExtendedness\_flag} clause is not applied for DP2. The bright-end saturation cuts ($g \geq 15.70$, $r \geq 15.08$, $i \geq 15.19$; Section~\ref{sec:query}) are applied in code after the query in both releases. The DP2 query (COSMOS center shown) is:

\begin{verbatim}
SELECT objectId, coord_ra, coord_dec, tract, patch,
  ebv, refBand,
  refExtendedness, refSizeExtendedness,
  griz_model_extendedness, g_sizeExtendedness,
  sersic_reff_major,
  i_ixx, i_iyy, i_ixy, i_kronRad, i_blendedness,
  u_psfMag, g_psfMag, r_psfMag,
  i_psfMag, z_psfMag, y_psfMag,
  u_psfMagErr, g_psfMagErr, r_psfMagErr,
  i_psfMagErr, z_psfMagErr, y_psfMagErr,
  u_cModelMag, g_cModelMag, r_cModelMag,
  i_cModelMag, z_cModelMag, y_cModelMag,
  u_cModelMagErr, g_cModelMagErr, r_cModelMagErr,
  i_cModelMagErr, z_cModelMagErr, y_cModelMagErr,
  u_inputCount, g_inputCount, r_inputCount,
  i_inputCount, z_inputCount, y_inputCount,
  u_extendedness, g_extendedness, r_extendedness,
  i_extendedness, z_extendedness, y_extendedness
FROM dp2.Object
WHERE CONTAINS(
  POINT('ICRS', coord_ra, coord_dec),
  CIRCLE('ICRS', 150.10, 2.10, 1.0)
) = 1
  AND u_psfFlux_flag = 0
  AND g_psfFlux_flag = 0
  AND r_psfFlux_flag = 0
  AND i_psfFlux_flag = 0
  AND z_psfFlux_flag = 0
  AND y_psfFlux_flag = 0
  AND i_cModel_flag = 0
  AND i_pixelFlags_saturatedCenter = 0
  AND i_pixelFlags_crCenter = 0
  AND i_pixelFlags_interpolatedCenter = 0
  AND i_pixelFlags_sensor_edgeCenter = 0
  AND i_pixelFlags_clippedCenter = 0
\end{verbatim}
The survey-homogeneity cut ($i$\_inputCount $\geq 0.85 \times$ the field median) is then applied in code (Section~\ref{sec:dp2}).

\section{Permutation Importance} \label{sec:perm_fig}

\begin{figure}
\centering
\includegraphics[width=\columnwidth]{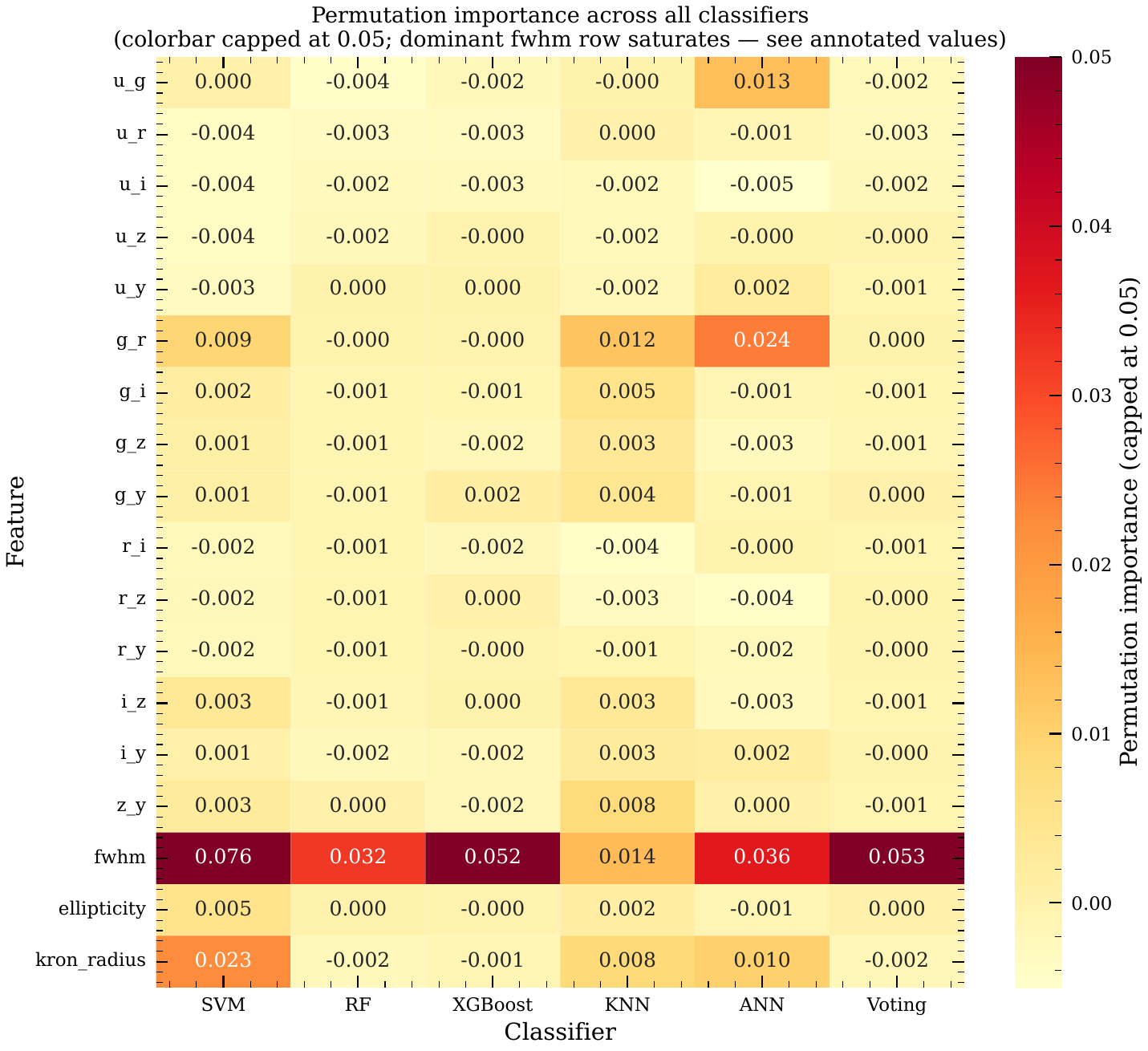}
\caption{Permutation importance heatmap across all six classifiers.
FWHM, $g\!-\!r$, and Kron radius are consistently the most important features. See details in Section~\ref{sec:importance}}
\label{fig:perm_importance}
\end{figure}

Figure~\ref{fig:perm_importance} shows the model-agnostic permutation importance (Section~\ref{sec:importance}) across all six classifiers for the ECDFS DP1 training set. Each cell is the mean decrease in macro-$F_{1}$ when a single feature's values are randomly shuffled, which breaks that feature's association with the label; larger values indicate greater reliance on the feature. FWHM is the dominant discriminator for every method, with Kron radius and $(g-r)$ the leading secondary features.

\section{Confusion Matrices and Performance Comparison} \label{sec:CM_roc_fig}

\begin{figure*}[t]
\centering
\includegraphics[width=\textwidth]{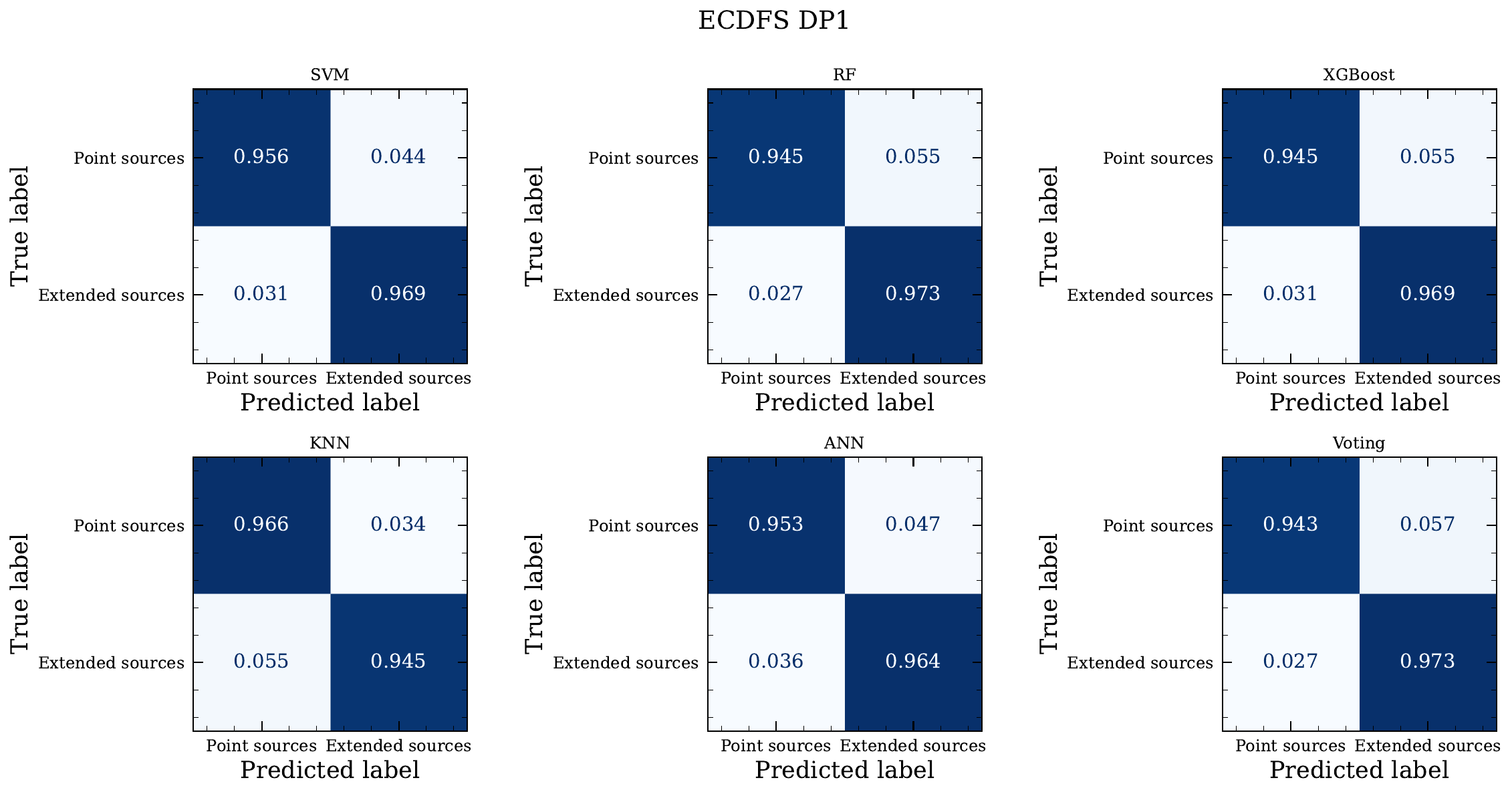}\\
\includegraphics[trim=0cm 0cm 0cm -1.2cm,clip,width=\textwidth]{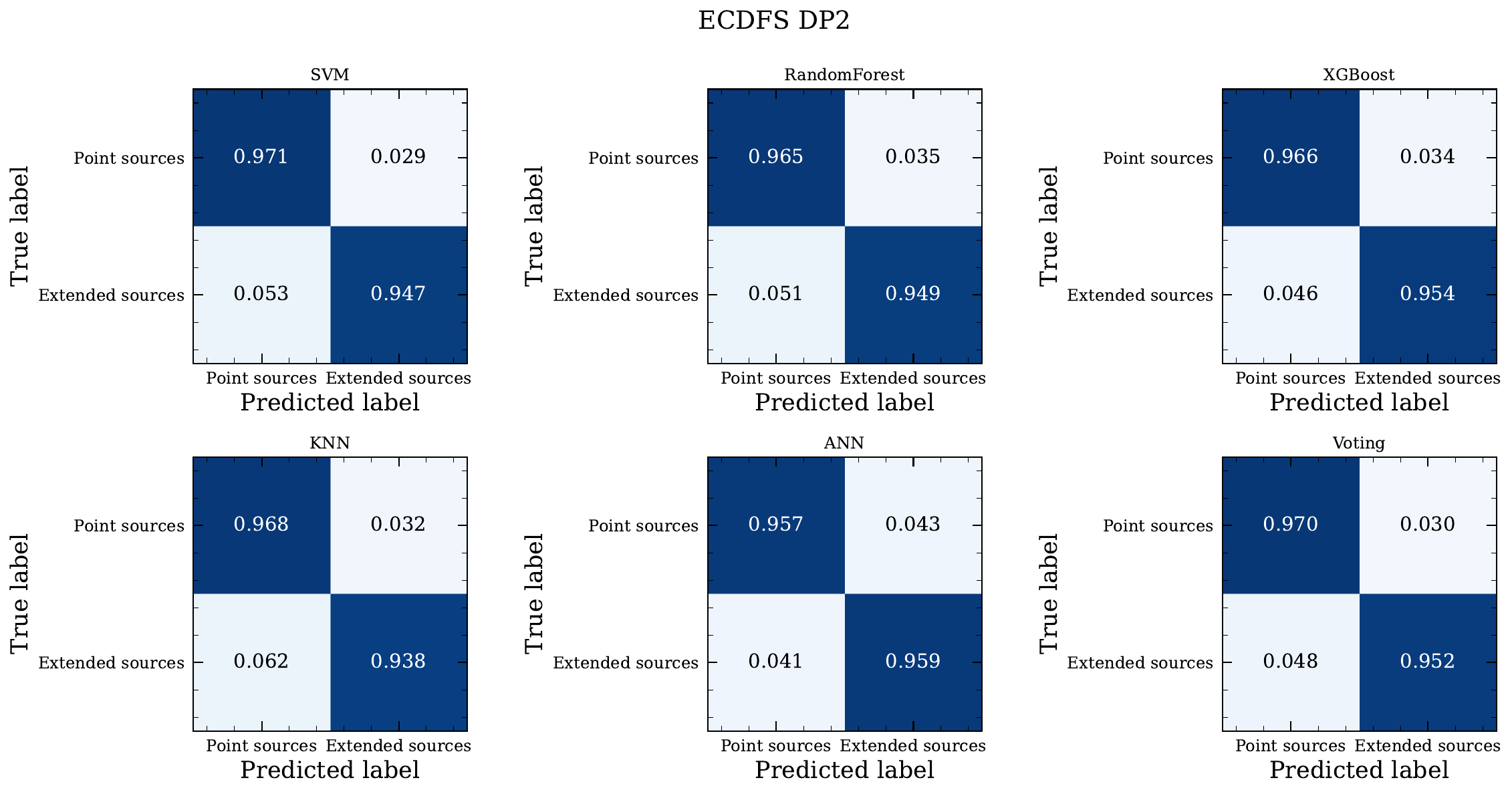}\\
\caption{Normalized confusion matrices for all six classifiers \newtext{for DP1 (top panels) and DP2 (bottom panels)}.
Each row sums to 1.0 and represents the true class.
The diagonal shows the recall (true positive rate) for each class.}
\label{fig:confusion}
\end{figure*}

Figure~\ref{fig:confusion} shows the normalized confusion matrices for all six classifiers in ECDFS, for DP1 (top) and DP2 (bottom). Each row is normalized to its true class, so the diagonal entries are the per-class recall discussed in Section~\ref{sec:confusion_roc}; the dominant off-diagonal term is extended sources misclassified as point sources, which drives the point-source precision penalty under the natural class imbalance.

\section{Application to other Fields} \label{sec:crossfield_fig}

\begin{figure*}[t]
\centering
\includegraphics[width=\columnwidth]{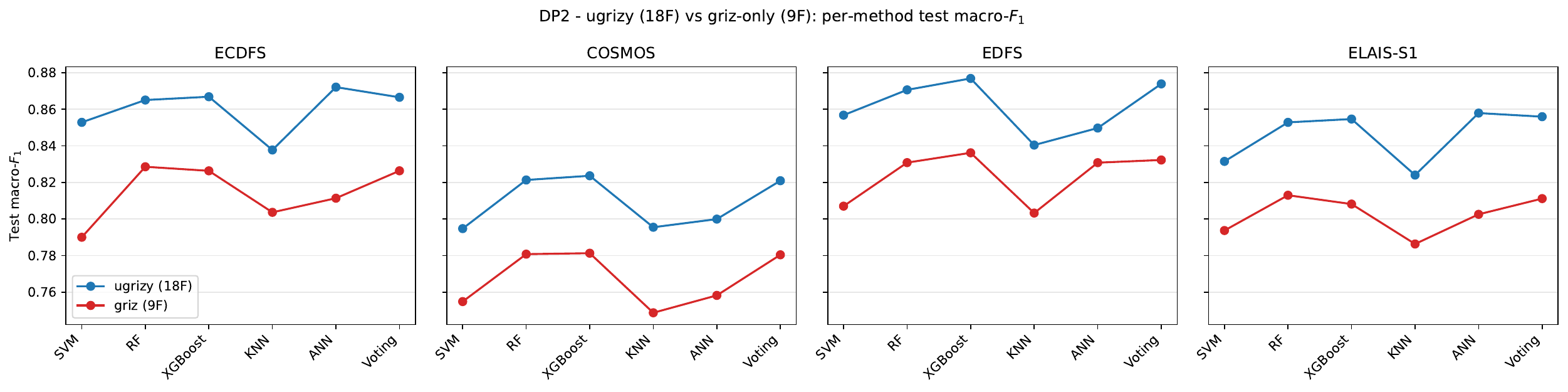}
\caption{DP2: per-method test macro-$F_{1}$ for the six classifiers, comparing the full $ugrizy$ feature set (18F, blue) with a griz-only ablation (9F, red), one subplot per DP2 field. Dropping $u$ and $y$ lowers the macro-$F_{1}$ by a nearly uniform $\sim 4\%$ across all fields and methods; the field ordering is preserved, showing that the cross-field $F_{1}$ differences are population- and depth-driven rather than a $u$/$y$ feature-richness effect.}
\label{fig:ugrizy_vs_grizonly_f1}
\end{figure*}

\begin{figure*}[t]
\centering
\includegraphics[width=0.8\textwidth]{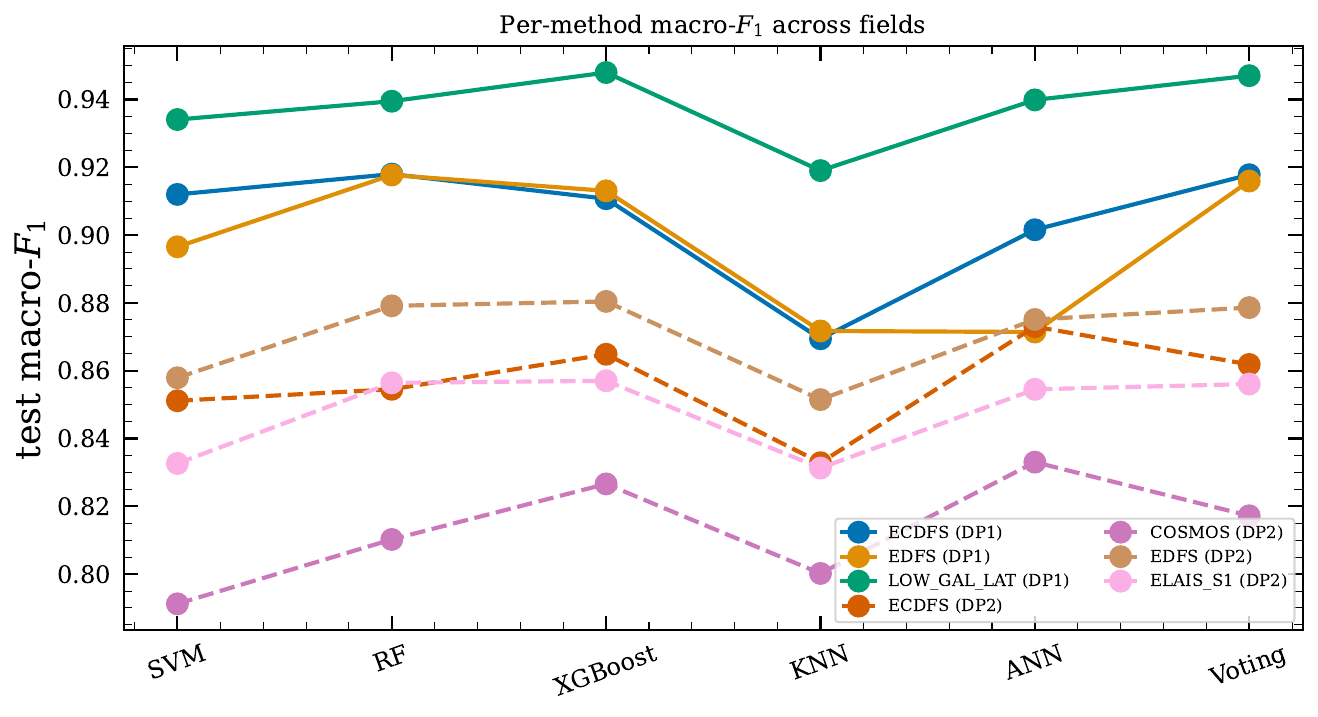}
\caption{Per-method test macro-$F_{1}$ across all seven labeled fields, with the three DP1 fields drawn as solid lines and the four DP2 fields as dashed lines. The DP1 fields (ECDFS, EDFS, Low Galactic Latitude) sit above the DP2 fields, consistent with the shallower commissioning-era DP2 depth, and the method ordering (tree ensembles best, KNN weakest) is preserved across releases and fields.}
\label{fig:crossfield_f1}
\end{figure*}

This appendix collects the cross-field comparison figures. Figure~\ref{fig:ugrizy_vs_grizonly_f1} compares the full \ugrizy{} (18F) feature set against a $griz$-only (9F) ablation across the four DP2 fields, one panel per field (Section~\ref{sec:xfield}): dropping the $u$ and $y$ bands lowers the macro-$F_{1}$ by a small, roughly uniform amount ($\sim 0.04$) without changing the ranking of methods or fields. Figure~\ref{fig:crossfield_f1} shows the per-method test macro-$F_{1}$ across all seven labeled fields, with the three DP1 fields drawn as solid lines and the four DP2 fields as dashed lines.

\end{appendix}

\bibliographystyle{aasjournalv7}
\bibliography{main}

\end{document}